\documentclass[a4paper,11pt]{article}
\usepackage{jheppub_mod}
\usepackage[T1]{fontenc}

\usepackage{hyperref}
\usepackage{graphicx}
\usepackage{amsmath,amssymb,mathtools,slashed}
\usepackage{booktabs,tabulary}
\usepackage{dsfont}
\usepackage{color}
\usepackage{multirow}
\usepackage{subcaption}
\usepackage{braket}

\newcommand{\mpi}{M_\pi}

\newcommand{\beq}{\begin{equation}}
\newcommand{\eeq}{\end{equation}}
\newcommand{\diff}{\text{d}}
\newcommand{\eps}{\epsilon}

\newcommand{\Order}{\mathcal{O}}

\newcommand{\GeV}{\,\text{GeV}}
\newcommand{\MeV}{\,\text{MeV}}
\newcommand{\keV}{\,\text{keV}}
\renewcommand{\Im}{\text{Im}\,}
\renewcommand{\Re}{\text{Re}\,}

\newcommand{\etapp}{\eta^{(\prime)}}
\newcommand{\sm}{s_\text{m}}
\newcommand{\Br}{\text{Br}}

\allowdisplaybreaks[1]

\title{Dispersion relation for hadronic light-by-light scattering: $\boldsymbol{\eta}$ and $\boldsymbol{\eta'}$ poles}

\author[a]{Simon Holz,}
\author[a]{Martin Hoferichter,}
\author[a,b]{Bai-Long Hoid,}
\author[c]{and Bastian Kubis}

\affiliation[a]{
Albert Einstein Center for Fundamental Physics, Institute for Theoretical Physics, University of Bern, Sidlerstrasse 5, 3012 Bern, Switzerland}

\affiliation[b]{
Institut f\"ur Kernphysik and PRISMA$^+$  Cluster of Excellence, Johannes Gutenberg Universit\"at,  55099 Mainz, Germany} 

\affiliation[c]{
Helmholtz-Institut f\"ur Strahlen- und Kernphysik (Theorie) and \\
Bethe Center for Theoretical Physics, Universit\"at Bonn, 53115 Bonn, Germany}

\emailAdd{holz@itp.unibe.ch}
\emailAdd{hoferichter@itp.unibe.ch}
\emailAdd{longbai@itp.unibe.ch}
\emailAdd{kubis@hiskp.uni-bonn.de}

\abstract{The pseudoscalar-pole
contributions to hadronic light-by-light scattering are determined by the respective transition form factors (TFFs) into two virtual photons. These TFFs constitute complicated functions of the photon virtualities that, in turn, can be reconstructed in a dispersive approach from their discontinuities. In this work, we present such an analysis for the $\etapp$ TFFs, implementing a number of constraints from both experiment and theory:
 normalizations from the $\etapp\to\gamma\gamma$ decay widths, unitarity constraints from the $\etapp\to\pi^+\pi^-\gamma$ spectra, chiral symmetry for the $\etapp\to2(\pi^+\pi^-)$ amplitudes, vector-meson couplings, singly-virtual data from $e^+e^-\to e^+e^-\etapp$,  and the asymptotic behavior predicted by the light-cone expansion.
 In particular, we account for the leading left-hand-cut singularity by including effects from the $a_2$ resonance, necessitating the solution of an inhomogeneous Muskhelishvili--Omn\`es problem via a carefully chosen path deformation.
 The resulting TFFs allow us to evaluate the $\etapp$-pole contributions to the anomalous magnetic moment of the muon,
 $a_\mu^{\eta\text{-pole}}=14.7(9)\times 10^{-11}$ and $a_\mu^{\eta'\text{-pole}}=13.5(7)\times 10^{-11}$,
 completing a dedicated program for the lowest-lying pseudoscalar intermediate states in a dispersive approach to hadronic light-by-light scattering,
  $a_\mu^{\text{PS-poles}}=91.2^{+2.9}_{-2.4}\times 10^{-11}$.}

\begin{document} 

\maketitle

\section{Introduction}
\label{sec:intro}

The decay of a pseudoscalar meson $P$ into virtual photons, $P(q_1+q_2)\to\gamma^*(q_1,\mu)\gamma^*(q_2,\nu)$, is described by a transition form factor (TFF) $F_{P\gamma^*\gamma^*}(q_1^2,q_2^2)$ via the matrix element
\beq
\label{TFF_def}
i \int  \text{d}^4x\, e^{iq_1 \cdot x} \langle 0 | T\{ j_{\mu}(x)j_{\nu}(0) \}  | P(q_1+q_2) \rangle
=  \epsilon_{\mu\nu\rho\sigma}q_1^{\rho}q_2^{\sigma} F_{P\gamma^*\gamma^*}(q_1^2,q_2^2),
\eeq
where
\beq
j^\mu(x)=\bar q(x){\mathcal Q}\gamma^\mu q(x),\qquad q=(u,d,s)^T,
\qquad
{\mathcal Q} = \frac{1}{3}\text{diag}(2,-1,-1),
\eeq
denotes the electromagnetic current and $\eps^{0123}=+1$. The normalizations of the TFFs are directly related to the two-photon decay widths
\beq
\Gamma[P\to\gamma\gamma]=\frac{\pi\alpha^2M_P^3}{4}F_{P\gamma\gamma}^2,\qquad
F_{P\gamma\gamma}=F_{P\gamma^*\gamma^*}(0,0), \qquad \alpha=\frac{e^2}{4\pi}.
\eeq
Due to the pseudo-Goldstone-boson nature of the $\pi^0$, the Wess--Zumino--Witten (WZW) anomaly~\cite{Wess:1971yu,Witten:1983tw} implies a powerful low-energy theorem
\beq
\label{Fpigg_WZW}
F_{\pi\gamma\gamma}^\text{WZW}=\frac{1}{4\pi^2 F_\pi}=0.2744(3)\GeV^{-1},
\eeq
with pion decay constant $F_\pi=92.32(10)\MeV$~\cite{ParticleDataGroup:2024cfk}, which already agrees so well with experiment, $F^\text{exp}_{\pi\gamma\gamma} =0.2754(21)\GeV^{-1}$ as measured by PrimEx-II~\cite{PrimEx-II:2020jwd} via the Primakoff process, that higher-order chiral corrections tend to generate a tension~\cite{Bijnens:1988kx,Goity:2002nn,Ananthanarayan:2002kj,Kampf:2009tk,Gerardin:2019vio}. For $\etapp$, the normalizations are known to~\cite{ParticleDataGroup:2024cfk}
\beq
\label{F_etagg_etapgg}
F^\text{exp}_{\eta\gamma\gamma}=0.2736(48)\GeV^{-1},\qquad
F^\text{exp}_{\eta'\gamma\gamma}=0.3437(55)\GeV^{-1},
\eeq
derived from $e^+e^-\to e^+e^-\eta$~\cite{JADE:1985biu,CrystalBall:1988xvy,Roe:1989qy,Baru:1990pc,KLOE-2:2012lws} and the global $\eta'$ fit of the Review of Particle Physics (RPP)~\cite{ParticleDataGroup:2024cfk}, respectively, the latter being consistent with the average of direct determinations from $e^+e^-\to e^+e^-\eta'$~\cite{CrystalBall:1988xvy,TPCTwoGamma:1988izb,Roe:1989qy,Butler:1990vv,Baru:1990pc,CELLO:1990klc,CrystalBall:1991zkb,L3:1997ocz}. In these cases, a WZW interpretation as simple as Eq.~\eqref{Fpigg_WZW} is not available, since $\eta$--$\eta'$ mixing has to be taken into account, to the extent that the experimental normalizations~\eqref{F_etagg_etapgg} actually constitute a relatively clean way to help determine decay constants and mixing parameters~\cite{Escribano:2015yup,Gan:2020aco}. Accordingly, for our study Eq.~\eqref{F_etagg_etapgg} serves as an important constraint for the TFF normalizations, independent of assumptions on the $\eta$--$\eta'$ mixing pattern.

Beyond the normalizations, the TFFs fulfill a number of constraints that help reconstruct their momentum dependence. Asymptotically, the behavior of the TFFs is predicted by the light-cone expansion~\cite{Lepage:1979zb,Lepage:1980fj,Brodsky:1981rp,Novikov:1983jt,Nesterenko:1982dn,Gorsky:1987idk}, while in between the singularities of the TFFs allow one to infer the momentum dependence via dispersion relations. This program was carried out in great detail for the $\pi^0$ in Refs.~\cite{Schneider:2012ez,Hoferichter:2012pm,Hoferichter:2014vra,Hoferichter:2018dmo,Hoferichter:2018kwz,Hoferichter:2021lct}, and steady progress towards a similarly comprehensive analysis for $\etapp$ was achieved in
Refs.~\cite{Stollenwerk:2011zz,Hanhart:2013vba,Kubis:2015sga,Holz:2015tcg,Holz:2022hwz,Holz:2022smu}, culminating in Ref.~\cite{Holz:2024lom}. Here, we provide a detailed account of this calculation.

The primary motivation for such detailed studies of the $P=\pi^0,\eta,\eta'$ TFFs derives from hadronic light-by-light (HLbL) scattering, given that the pseudoscalar poles constitute the leading singularities of the HLbL tensor, whose strength is determined by the TFFs. A robust evaluation of the $\etapp$ TFFs is thus an essential ingredient for a complete dispersive analysis of HLbL scattering~\cite{Hoferichter:2024vbu,Hoferichter:2024bae}, to consolidate and improve upon the previous white-paper consensus $a_\mu^\text{HLbL} = 92(19) \times 10^{-11}$~\cite{Aoyama:2020ynm,Melnikov:2003xd,Masjuan:2017tvw,Colangelo:2017qdm,Colangelo:2017fiz,Hoferichter:2018dmo,Hoferichter:2018kwz,Gerardin:2019vio,Bijnens:2019ghy,Colangelo:2019lpu,Colangelo:2019uex,Pauk:2014rta,Danilkin:2016hnh,Jegerlehner:2017gek,Knecht:2018sci,Eichmann:2019bqf,Roig:2019reh}. In particular, to not only match the current experimental precision~\cite{Muong-2:2023cdq,Muong-2:2024hpx},
but also the further advances expected from the final result of the Fermilab experiment~\cite{Muong-2:2015xgu}, HLbL scattering requires at least a two-fold improvement in precision as well. In addition to efforts aimed at
 subleading effects from hadronic states at intermediate energies~\cite{Hoferichter:2020lap,Zanke:2021wiq,Danilkin:2021icn,Stamen:2022uqh,Ludtke:2023hvz,Hoferichter:2023tgp,Hoferichter:2024fsj,Ludtke:2024ase,Deineka:2024mzt,Hoferichter:2025yih}, higher-order short-distance constraints~\cite{Bijnens:2020xnl,Bijnens:2021jqo,Bijnens:2022itw,Bijnens:2024jgh},
 and the matching between hadronic and short-distance realizations~\cite{Leutgeb:2019gbz,Cappiello:2019hwh,Knecht:2020xyr,Masjuan:2020jsf,Ludtke:2020moa,Colangelo:2021nkr,Leutgeb:2021mpu,Leutgeb:2022lqw,Colangelo:2024xfh},
completing a dispersive evaluation of the pseudoscalar poles is thus imperative. These efforts are complementary to recent calculations in lattice QCD~\cite{Blum:2019ugy,Chao:2021tvp,Chao:2022xzg,Blum:2023vlm,Fodor:2024jyn}, and should proceed in parallel to ongoing work to try and resolve the complicated situation regarding hadronic vacuum polarization~\cite{Colangelo:2022jxc}, i.e., the tensions among data-driven evaluations~\cite{Davier:2017zfy,Keshavarzi:2018mgv,Colangelo:2018mtw,Hoferichter:2019gzf,Davier:2019can,Keshavarzi:2019abf,Hoid:2020xjs,Crivellin:2020zul,Keshavarzi:2020bfy,Malaescu:2020zuc,Colangelo:2020lcg,Stamen:2022uqh,Colangelo:2022vok,Colangelo:2022prz,Hoferichter:2022iqe,Hoferichter:2023sli,Hoferichter:2023bjm,Stoffer:2023gba,Davier:2023fpl,CMD-3:2023alj,CMD-3:2023rfe,Leplumey:2025kvv}  and with lattice QCD~\cite{Borsanyi:2020mff,Ce:2022kxy,ExtendedTwistedMass:2022jpw,FermilabLatticeHPQCD:2023jof,RBC:2023pvn,Boccaletti:2024guq,Blum:2024drk,Djukanovic:2024cmq,Bazavov:2024eou}, including renewed scrutiny of
radiative corrections~\cite{Campanario:2019mjh,Ignatov:2022iou,Colangelo:2022lzg,Monnard:2021pvm,Abbiendi:2022liz,BaBar:2023xiy,Aliberti:2024fpq}. Together with the well-established  QED~\cite{Aoyama:2012wk,Aoyama:2019ryr} and electroweak~\cite{Czarnecki:2002nt,Gnendiger:2013pva} contributions as well as higher-order hadronic effects~\cite{Calmet:1976kd,Kurz:2014wya,Colangelo:2014qya,Hoferichter:2021wyj}, these community-wide efforts are necessary to bring the leading hadronic contributions under control.

For the $\etapp$ poles, the current situation is less consolidated than for the $\pi^0$ pole, which was estimated using a dispersive approach~\cite{Hoferichter:2018dmo,Hoferichter:2018kwz}, Canterbury approximants (CA)~\cite{Masjuan:2017tvw}, and within lattice QCD~\cite{Gerardin:2019vio}.\footnote{Work is ongoing to clarify a potential deficit
between more recent lattice-QCD calculations~\cite{Christ:2022rho,Gerardin:2023naa,ExtendedTwistedMass:2023hin,Lin:2024khg} and the PrimEx-II normalization~\cite{PrimEx-II:2020jwd}.} While first lattice-QCD calculations have become available for $\etapp$ as well~\cite{ExtendedTwistedMass:2022ofm,Gerardin:2023naa}, at present the most precise results can be expected relying on as much experimental input as possible, for which a dispersive approach is ideally suited. After reviewing the master formula for pseudoscalar poles in Sec.~\ref{sec:pole}, we lay out the dispersive formalism in detail in Sec.~\ref{sec:DR}, starting with the isovector TFFs and their relation to the
$\etapp\to\pi^+\pi^-\gamma$ spectra, the $\eta'\to2(\pi^+\pi^-)$ amplitude, and the solution of the inhomogeneous Muskhelishvili--Omn\`es (MO) problem in the presence of an $a_2$ left-hand cut. Isoscalar contributions and effective poles are discussed in Sec.~\ref{sec:isoscalar}, the latter serving as a means to implement the effects of higher intermediate states not explicitly included, and thereby interpolate to the asymptotic behavior, to which we match in Sec.~\ref{sec:SDC}.
 The main numerical results are reported in Sec.~\ref{sec:num}, before we conclude in Sec.~\ref{sec:summary}.  Further details of the calculation are collected in the appendices.

\section{Pseudoscalar-pole contributions to $\boldsymbol{a_\mu}$}
\label{sec:pole}

The general master formula for the HLbL contributions to $a_\mu$ can be written in the form~\cite{Colangelo:2015ama,Colangelo:2017fiz}
\begin{equation}
\label{eq:master_formula}
a_\mu^{\text{HLbL}} = \frac{2\alpha^3}{3\pi^2}\int_0^\infty \diff Q_1\int_0^\infty \diff Q_2\int_{-1}^1 \diff \tau\sqrt{1-\tau^2}\,Q_1^3\,Q_2^3\sum_{i=1}^{12}T_i(Q_1,Q_2,Q_3)\,\bar{\Pi}_i(Q_1,Q_2,Q_3),
\end{equation}
where $\tau$ is the cosine of the remaining angle between $Q_1$ and $Q_2$, and $Q_3^2 = Q_1^2 + 2 Q_1 Q_2 \tau + Q_2^2$. This decomposition~\eqref{eq:master_formula} isolates the kinematic aspects of the HLbL tensor into known kernel functions $T_i$, while the dynamical content of the theory is contained in the scalar functions $\bar\Pi_i$. In particular, in a dispersive approach to HLbL scattering~\cite{Colangelo:2014dfa,Colangelo:2014pva,Colangelo:2015ama,Hoferichter:2013ama,Colangelo:2017fiz}, the scalar functions are reconstructed via their singularities, the leading ones originating from pseudoscalar poles according to
\begin{align}
\label{Eq:Pi_pole}
\bar\Pi_1(Q_1,Q_2,\tau)&=-\frac{F_{P\gamma^*\gamma^*}(-Q_1^2,-Q_2^2)F_{P\gamma^*\gamma^*}(-Q_3^2,0)}{Q_3^2+M_P^2}, \notag\\
\bar \Pi_2(Q_1,Q_2,\tau)&=-\frac{F_{P\gamma^*\gamma^*}(-Q_1^2,-Q_3^2)F_{P\gamma^*\gamma^*}(-Q_2^2,0)}{Q_2^2+M_P^2},
\end{align}
with TFFs as defined in Eq.~\eqref{TFF_def}. This form of the pseudoscalar-pole contributions can be obtained from standard techniques~\cite{Knecht:2001qg,Knecht:2001qf,Jegerlehner:2009ry}, integrating over the angles using Gegenbauer polynomials~\cite{Rosner:1967zz,Levine:1974xh,Levine:1979uz}. More precisely, the form in Eq.~\eqref{Eq:Pi_pole} follows when performing dispersion relations in four-point kinematics, before taking the external photon momentum to zero, while in triangle kinematics the arguments of the singly-virtual TFFs change according to $Q_i^2\to-M_P^2$~\cite{Colangelo:2019uex,Knecht:2020xyr,Ludtke:2023hvz}.

\section{Dispersion relations for the isovector transition form factors}
\label{sec:DR}

\subsection{Overview of dispersive approach}

\begin{figure}[tb]
    \centering
    \includegraphics[width=.8\linewidth]{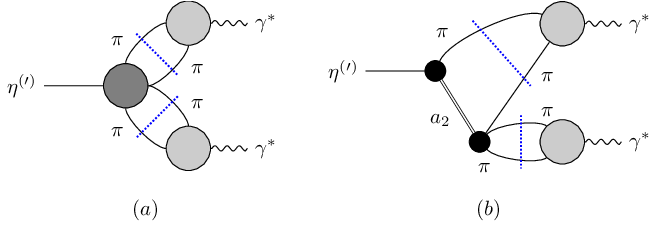}
    \caption{Isovector $\eta$ and $\eta'$ TFFs $(a)$ in a diagrammtic representation. The left-hand-cut contribution due to exchange of the $a_2(1320)$ tensor meson is shown in $(b)$. Unitarity cuts are indicated by the dotted blue lines.}
    \label{Fig:tffi1_general}
 \end{figure}
The QCD vertex function of the decay $P(q_1+q_2)\to\gamma^*(q_1,\mu)\gamma^*(q_2,\nu)$ defines the TFF as anticipated in Eq.~\eqref{TFF_def}. For the evaluation of the HLbL integral~\eqref{eq:master_formula} we need to reconstruct its full doubly-virtual dependence, for which the following decomposition proves useful:
 \beq
     F_{\etapp \gamma^*\gamma^*}(q_1^2,q_2^2) = F_{\etapp}^{(I=1)}(q_1^2,q_2^2) + F_{\etapp}^{(I=0)}(q_1^2,q_2^2) + F_{\etapp}^{\text{eff}}(q_1^2,q_2^2) + F_{\etapp}^{\text{asym}}(q_1^2,q_2^2). \label{Eq:tff_compl}
 \eeq
 In contrast to the $\pi^0$, the vanishing isospin $I=0$ of $\etapp$ implies that either both photons have to be isovector or both isoscalar. The corresponding low-energy contributions are denoted by $F_{\etapp}^{(I=1)}$ and $F_{\etapp}^{(I=0)}$ in Eq.~\eqref{Eq:tff_compl}, respectively. In the remainder of this section, starting from the underlying $\eta^\prime \to2(\pi^+\pi^-)$ decay amplitude, we present a detailed dispersive analysis of $F_{\etapp}^{(I=1)}$, which, numerically, constitutes the largest contribution. The dispersive analysis proceeds via the diagram shown in Fig.~\ref{Fig:tffi1_general}$(a)$ and, additionally, takes into account factorization-breaking effects in the dependence of the TFFs on the two photon virtualities $q_{1/2}^2$ via the left-hand-cut contribution shown in  Fig.~\ref{Fig:tffi1_general}$(b)$. The effective poles, represented by $F_{\etapp}^{\text{eff}}$, are introduced to impose the correct normalizations and to interpolate to the asymptotic region; they are discussed together with $F_{\etapp}^{(I=0)}$ in Sec.~\ref{sec:isoscalar}. Finally, $F_{\etapp}^{\text{asym}}$ incorporates the leading asymptotic behavior from the light-cone expansion, as discussed in Sec.~\ref{sec:SDC}.
 
\subsection[$\eta^\prime \to2(\pi^+\pi^-)$ amplitude]{$\boldsymbol{\eta^\prime \to2(\pi^+\pi^-)}$ amplitude}

	As an odd number of pseudoscalar mesons is involved, the decay amplitude for $\eta'(q)$$\to\allowbreak\pi^+(p_1)\allowbreak\pi^-(p_2)\allowbreak\pi^+(p_3)\allowbreak\pi^-(p_4)$ can be written in terms of the Levi-Civita symbol and a scalar function $\mathcal{F}$,
	\begin{equation}
	\label{Eq:eta4pi_general}
		\mathcal{M}_{\eta' \to 4 \pi} = \epsilon_{\mu \nu \rho \sigma} p_1^\mu p_2^\nu p_3^\rho p_4^\sigma \mathcal{F}(s_{12},\, s_{13},\, s_{14},\, s_{23},\, s_{24},\, s_{34} ),
	\end{equation}
	where the Mandelstam variables are defined as
	\begin{equation}
		s_{ij} = (p_i + p_j)^2\quad\text{with}\ i,j \in \lbrace 1,\ldots,4 \rbrace.
	\end{equation}
	If two pions in the final state were in a relative $S$-wave, one way to conserve total angular momentum would be for the remaining two pions in the final state  to also be found in a relative $S$-wave. However, the parity $P_{\pi\pi}$ of a two-pion system is determined by $P_{\pi\pi} = P_\pi^2 (-1)^L = (-1)^L=+1$, where $L$ is the angular momentum of the two pion system.  The $\eta'$ carries a negative parity eigenvalue, hence, parity conservation would be violated in this case. However, if one of the two-pion systems were in a relative $P$-wave, Bose symmetry would demand this pion system to be in a state of odd isospin under strong interaction, with $I=1$ being the only available possibility. Since the $\eta'$ carries isospin $I=0$, the other two-pion system in the final state would also need to carry $I=1$ and therefore be in a relative $P$-wave. These two $P$-wave pion pairs can then, with a relative angular momentum of $1$ between the two systems, be coupled to $J^{PC} = 0^{-+}$.

	In order to examine the left-hand cut structure of the decay, crossing symmetry may be employed: in the scattering process $\eta' \pi \to 3\pi$ the lowest hadronic intermediate state would be $\pi\eta$, where in the $S$-wave the transition to $3 \pi$ would be forbidden by parity and in the $P$-wave the state would exhibit exotic quantum numbers $I^G(J^{PC})=1^- (1^{-+})$. The lowest partial wave that receives resonant enhancement is the $D$-wave with quantum numbers $I^G(J^{PC})=1^- (2^{++})$ with the lowest corresponding resonance being the $a_2(1320)$.

	\begin{figure}[t]
        \centering
        \includegraphics[width=.9\linewidth]{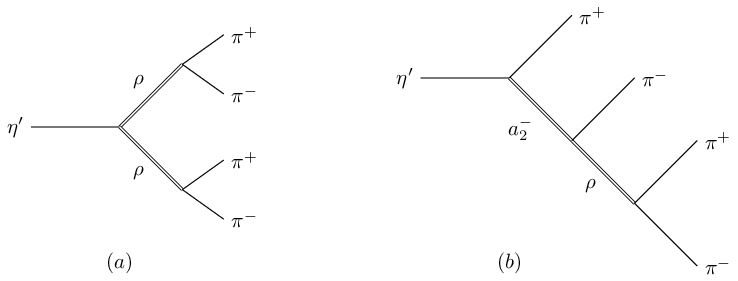}
		\caption{Sample diagrams of contributions to the decay $\eta' \to 2(\pi^+ \pi^-)$ with the leading contribution in the HLS scheme $(a)$ and the left-hand-cut contribution due to the $a_2(1320)$ $(b)$.}
		\label{fig:eta4pi_feynman}
	\end{figure}

	In Ref.~\cite{Guo:2011ir} the amplitude for $\eta' \to 2 (\pi^+ \pi^-)$ was first examined within chiral perturbation theory (ChPT) by virtue of the anomalous WZW term~\cite{Wess:1971yu,Witten:1983tw}. While there is no direct contribution to the process at leading order $\mathcal{O}(p^4)$, at $\mathcal{O}(p^6)$ diagrams involving kaon loops and counterterms derived from a sixth-order Lagrangian of odd intrinsic parity~\cite{Bijnens:2001bb} contribute. Additionally, employing a hidden local symmetry (HLS) model~\cite{Fujiwara:1984mp,Bando:1987br} the authors of Ref.~\cite{Guo:2011ir} found the amplitude to be dominated by $\rho$-meson-exchange contributions shown in Fig.~\ref{fig:eta4pi_feynman}$(a)$.

	The left-hand-cut contribution, showcased in Fig.~\ref{fig:eta4pi_feynman}$(b)$, can be incorporated by means of phenomenological resonance Lagrangians: the interaction term of a tensor meson and two pseudoscalars as well as the tensor meson propagator can be found in Ref.~\cite{Ecker:2007us}, while the model for the tensor--vector--pseudoscalar interaction is featured in Ref.~\cite{Giacosa:2005bw}.	Finally the interaction of a vector meson with two pseudoscalars is taken from Ref.~\cite{Bando:1987br}.
 Employing the HLS scheme and the phenomenological Lagrangians mentioned above, see App.~\ref{app:Feynman}, the scalar function in the decay amplitude of $\eta' \to 2(\pi^+ \pi^-)$ in Eq.~\eqref{Eq:eta4pi_general} can be written as $\mathcal{F} = \mathcal{F}_\text{HLS}+ \mathcal{F}_{a_2}$, where\footnote{The amplitude for the $\eta$ would be larger by a factor $\sqrt{2}$.}
	 \begin{equation}
	 	\mathcal{F}_\text{HLS}(s_{12},\, s_{13},\, s_{14},\, s_{23},\, s_{24},\, s_{34} ) = \frac{\sqrt{3}}{8 \pi^2 F_\pi^5} \left[\frac{M_\rho^4}{D_\rho(s_{12})D_\rho(s_{34})} - \frac{M_\rho^4}{D_\rho(s_{14})D_\rho(s_{23})} \right],
	 \end{equation}
	 with $D_\rho(s) = M_\rho^2 - s - i M_\rho \Gamma_\rho(s)$ and the energy-dependent width of the $\rho$, $\Gamma_\rho(s)$ (the precise form of which is never required below), and
		\begin{align}
			\mathcal{F}_{a_2}(s_{12}, s_{13}, s_{14}, s_{23}, s_{24}, s_{34}) &= \left(\hat{\mathcal{F}}(s_{12},s_{23},s_{24},s_{34}) + \hat{\mathcal{F}}(s_{12},s_{13},s_{14},s_{34})\right)\frac{c_{a_2}}{D_\rho(s_{34})} \notag\\
			&+\left(\hat{\mathcal{F}}(s_{34},s_{14},s_{24},s_{12}) + \hat{\mathcal{F}}(s_{34},s_{23},s_{13},s_{12})\right)\frac{c_{a_2}}{D_\rho(s_{12})} \notag\\
			&-\left(\hat{\mathcal{F}}(s_{14},s_{34},s_{24},s_{23}) + \hat{\mathcal{F}}(s_{14},s_{12},s_{13},s_{23})\right)\frac{c_{a_2}}{D_\rho(s_{23})} \notag\\
			&-\left(\hat{\mathcal{F}}(s_{23},s_{13},s_{34},s_{14}) + \hat{\mathcal{F}}(s_{23},s_{12},s_{24},s_{14})\right)\frac{c_{a_2}}{D_\rho(s_{14})}, \label{Eq:Fa2_Drho}
		\end{align}
	where the constant $c_{a_2}=\frac{2}{\sqrt{3}} c_\text{TPP} c_\text{VPP} c_\text{TPV}$ collects the couplings from the different  Lagrangian interactions and
	\begin{equation}
			\hat{\mathcal{F}}(s_{12},s_{23},s_{24},s_{34}) = \frac{(M_{\eta'}^2 - M_\pi^2)(s_{23} + s_{24} - 2M_\pi^2) - M_{a_2}^2(2s_{12}+s_{23}+s_{24}-6 M_\pi^2)}{2M_{a_2}^2(M_{a_2}^2+3M_\pi^2-s_{23}-s_{24}-s_{34})}.
	\end{equation}
	In order to improve on the treatment of pion final-state interaction in the decay, we replace the $\rho$-propagator by the Omn\`es function~\cite{Omnes:1958hv,Plenter:2017} 
	\begin{equation}
		\frac{M_\rho^2}{D_\rho(s)} \to \Omega(s)=\exp\Bigg\{\frac{s}{\pi}\int_{4\mpi^2}^\infty \text{d} s'\frac{\delta^1_1(s')}{s'(s'-s)}\Bigg\},
	\end{equation}
	where $\delta^1_1(s)$ denotes the $P$-wave $\pi\pi$ phase shift.
	Of course this replacement only gives a proper prescription for the pairwise rescattering of the two pions originating from the $\rho$ lines in Fig.~\ref{fig:eta4pi_feynman} and not a full dispersive description of the four-particle final-state interaction. In order to include the two pions coupling to the intermediate $a_2$ resonance in the final-state-interaction approximation, we consider the amplitude for the (fictional) decay $\eta' \to \pi^+ \pi^- \rho^*$.

	\begin{figure}[t]
        \centering
        \includegraphics[width=.9\linewidth]{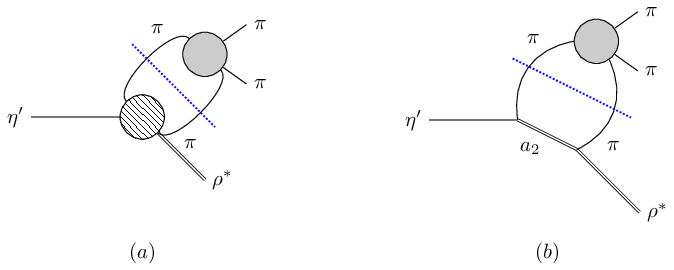}
		\caption{Fictional decay $\eta' \to \pi^+ \pi^- \rho^*$ in general form $(a)$ and its contribution due to the $a_2(1320)$ tensor meson $(b)$. $P$-wave rescattering of the pion pair is denoted by the gray blob. The dotted blue lines indicate the unitarity cut.}
		\label{fig:etatopipirho}
	\end{figure}

This amplitude, describing the decay $\eta'(q) \to \pi^+(p_1) \pi^-(p_2) \rho^*(k)$ with $a_2$ contribution as shown in Fig.~\ref{fig:etatopipirho}, can be written in terms of a scalar function
	\begin{equation}
	\label{Eq:amp_etapipirho}
		\mathcal{M}_{\eta' \to \pi \pi \rho} = \epsilon_{\mu \nu \alpha \beta} \epsilon^{\mu \ast}(k) p_1^\nu p_2^\alpha q^\beta \mathcal{F}^{\eta' \to \pi \pi \rho}(s,\, t,\, u,\, k^2),
	\end{equation}
	where $\epsilon^{\mu}(k)$ is the polarization vector of the outgoing $\rho$ and the Mandelstam variables are defined by
	\begin{equation}
		s=(q-p_1)^2, \qquad t = (p_1+p_2)^2, \qquad u = (q-p_2)^2.
	\end{equation}
    Making use of the phenomenological Lagrangian, the scalar function in Eq.~\eqref{Eq:amp_etapipirho} can be decomposed into
    \begin{equation}
        \label{Eq:etapipirho_noproj}
		\mathcal{F}^{\eta' \to \pi \pi \rho}(s,\, t,\, u,\, k^2)  = \tilde{\mathcal{F}}^{\eta'}(t,k^2) + G^{\eta'}(s,t,u,k^2) + G^{\eta'}(u,t,s,k^2),
	\end{equation}
	where $G$ describes the tree-level $a_2$ contribution via
	\begin{equation}
	\label{Eq:G_not_int}
		G^{\eta'}(s,t,u,k^2) = \frac{c_{\eta' \pi\pi\rho}^{a_2}}{M_{a_2}^2-s} \left(t-u+M_{\eta'}^2-M_\pi^2-\frac{(s+M_\pi^2-k^2)(M_{\eta'}^2-M_\pi^2)}{M_{a_2}^2}\right),
	\end{equation}
	with coupling $c_{\eta'\pi\pi\rho}^{a_2} = c_{a_2 \eta' \pi} c_{\text{TPV}}/\sqrt{3}$, see App.~\ref{app:Feynman} for details. The partial-wave expansion of the scalar function is carried out in terms of the derivatives of the Legendre polynomials, since the amplitude involves three pseudoscalar and one vector particle~\cite{Jacob:1959at}
	\begin{equation}
	\label{Eq:etapipirho_partialwdec}
		\mathcal{F}^{\eta' \to \pi \pi \rho}(s,t,u,k^2) = \sum\limits_{\text{odd } l} P_l'(z_t) f_l^{\eta'\to \pi\pi\rho}(t,k^2),
	\end{equation}
	where $z_t$ is the cosine of the $t$-channel center-of-mass angle $\theta_t$ given by
	\begin{equation}
		z_t=\cos \theta_t = \frac{s-u}{\sigma_\pi(t) \sqrt{\lambda(t,k^2,M_{\eta'}^2)}},
	\end{equation}
	with the K\"all\'en function defined as $\lambda(a,b,c)=a^2+b^2+c^2-2(ab+ac+bc)$ and $\sigma_\pi(t)=\sqrt{1-4\mpi^2/t}$. The sum in the partial-wave expansion only runs over odd angular momenta, because the pion system is in an $I=1$ state. The $P$-wave can be obtained by projecting it out via
	\begin{equation}
	\label{Eq:f1etapipirho}
		f_1^{\eta'}(t,k^2) = \frac{3}{4} \int_{-1}^{1} \text{d} z_t \, (1-z_t^2)\mathcal{F}^{\eta' \to \pi \pi \rho}(s,t,u,k^2)= \tilde{\mathcal{F}}^{\eta'}(t,k^2) + \hat{G}^{\eta'}(t,k^2),
	\end{equation}
	where in the physical decay region defined by the variables $4M_\pi^2 < t < M_{\eta'}^2$ and $0 < k^2 < (M_{\eta'} - \sqrt{t})^2$ the $P$-wave projection of the $a_2$ tree-level contributions is given by~\cite{Holz:2015tcg}
	\begin{align}
	\label{Eq:hat_decayregion}
			\hat{G}^{\eta'}(t,k^2)&\equiv\frac{3}{4} \int_{-1}^{1} \text{d} z_t \, (1-z_t^2) \Big[ G^{\eta'}(s,t,u,k^2) + G^{\eta'}(u,t,s,k^2) \Big]\notag\\
			&=2 c_{\eta'\pi\pi\rho}^{a_2} \Bigg(\frac{M_{\eta'}^2-M_\pi^2}{M_{a_2}^2} - 1 + \Bigg[M_{a_2}^2 - M_{\eta'}^2 - 2M_\pi^2 - k^2 +2t  \notag\\
			&  \qquad \quad +\frac{(k^2-M_\pi^2)(M_{\eta'}^2-M_\pi^2)}{M_{a_2}^2}\Bigg] \frac{3Q\big[y(t,k^2)\big]}{2M_{a_2}^2-M_{\eta'}^2-2M_\pi^2-k^2+t}	\Bigg),
	\end{align}
	with the function $Q$ in terms of $y=y(t,k^2)$ defined as
	\begin{equation}
	\label{Eq:Qandyfn}
		Q(y)=y\left(\frac{1}{2}(1-y^2)\log \frac{y+1}{y-1} + y\right),\qquad  y(t,k^2)=\frac{2M_{a_2}^2-M_{\eta'}^2-2M_\pi^2-k^2+t}{\sigma_\pi(t)\sqrt{\lambda(t,k^2,M_{\eta'}^2)}}.
	\end{equation}
	This derivation follows when applying
	 the $a_2$ left-hand-cut model to $\gamma^* \to \eta' \pi^+ \pi^-$.
	By means of the unitarity relation and the $\pi \pi$ $P$-wave scattering amplitude, the imaginary part of the $\eta' \to \pi^+ \pi^- \rho^*$ amplitude is governed by the following relation (since $\hat{G}^{\eta'}$ is real in the physical decay region):
	\begin{equation}
	\label{Eq:inhom_Omnes}
		\operatorname{Im} \tilde{\mathcal{F}}^{\eta'}(t,k^2) = f^{\eta'}_1(t,k^2) \sin \delta_1^1(t) e^{-i \delta_1^1(t)} = \left( \tilde{\mathcal{F}}^{\eta'}(t,k^2) + \hat{G}^{\eta'}(t,k^2)\right)\sin \delta_1^1(t) e^{-i \delta_1^1(t)}.
	\end{equation}
	This equation poses an inhomogeneous MO problem, where the inhomogeneity $\hat{G}^{\eta'}(t,k^2)$ is known by usage of the phenomenological Lagrangians mentioned above. Its solution can be expressed in terms of the hat function $\hat G^{\eta'}(t,k^2)$:
	\begin{equation}
	\label{Eq:etapipirho_inhom}
		\tilde{\mathcal{F}}^{\eta'}(t,k^2)=\bigg[P(t) + \frac{t^2}{\pi} D^{\eta'}(t,k^2) \bigg] \Omega(t),
	\end{equation}
    with
    \begin{equation}
    \label{Eq:def_Dint}
        D^{\eta'}(t,k^2) = \int_{4M_\pi^2}^{\infty} \frac{\text{d} \tau}{\tau^2} \frac{\hat{G}^{\eta'}(\tau,k^2) \sin \delta_1^1(\tau)}{(\tau - t - i\epsilon) |\Omega(\tau)|},
    \end{equation}
	where due to the asymptotic behavior of the integrand two subtractions have been carried out to render the integral convergent, i.e., $P(t)$ is a first-order polynomial. The scalar function in the $\eta' \to 2(\pi^+ \pi^-)$ amplitude of Eq.~\eqref{Eq:eta4pi_general} can be expressed in terms of the $\eta' \to \pi^+ \pi^- \rho^*$ scalar function by
	\begin{align}
	\label{Eq:eta4pi_rhodecay}
			&\mathcal{F}\big(s_{12},\, s_{13},\, s_{14},\, s_{23},\, s_{24},\, s_{34} \big) \notag\\
			&=\mathcal{F}^{\eta' \to \pi \pi \rho}\big(s_{23} + s_{24} + s_{34} -3M_\pi^2,\, s_{12},\, s_{13}+s_{14}+s_{34} - 3M_\pi^2,\, s_{34}\big) \Omega(s_{34})\notag\\
			&+\mathcal{F}^{\eta' \to \pi \pi \rho}\big(s_{12} + s_{14} + s_{24} -3M_\pi^2,\, s_{34},\, s_{12}+s_{13}+s_{23} - 3M_\pi^2,\, s_{12}\big) \Omega(s_{12})\notag\\
			&-\mathcal{F}^{\eta' \to \pi \pi \rho}\big(s_{12} + s_{14} + s_{24} -3M_\pi^2,\, s_{23},\, s_{13}+s_{14}+s_{34} - 3M_\pi^2,\, s_{14}\big) \Omega(s_{14})\notag\\
			&-\mathcal{F}^{\eta' \to \pi \pi \rho}\big(s_{23} + s_{24} + s_{34} -3M_\pi^2,\, s_{14},\, s_{12}+s_{13}+s_{23} - 3M_\pi^2,\, s_{23}\big) \Omega(s_{23}).
	\end{align}
    Note that the coupling to the left-hand-cut contribution, contained in the definition of $\hat{G}^{\eta'}$, needs to be adapted to $c_{\eta'\pi\pi\rho}^{a_2}\to  2 c_{\eta'\pi\pi\rho}^{a_2} c_\text{VPP} / M_{\rho}^2 \equiv c_{\eta' 4 \pi}$ in this expression, to account for the fact that the external $\rho$ is now resolved into $\pi^+\pi^-$.
    The analysis of $\eta' \to 2(\pi^+ \pi^-)$ decays in ChPT of Ref.~\cite{Guo:2011ir} shows that these amplitudes contain no terms at leading order (in the anomalous sector) $\mathcal{O}(p^4)$, and 
   the structure of the higher-order contributions
    at $\mathcal{O}(p^6)$, also when matched to $\rho$-exchange contributions, is such that the function $\mathcal{F}^{\eta' \to \pi \pi \rho}$ in the decomposition~\eqref{Eq:eta4pi_rhodecay} is linear in its last argument.  Hence, we incorporate the chiral constraint on the dispersive representation as follows: the constant term in the subtraction polynomial in Eq.~\eqref{Eq:etapipirho_inhom} needs to vanish
    \begin{equation}
        P(t) \mapsto P_s (t)=A t,
    \end{equation}
    and, furthermore, the left-hand-cut contribution needs to be modified via
    \begin{equation}
    \label{Ghats}
        \hat{G}^{\eta'}(t,k^2) \mapsto \hat{G}_s^{\eta'}(t,k^2) = \hat{G}^{\eta'}(t,k^2) - \hat{G}^{\eta'}(0,k^2).
    \end{equation}
    In the following we work with this subtracted version of the amplitude in Eqs.~\eqref{Eq:etapipirho_noproj} and~\eqref{Eq:f1etapipirho}.

    Additionally, since in the phenomenological model used to evaluate the diagrams of Fig.~\ref{fig:etatopipirho}, the $a_2$ tensor meson is assumed to have no width, the procedure described in Sec.~3.2 of Ref.~\cite{Holz:2015tcg} can be applied here to approximate finite-width effects by means of smearing out the resonance with the help of dispersively improved Breit--Wigner (BW) functions~\cite{Lomon:2012pn,Moussallam:2013una,Zanke:2021wiq,Crivellin:2022gfu}. In the opposite direction, we performed a number of cross checks to ensure that our description of the $a_2$ contribution reduces to the results of Ref.~\cite{Kubis:2015sga} in the appropriate narrow-width limits, see App.~\ref{app:cross_check}.

    As can be seen from Eq.~\eqref{Eq:etapipirho_inhom}, we need to evaluate $\hat{G}^{\eta'}(\tau,k^2)$ for $\tau,\, k^2 \in [4 M_\pi^2,\, \infty)$. However, the hat function exhibits a branch cut between $\tau \in [(M_{\eta'}-\sqrt{k^2})^2,\,(M_{\eta'}+\sqrt{k^2})^2]$, which complicates the direct numerical evaluation of the integral immensely. In the following, we therefore present a solution strategy that avoids crossing this branch cut~\cite{Holz:2022smu}.

    \subsection{Solution of the inhomogeneous Muskhelishvili--Omn\`es problem}
    
    \begin{figure}[t]
		\centering
		\includegraphics[width=0.75\textwidth]{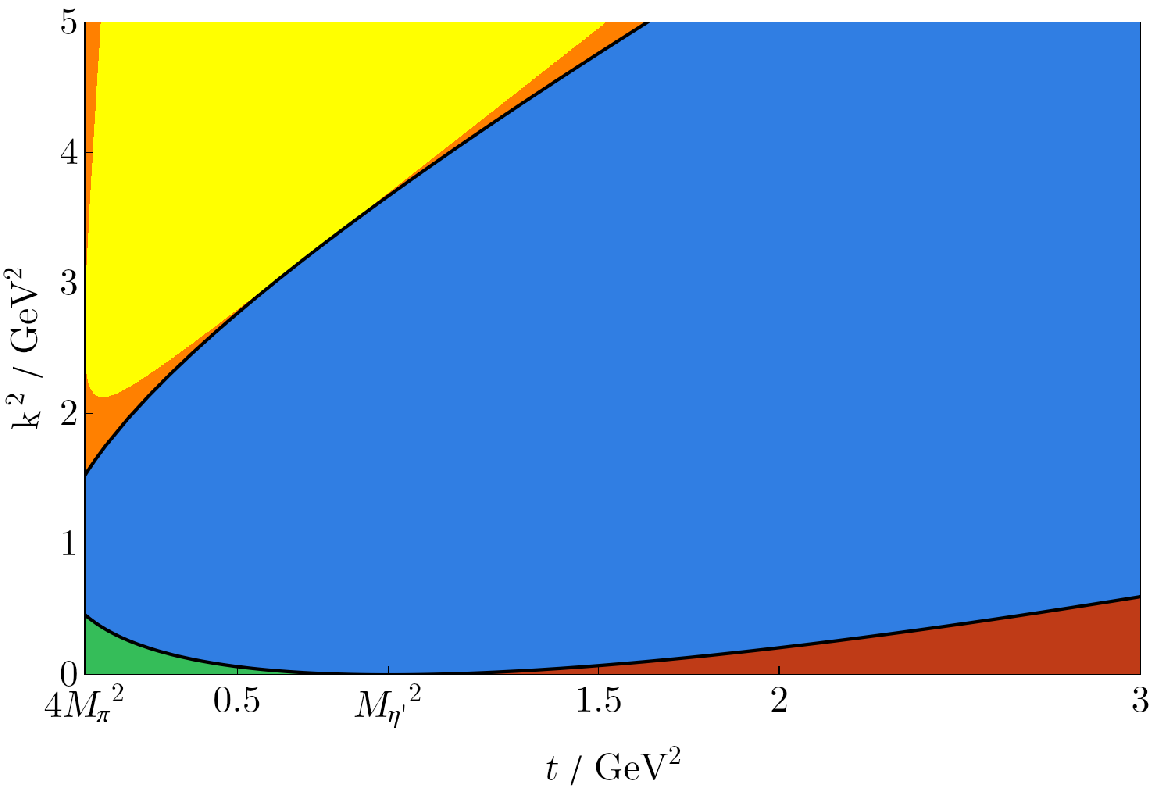}
		\caption{Kinematic (real valued) $t$--$k^2$ plane. The decay region of $\eta' \to \pi \pi \rho^*$ is marked in green, the scattering region of $\rho^* \eta' \to \pi \pi$ in red. The orange/yellow colored areas mark the decay region of $\rho^* \to \eta' \pi \pi$, where the yellow area within shows the region in which the $a_2(1320)$ intermediate state is allowed to go on-shell. The blue area indicates the unphysical region.}
		\label{fig:etapipirho_regions}
	\end{figure}

	In order to evaluate Eq.~\eqref{Eq:etapipirho_inhom} numerically, we follow  a strategy inspired by the methods of Ref.~\cite{Gasser:2018qtg}: the integration path is changed in such a way that all branch points and branch cuts of the integral are avoided. In order to do so, two immediate issues need to be addressed: the analytic continuation of the hat function $\hat{G}^{\eta'}$ into the complex plane and the input of the $\pi \pi$ $P$-wave phase shift $\delta_1^1$, which is only observable by experiment on the real axis. For the sake of simplicity, we refer to the case of the $\eta'$ for this part.

	\begin{figure}[t!]
		\centering
		\includegraphics[width=\linewidth]{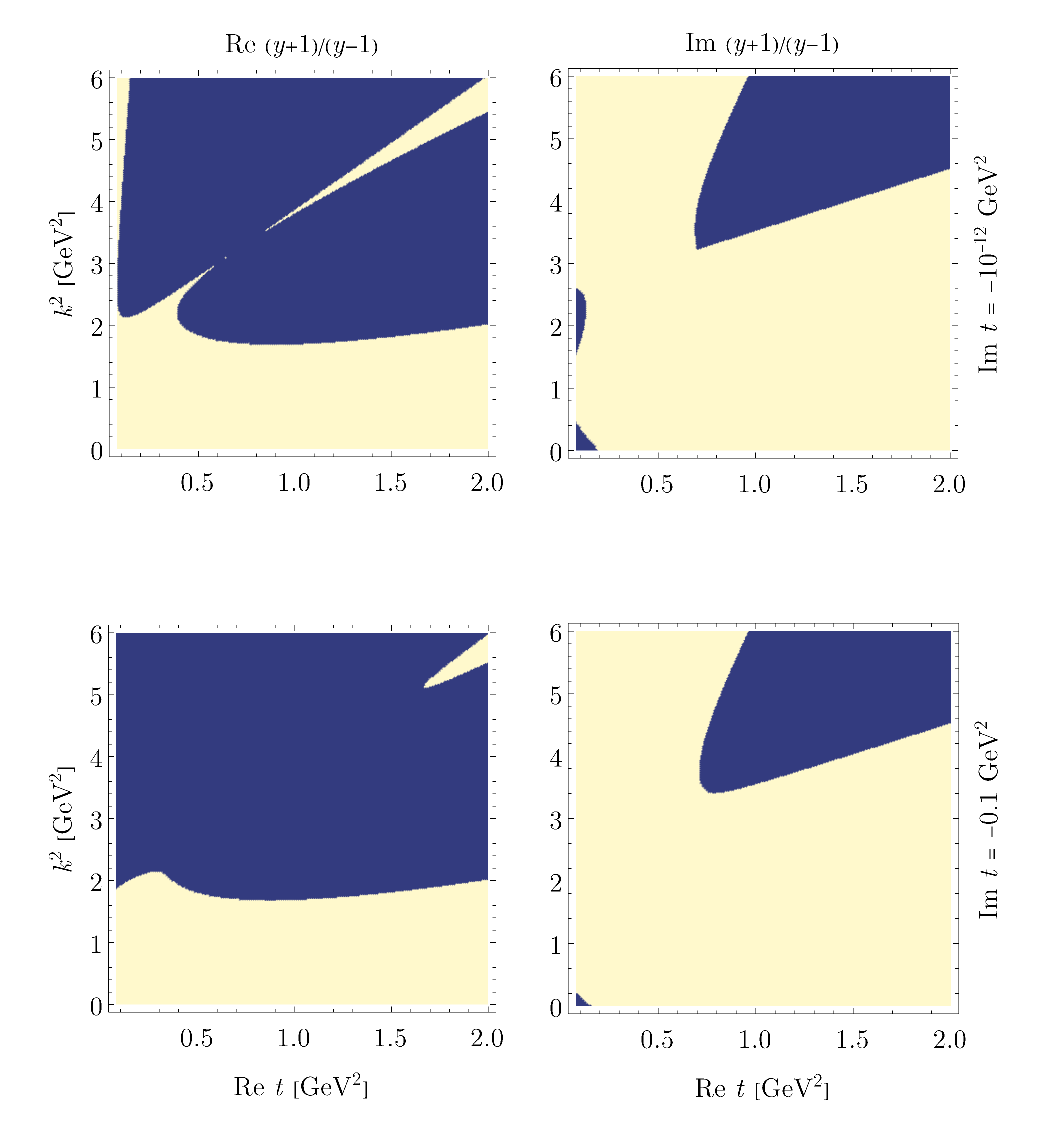}
		\caption{Study of the argument of the logarithm appearing in $Q(y)$ in Eq.~\eqref{Eq:Qandyfn}, $\log\big[(y(t,k^2)+1)/(y(t,k^2)-1)\big]$, for $\operatorname{Im}\, t \in \lbrace -10^{-12},\, -0.1 \rbrace \GeV^2$. In these density plots, blue color signifies a negative sign of the quantity, yellow color stands for positive sign. The real part is shown on the left-hand side, the imaginary part on the right-hand side.}
		\label{Fig:yfrac_density}
	\end{figure}

	The sources of the branch cuts of $\hat{G}^{\eta'}(t,k^2)$ located at $t \in \big[\big(M_{\eta'}-\sqrt{k^2}\big)^2 ,\, \big(M_{\eta'} + \sqrt{k^2}\big)^2\big]$ and $t \in [0,\, 4M_\pi^2]$ are to be found in the square root of the K\"all\'en function $\sqrt{\lambda(t,k^2,M_{\eta'}^2)}$ and the two-pion phase space $\sigma_\pi(t)$, respectively. The analytic continuation of $\sqrt{\lambda(t,k^2,M_{\eta'}^2)}$ in both $t$ and $k^2$ is not feasible by considering both variables to be complex at the same time. Rather, one needs to analytically continue one variable at a time into the complex plane. It should be noted that the physical regions of the decay $\eta' \to \pi \pi \rho^*$---for $t \in \big(4M_\pi^2,\,\big(M_{\eta'}-\sqrt{k^2}\big)^2\big)$ and $k^2 \in \big(0,\,\big(M_{\eta'}-\sqrt{t}\big)^2\big)$---and scattering process $\rho^* \eta' \to \pi \pi$---for $t \in \big(\big(M_{\eta'}+\sqrt{k^2}\big)^2,\infty\big)$ and $k^2 \in \big(0,\,\big(M_{\eta'}-\sqrt{t}\big)^2\big)$---are direct neighbors of each other in the kinematical $t$--$k^2$ plane at the point $t=M_{\eta'}^2$, see Fig.~\ref{fig:etapipirho_regions}. In both regions the partial-wave decomposition of Eq.~\eqref{Eq:etapipirho_partialwdec} is well defined and by the angular integration of Eq.~\eqref{Eq:hat_decayregion} $\hat{G}$ can be obtained. In order for those two regions to be connected, $\sqrt{\lambda}$ should factorize in $t$, i.e., the prescription
	\begin{equation}
	\label{Eq:sqlam_t_factorize}
		\sqrt{\lambda\big(t,k^2,M_{\eta'}^2\big)} \rightarrow i \sqrt{\big(M_{\eta'}-\sqrt{k^2}\big)^2-t}\,\sqrt{t - \big(M_{\eta'}+\sqrt{k^2}\big)^2},
	\end{equation}
   where the branch cuts run from the upper/lower branch point at $t=\big(M_{\eta'}\pm\sqrt{k^2}\big)^2$ along the real $t$-axis towards positive/negative infinity, is viable.

	\begin{figure}[t!]
		\centering
		\includegraphics[width=\linewidth]{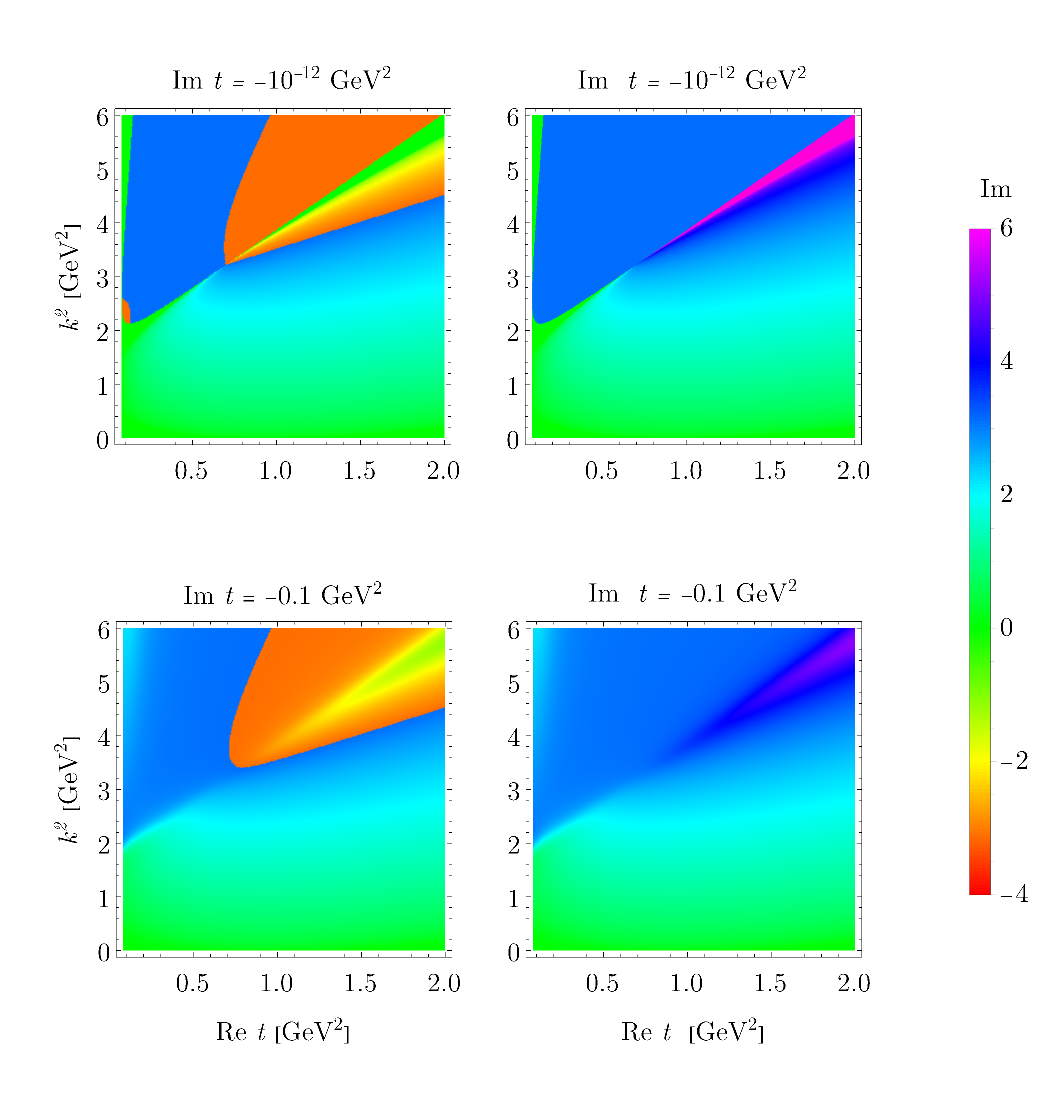}
		\caption{Imaginary part of the logarithm appearing in $Q(y)$ in Eq.~\eqref{Eq:Qandyfn}, $\log \big[ (y(t,k^2)+1)/(y(t,k^2)-1) \big]$,  for different imaginary parts of the kinematical variable $t$. The plots on the left-hand side display the behavior of the principal value, on the right-hand side the analytically continued logarithm is shown.}
		\label{Fig:Ghat_ImLog}
	\end{figure}

	The logarithm $\log \big[(y(t,k^2)+1)/(y(t,k^2)-1)\big]$ appearing in the $Q$-function of Eq.~\eqref{Eq:Qandyfn} requires special attention as to in which cases a different sheet needs to be selected. With the prescription of factorizing the roots of $\sqrt{\lambda}$ in the $t$ variable, as shown in Eq.~\eqref{Eq:sqlam_t_factorize}, $t$ can be given an infinitesimal imaginary part in order to find an analytic continuation of $\hat{G}^{\eta'}(t,k^2)$ in the complex plane.
	In Fig.~\ref{Fig:yfrac_density} the sign-behavior of the logarithm's argument in $\hat{G}^{\eta'}$ is shown for two values of the imaginary part of $t$ in the $t$--$k^2$ plane. In the complex plane, the branch cut of the logarithm is conventionally taken to extend from the branch point at the origin along the negative real axis.
	In cases where the trajectory of the logarithm's argument crosses the branch cut in the complex plane, the appropriate Riemann sheet needs to be selected. In terms of Fig.~\ref{Fig:yfrac_density} this means that a prescription for analytically continuing the logarithm is necessary whenever the real part of the argument is negative (blue regions) and $\operatorname{Im}\big[ (y(t,k^2)+1)/(y(t,k^2)-1)\big]$ changes sign. As can be seen in the left column of Fig.~\ref{Fig:Ghat_ImLog}, the imaginary part of the logarithm's principal value exhibits several discontinuities in the $t$--$k^2$ plane. A conspicuous discontinuity appears above the point $t=(M_{\eta'}^2+M_\pi^2-M_{a_2}^2)^2/M_{\eta'}^2 \approx 0.7 \GeV^2$, $k^2=(M_{a_2}^2 - M_{\pi}^2)^2/M_{\eta'}^2 \approx 3.2 \GeV^2$, the point where the borders of the decay region of $\rho^* \to \eta' \pi \pi$ and the $a_2$ on-shell region coincide, as seen in Fig.~\ref{fig:etapipirho_regions}. Therefore we take the prescription
	\begin{equation}
		\log \frac{y(t,k^2)+1}{y(t,k^2)-1} = \log \left| \frac{y(t,k^2)+1}{y(t,k^2)-1} \right| + i \arg \frac{y(t,k^2)+1}{y(t,k^2)-1} + 2 \pi i \theta(t, k^2) ,
	\end{equation}
	where $\arg x\in(-\pi,\pi]$ and
	\begin{equation}
		\theta(t, k^2) =
		\begin{cases}
		1 , &  \text{for}\quad \operatorname{Im} \frac{y+1}{y-1} < 0\ \land\ \left(k^2 > \frac{(M_{a_2}^2 - M_{\pi}^2)^2}{M_{\eta'}^2}\ \lor\  \operatorname{Re} \frac{y+1}{y-1} < 0\right) , \\
		0 , & \text{otherwise} ,
		\end{cases}
	\end{equation}
	as analytic continuation of the logarithm in Eq.~\eqref{Eq:Qandyfn} for $\operatorname{Im}\ t <0$. As can be observed in the right column of Fig.~\ref{Fig:Ghat_ImLog}, the resulting logarithm does not contain unphysical discontinuities anymore. This prescription is only valid for nonzero imaginary parts of $t$ and evaluation on the real axis should be understood as approaching the real axis infinitesimally from below. With an increase in the imaginary part of $t$, the imaginary part of the logarithm displays a smoother behavior, as illustrated in Fig.~\ref{Fig:Ghat_ImLog}. By means of the Schwarz reflection principle, for $\operatorname{Im}\ t >0$,  the function $\theta(t, k^2)$ would appear as
	\begin{equation}
		\theta(t, k^2) =
		\begin{cases}
		-1 , &  \text{for}\quad \operatorname{Im} \frac{y+1}{y-1} > 0\ \land\ \left(k^2 > \frac{(M_{a_2}^2 - M_{\pi}^2)^2}{M_{\eta'}^2}\ \lor\  \operatorname{Re} \frac{y+1}{y-1} < 0\right) , \\
		0  , & \text{otherwise} .
		\end{cases}
	\end{equation}

	In order to deform the contour of the integration in Eq.~\eqref{Eq:etapipirho_inhom}, it is necessary to evaluate the integrand in the complex plane. However, the $\pi \pi$ $P$-wave phase shift $\delta_1^1(s)$ is only defined for real $s$ above threshold. By rewriting the integrand in Eq.~\eqref{Eq:etapipirho_inhom} in terms of the $\pi \pi$ $P$-wave scattering amplitude $t_1^1$~\cite{Hoferichter:2013ama,Niehus:2019nkl,Niehus:2021iin},
	\begin{align}
	\label{Eq:phase_rewrite}
	    \frac{\hat{G}^{\eta'}(\tau,k^2) \sin \delta_1^1(\tau)}{\tau^2(\tau - t) |\Omega(\tau)|} &= \frac{\hat{G}^{\eta'}(\tau,k^2)}{\tau^2(\tau - t)} \frac{\sin \delta_1^1(\tau) \sigma_\pi(\tau) e^{i\delta_1^1(\tau)}}{|\Omega(\tau)|e^{i\delta_1^1(\tau)}\sigma_\pi(\tau)} \notag\\
	    & = \frac{\hat{G}^{\eta'}(\tau,k^2)}{\tau^2(\tau - t)} \,  \Omega^{-1}(\tau) \sigma_\pi(\tau) \, t_1^1(\tau),
	    \end{align}
	the unitarized inverse-amplitude-method (IAM) amplitude $t_\text{IAM}(s)$, with input of the ChPT amplitudes of chiral orders $\mathcal{O}(p^2)$ and $\mathcal{O}(p^4)$ supplemented by the $\mathcal{O}(p^6)$ inspired contact terms, provides a sufficiently accurate analytic expression that can be evaluated for complex arguments in a straightforward manner, see App.~\ref{app:phaseshift} for details.

	\begin{figure}[t]
		\centering
		\includegraphics[width=\linewidth]{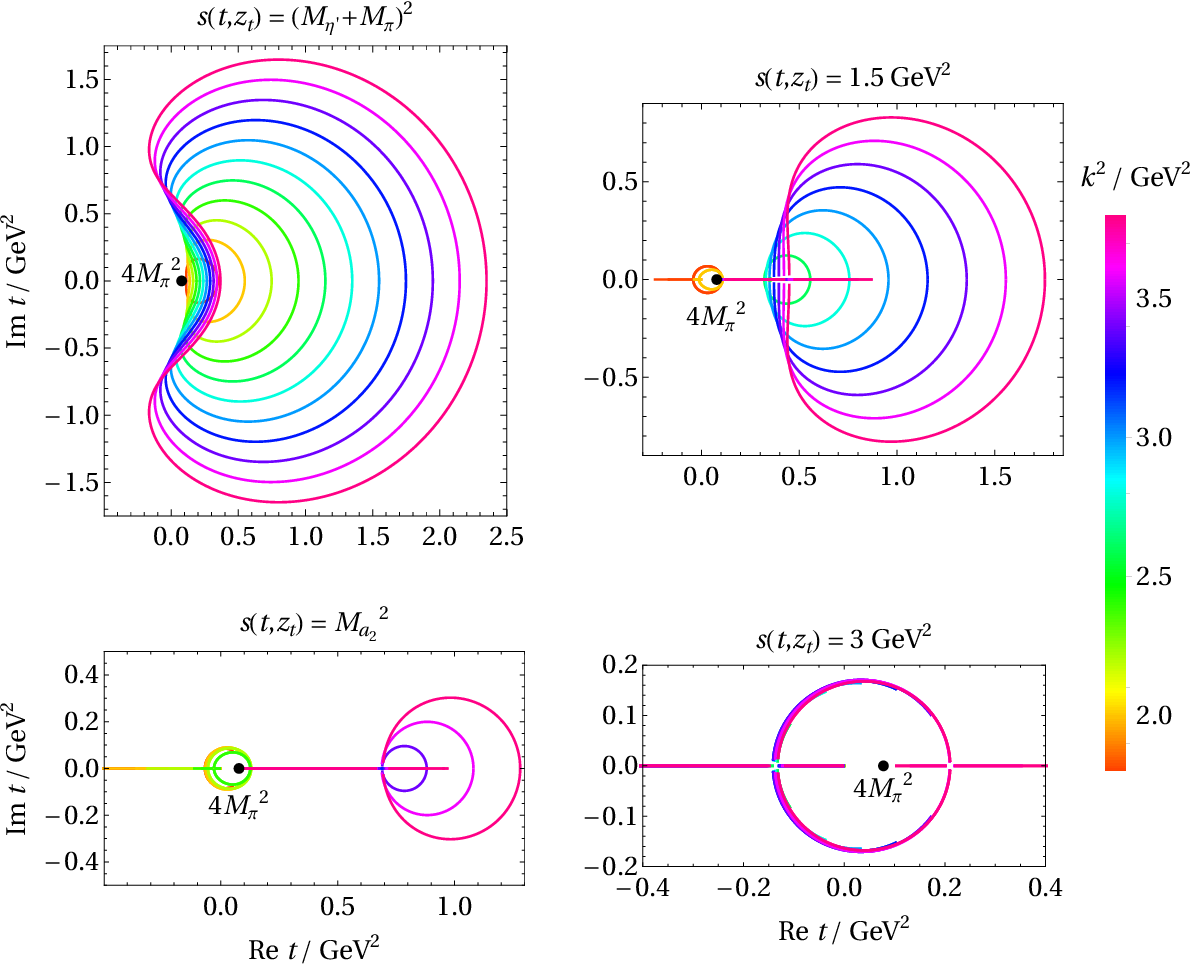}
		\caption{Critical regions in the complex $t$-plane for angles $z_t \in [-1,\,1]$ for different positions on the branch cut in the $s$-channel. The invariant mass square $k^2$ of the virtual $\rho^*$ is varied. Starting point of the inhomogeneous dispersion integral is $t=4M_\pi^2$.}
		\label{Fig:forb_reg_kvar}
	\end{figure}

    The method developed in Ref.~\cite{Gasser:2018qtg} to circumvent singularities in the angular average (the expression corresponding to Eq.~\eqref{Eq:f1etapipirho}) and the dispersion integral (corresponding to Eq.~\eqref{Eq:etapipirho_inhom}) was originally applied in an iterative manner while deforming the path of integration in the dispersion integral. Here, in the case at hand, the inhomogeneity $\hat{G}$ is given by means of a phenomenological model and thus no iterative computation of angular average and dispersion integral is necessary. Furthermore, since in Ref.~\cite{Gasser:2018qtg} the decay $\eta \to 3 \pi$ is considered, the branch cuts in the $s$ and $t$ channels are uniform. For $\eta' \to \pi^+ \pi^- \rho^*$, one does not only need to consider the more complicated branch cut structure, but also the variable mass of the ``virtual'' $\rho^*$. In this case the branch point in the $t$-channel lies at $t_\text{thr}=4M_\pi^2$, while in the $s$-channel the elastic threshold is located at $s_\text{thr}=(M_{\eta'}+M_\pi)^2$.

	\begin{figure}[t]
		\centering
		\includegraphics[width=\linewidth]{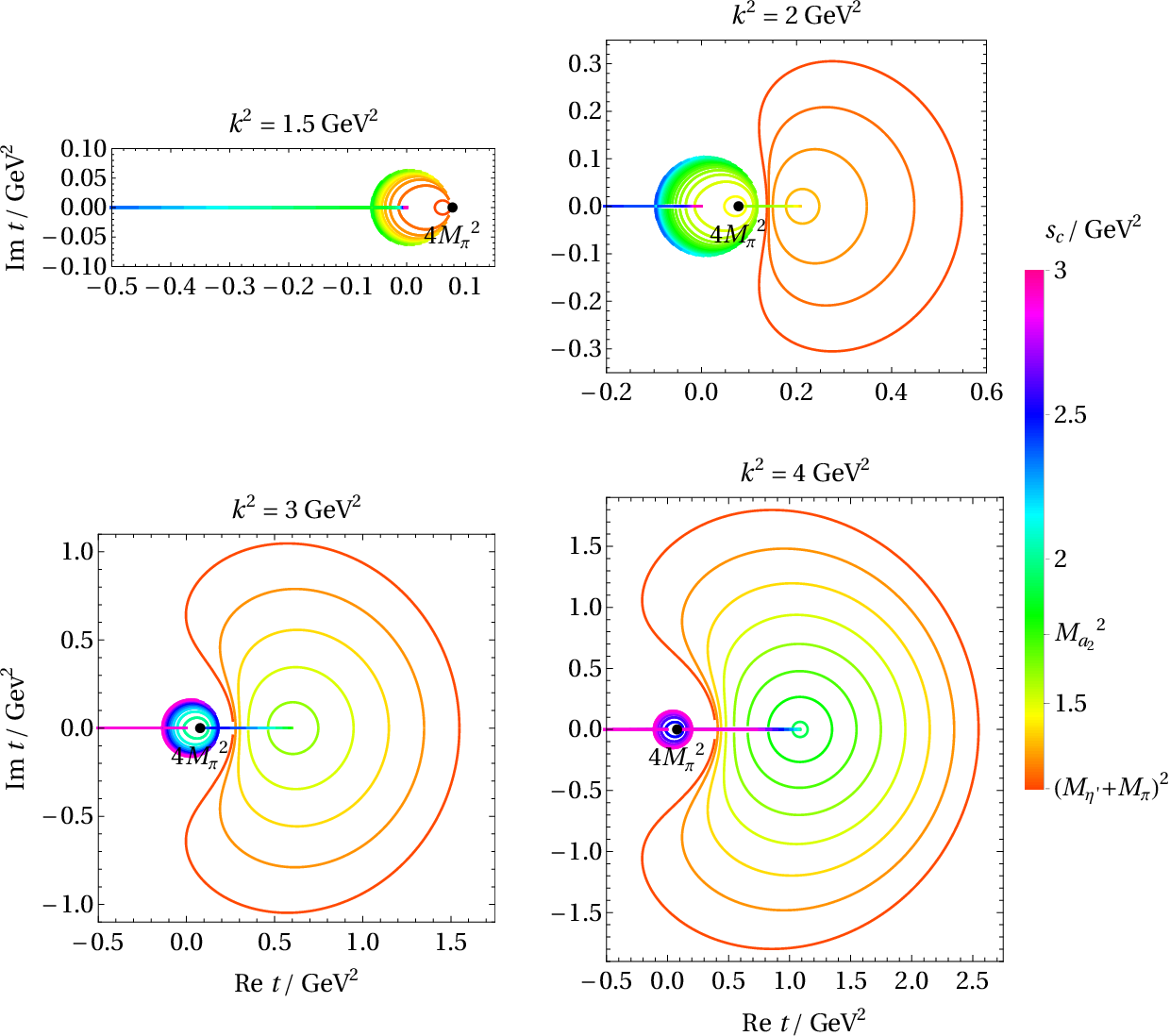}
		\caption{Critical regions in the complex $t$-plane for angles $z_t \in [-1,\,1]$ for different invariant mass squares $k^2$ of the virtual $\rho^*$. The position on the branch cut $s_c$ on the $s$-channel branch cut is varied. Starting point of the inhomogeneous dispersion integral is $t=4M_\pi^2$.}
		\label{Fig:forb_reg_svar}
	\end{figure}

	The critical regions~\cite{Gasser:2018qtg} that should be avoided by deforming the path of integration in the dispersion integral are specified by the condition
	\begin{equation}
	\label{Eq:Critregions_cond}
		s(t,z_t) = s_c, \quad s_c \in \big[\big(M_{\eta'}+M_\pi\big)^2,\, \infty\big),
	\end{equation}
	where $s_c$ is positioned on the $s$-channel branch cut and 
	\begin{equation}
		s(t,z_t) = \frac{1}{2} \left( 2 M_\pi^2 + M_{\eta'}^2 + k^2 - t + z_t \sigma_\pi(t) \sqrt{\lambda(t,k^2,M_{\eta'}^2)} \right).
	\end{equation}
	Solving Eq.~\eqref{Eq:Critregions_cond} for $t$ yields three independent solutions, out of which one is manifestly real and the two remaining ones are the complex conjugate of each other. Plots of these solutions with different parameter variations are provided in Figs.~\ref{Fig:forb_reg_kvar}~and~\ref{Fig:forb_reg_svar}. In the case of Ref.~\cite{Gasser:2018qtg}, in the iterative calculation of $\eta \to 3\pi$, the branch points in the $t$ and $s$ channels are the same. Furthermore, due to the iterative nature of the calculation, the cut originating at this branch point would be deformed together with the path of integration in the dispersion integral. In the case at hand, however, through usage of the $a_2$ amplitudes for the inhomogeneity, the dispersion integral does not serve as input for the angular integral. Through the model of $a_2$ exchange, the inhomogeneous part in the angular integral exhibits a pole at $s=M_{a_2}^2$, see Eq.~\eqref{Eq:G_not_int}. However, due to the elastic rescattering in the $s$-channel, values beyond $s=(M_{\eta'}+M_\pi)^2$ must be avoided. Observing the critical regions in Figs.~\ref{Fig:forb_reg_kvar}~and~\ref{Fig:forb_reg_svar}, it is apparent that this procedure fails for a number of kinematical configurations. In cases in which the critical regions fully surround the starting point of the dispersion integral at $4 M_\pi^2$, it is not possible to find a path of integration that does not cross these critical regions.

\begin{figure}[t]
		\centering
        \includegraphics[width=.75\textwidth]{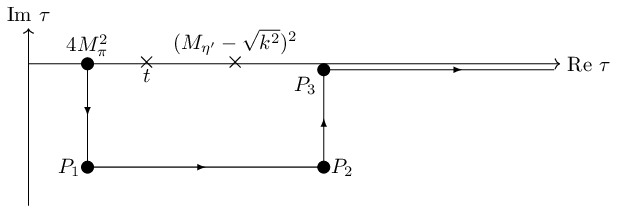}
		\caption{Deformed path of integration in the dispersion integral of Eq.~\eqref{Eq:etapipirho_inhom}. The Cauchy singularity of the integrand lies infinitesimally above the real $\tau$ axis at $t$ and the pseudo-threshold singularity at $\big(M_{\eta'}-\sqrt{k^2}\big)^2$.}
		\label{Fig:deformed_integration}
	\end{figure}

	For the present application, it is not necessary to evade all these critical regions, in principle, also an integration along the real axis should be possible. However, we were not able to obtain stable results in this way, and therefore consider a path deformation motivated by the previous discussion  as shown in Fig.~\ref{Fig:deformed_integration}.  In particular, deforming the path
    of integration of the dispersion integral then avoids the pseudo-threshold singularity, located at $t = \big(M_{\eta'} - \sqrt{k^2}\big)^2$, and allows us to reach a numerically stable result.\footnote{Otherwise, the integration around the pseudo-threshold would become particularly delicate when the analytic continuation requires a nonvanishing value of $\theta(t,k^2)$ in $Q(y)$, and would have to be performed with methods as described in Refs.~\cite{Stamen:2022eda,Mutke:2024tww}.} In order to choose a suitable integration path, the location of the singularities of the integrand including their infinitesimal imaginary parts are of importance. While the Cauchy singularity of the integrand in Eq.~\eqref{Eq:etapipirho_inhom} is located at $t + i \epsilon$ with $\epsilon>0$, the variable $k^2$ conventionally obtains an infinitesimal imaginary part by $k^2 \to k^2 + i \delta$ with $\delta>0$, due to its nature as a mass parameter~\cite{Gasser:2018qtg,Bronzan:1963mby}. Therefore, the pseudo-threshold singularity is located at
	\begin{equation}
		\big(M_{\eta'} - \sqrt{k^2}\big)^2 \to \big(M_{\eta'} - \sqrt{k^2}\big)^2 + i \delta \left(1 - \frac{M_{\eta'}}{\sqrt{k^2}}\right).
	\end{equation}
	Hence, for values of $k^2 > M_{\eta'}^2$, the singularity is located (like the Cauchy singularity) in the upper half plane. Therefore, in these cases, a deformation of the path of integration towards negative imaginary parts of the kinematic variable is justified, motivating the final path of integration shown in Fig.~\ref{Fig:deformed_integration}.
	
 \subsection[$\etapp \to \pi^+ \pi^- \gamma^*$ amplitude]{$\boldsymbol{\etapp \to \pi^+ \pi^- \gamma^*}$ amplitude}

 Since the spin structure of $\eta'(q) \to \pi^+(p_1) \pi^-(p_2) \gamma^*(k)$ is the same as in the decay $\eta' \to \pi^+ \pi^- \rho^*$ the matrix element can be written in terms of a scalar function in the same manner:
	\begin{equation}
	\label{Eq:amp_etapipigams}
		\mathcal{M}_{\eta' \to \pi \pi \gamma} = e \epsilon_{\mu \nu \alpha \beta} \epsilon^{\mu \ast}(k) p_1^\nu p_2^\alpha q^\beta \mathcal{F}_{\eta' \pi\pi\gamma}(s,\, t,\, u,\, k^2),
	\end{equation}
	where $\epsilon^{\mu}(k)$ is the polarization vector of the outgoing (virtual) photon, see Eq.~\eqref{Eq:amp_etapipirho}. The Mandelstam variables are defined by
	\begin{equation}
		s=(q-p_1)^2, \qquad t = (p_1+p_2)^2, \qquad u = (q-p_2)^2 .
	\end{equation}
	The discontinuity of the scalar function (in the $k^2$ variable) can be reconstructed from the $\eta'\to 2(\pi^+ \pi^-)$ amplitude and the pion vector form factor $F_\pi^V(k^2)$ by means of the unitarity relation
	\begin{equation}
		\operatorname{disc}\mathcal{M}_{\eta' \to \pi \pi \gamma} = i (2\pi)^4 \int \text{d} \Phi_2\, \delta^{(4)}(k_n - k) \mathcal{M}_{\gamma^*\to\pi\pi}^* \mathcal{M}_{\eta'\to 4\pi},
	\end{equation}
	in the approximation of taking only the lowest-lying intermediate states into account, where $\text{d} \Phi_2$ marks the two-particle phase-space integration and $k_n$ labels the intermediate momenta. The amplitude for $\gamma^*(k) \to \pi^-(p_3) \pi^+(p_4)$ written in terms of the pion vector form factor appears as
	\begin{equation}
		\mathcal{M}_{\gamma^*\to\pi\pi} = e \epsilon_\mu(k)(p_3 - p_4)^\mu F_\pi^V(k^2).
	\end{equation}
	Furthermore, the $\eta' \to 2(\pi^+ \pi^-)$ amplitude of Eq.~\eqref{Eq:eta4pi_general} with $a_2$ left-hand-cut contribution and description of final-state interaction, see Eq.~\eqref{Eq:eta4pi_rhodecay}, can be written as
	\begin{equation}
		\mathcal{M}_{\eta'\to 4\pi} = \epsilon_{\mu \nu \rho \sigma} p_1^\mu p_2^\nu p_3^\rho p_4^\sigma \left[f_1^{\eta'}(t,k^2) \Omega(k^2) + f_1^{\eta'}(k^2,t) \Omega(t)\right],
	\end{equation}
	where $f_1^{\eta'}(t,k^2)$ is the partial wave of Eq.~\eqref{Eq:f1etapipirho} and higher partial waves as well as crossed terms of the final-state interaction have been neglected. After considering the different momentum combinations of the intermediate pions and performing the two-particle phase-space integration, the discontinuity of the scalar function in Eq.~\eqref{Eq:amp_etapipigams} appears as
	\begin{equation}
	\label{etapipig_disc}
		\operatorname{disc}_{k^2} \mathcal{F}_{\eta' \pi\pi\gamma}(t,k^2) = i \frac{k^2 \sigma_\pi^3(k^2)}{48 \pi} \big[F_\pi^V(k^2)\big]^* \left[f_1^{\eta'}(t,k^2) \Omega(k^2) + f_1^{\eta'}(k^2,t) \Omega(t)\right],
	\end{equation}
	where the scalar function depends only on $t$ and $k^2$, see App.~\ref{app:deriv_disc_etapipig} for the derivation. The unsubtracted dispersion relation follows from this equation as
	\begin{equation}
	\label{Eq:etapipig_scfn}
		\mathcal{F}_{\eta' \pi\pi\gamma}(t,k^2) = \frac{1}{96 \pi^2} \int_{4 M_\pi^2}^{\infty} \text{d}x \ \frac{x \sigma_\pi^3(x)\big[F_\pi^V(x)\big]^* \left(f_1^{\eta'}(t,x) \Omega(x) + f_1^{\eta'}(x,t) \Omega(t)\right)}{x-k^2-i\epsilon}.
	\end{equation}
	In order to determine the input for the $\eta'$ TFF, the subtraction constants in this representation are fit to data for the real-photon decay spectrum of $\eta' \to \pi^+ \pi^- \gamma$ from BESIII~\cite{BESIII:2017kyd}, see Sec.~\ref{sec:fits}. Furthermore, the dispersion integrals, as in the equation above, are carried out up to an integral cutoff $\Lambda^2$, which is varied between $\Lambda\in\{ 1.5,\, 2.5 \}\GeV$ in the following numerical evaluation.

	A similar relation can be used for the determination of the $\eta$ TFF, even though the decay in the first step of this description, $\eta \to 2 (\pi^+ \pi^-)$, is kinematically forbidden. More specifically, in order to apply this description to the $\eta$ case, in Eqs.~\eqref{Eq:hat_decayregion}~and~\eqref{Eq:Qandyfn} the replacement $M_{\eta'} \to M_{\eta}$ needs to be performed in order to obtain the partial-wave amplitude $f_1^{\eta\to\pi\pi\rho}$. Additionally, the parameters in the subtraction polynomial appearing in the partial-wave amplitude are fit to the decay spectrum of $\eta \to \pi^+ \pi^- \gamma$ from KLOE~\cite{KLOE:2012rfx}.

    Since we utilize a subtracted representation of the $\etapp \to 2(\pi^+ \pi^-)$ amplitude as input, the dispersion relation of Eq.~\eqref{Eq:etapipig_scfn} would be divergent without the appropriate modifications. That is, the low-energy representation does not hold up to arbitrarily high energies and needs to be smoothly matched onto the expected asymptotic behavior. Thus, the following prescription to continue our low-energy description to values above a cut parameter $s_c$  is adopted to render the dispersion integral manifestly finite:
    \begin{align}
    \label{Eq:f1_contsc}
		f_1^{\etapp}(t,k^2) =
		\begin{cases}
			\Omega(t) \big[P(s_c) + \frac{s_c^2}{\pi} D^{\etapp}(s_c,k^2)\big] + \hat{G}^{\etapp}(s_c,k^2) \frac{s_c}{t}, &t>s_c \land k^2 < s_c, \\
			\Omega(t) \big[P(t) + \frac{t^2}{\pi} D^{\etapp}(t,s_c)\frac{s_c}{k^2}\big] + \hat{G}^{\etapp}(t,s_c) \frac{s_c}{k^2}, &t<s_c \land k^2>s_c,\\
			\Omega(t) \big[P(s_c) + \frac{s_c^2}{\pi} D^{\etapp}(s_c,s_c) \frac{s_c}{k^2}\big] + \hat{G}^{\etapp}(s_c,s_c) K(t,k^2), &t,k^2>s_c,
		\end{cases}
	\end{align}
    where
    \begin{equation}
		K(t,k^2) = \frac{s_c(t+s_c)(k^2+s_c)}{2 t k^2 (t + k^2)},
	\end{equation}
   and $D^{\etapp}$ is defined in Eq.~\eqref{Eq:def_Dint}. In practice this prescription forces $f_1^{\etapp}(t,k^2)$ to drop off like $1/t$ and the parts related to the inhomogeneity like $1/k^2$ above $s_c$. In particular, the prescription ensures that crossing between the four regions $(t,k^2)<s_c$; $t>s_c,k^2<s_c$; $t<s_c,k^2>s_c$; and $(t,k^2)>s_c$ is continuous. Treating the two arguments $t$ and $k^2$ on the same footing is done in view of the application to the Bose-symmetric TFFs, see Sec.~\ref{sec:space_like}. Finally, the procedure to account for finite-width effects of $a_2$ exchange, as outlined in Sec.~3.2 of Ref.~\cite{Holz:2015tcg}, dispersing $\hat{G}(t,k^2)$ and accordingly $D(t,k^2)$ around the $a_2$ mass parameter, is adopted here as well.

\subsection[Fits to $\etapp \to \pi^+ \pi^- \gamma$]{Fits to $\boldsymbol{\etapp \to \pi^+ \pi^- \gamma}$}
\label{sec:fits}

     As the $\etapp \to \pi^+ \pi^- \gamma$ scalar functions in Eq.~\eqref{Eq:etapipig_scfn} are based on the $\etapp \to 2(\pi^+ \pi^-)$ amplitude, chiral constraints on the latter need to be taken into account. Imposing the constraint observed at $\mathcal{O}(p^6)$ in the chiral expansion~\cite{Guo:2011ir} that the amplitudes vanish for $t=0$, we adapt Eq.~\eqref{Eq:f1etapipirho} to
     \begin{equation}
		\label{Eq:f1subtracted}
		f_1^{\etapp}(t,k^2) \to \left[ A t + \frac{t^2}{\pi} \int_{4 M_\pi^2}^{\Lambda^2} \, \frac{\diff \tau}{\tau^2} \frac{\hat{G}_s^{\etapp}(\tau,k^2) \sin \delta_1^1(\tau)}{(\tau-t-i\epsilon)|\Omega(\tau)|} \right] \Omega(t) + \hat{G}_s^{\etapp}(t,k^2),
	\end{equation}
	with $\hat G_s^{\etapp}$ defined in Eq.~\eqref{Ghats}.

\begin{table}[t]
	\renewcommand{\arraystretch}{1.3}
	    \centering
        	\begin{tabular}{l r r r r}
				\toprule
				$(s_c, \Lambda^2)\ [\text{GeV}^2]$ & $(1,\,2.25)$ & $(1.5,\,2.25)$ & $(1,\,6.25)$ & $(1.5,\,6.25)$\\ \midrule
				$A \ [\text{GeV}^{-7}]$ & 1895(33) & 1887(32) & 1864(32) & 1859(31)\\
				$c_{\eta'4\pi} \ [\text{GeV}^{-5}]$  & 416.1(5.9) & 406.0(6.0) & 417.1(6.0) & 406.3(6.2) \\
				$\chi^2/\text{dof}$ & 1.87 & 1.90 & 1.90 & 1.95  \\ \midrule
				$\Br\big[\eta' \to 2(\pi^+\pi^-)\big]\times 10^5$ & 5.31(13) & 5.20(12) & 5.20(12) & 5.10(12) \\ \bottomrule
        \end{tabular}
        \caption{Fit parameters and reduced $\chi^2$, normalized to the number of degrees of freedom ($\text{dof}$), in addition to the extracted branching fraction $\Br\big[\eta' \to 2(\pi^+\pi^-)\big]$ for the $\eta'\to\pi^+\pi^-\gamma$ fits to the BESIII data~\cite{BESIII:2017kyd}, with data points in the $\rho$--$\omega$-mixing region excluded as explained in the main text. The fit variants correspond to different cut parameters $s_c$ in the underlying $\eta' \to 2(\pi^+ \pi^-)$ amplitudes and different values for the integral cutoff $\Lambda^2$. The quoted errors reflect the fit uncertainties.}
        \label{Tab:etaPpipig_fits}
    \renewcommand{\arraystretch}{1.0}
    \end{table}
    
    Furthermore, the coupling multiplying the left-hand-cut contribution $\hat{G}_s^{\etapp}(t,k^2)$ and thus also the inhomogeneous dispersion integral $D^{\etapp}(t,k^2)$ can, in principle, be fixed from decay widths by means of phenomenological Lagrangians as outlined in App.~\ref{app:Feynman}. The decay widths $\Gamma[a_2 \to \etapp \pi]$ and $\Gamma[\rho\to\pi\pi]$ would be required in this approach. Additionally, one would need the values for $\Gamma[a_2 \to 3\pi]$ or $\Gamma[a_2 \to \pi \gamma]$, where both reactions would proceed via intermediate $\rho$ resonances, see App.~\ref{app:cross_check} for details. However, as demonstrated in App.~\ref{app:cross_check}, we do not obtain fully consistent results, e.g., comparing the extracted couplings for the $\etapp \to \pi^+ \pi^- \gamma$ amplitudes  (i) via the $a_2\to 3\pi$ decay with the strength of the left-hand-cut  coupling fixed on that level~\cite{Kubis:2015sga} via $\Gamma[a_2 \to \etapp \pi]$ and  (ii) via $\Gamma[a_2 \to \pi \gamma]$ due to an $a_2 \pi \gamma$ contact interaction. Part of the mismatch can be attributed to overlapping $\rho$ resonances in the $a_2\to3\pi$ Dalitz plot, which would need to be taken into account for a robust determination, but at the same time the uncertainties in measuring the radiative decay $a_2\to\pi\gamma$ are substantial as well.  
    
    For these reasons, we aim to determine the left-hand-cut couplings $c_{\etapp 4 \pi}$ phenomenologically, directly from the $\etapp\to\pi^+\pi^-\gamma$ spectra. For the $\eta'$, such a strategy gives a reasonable description of the data, see Table~\ref{Tab:etaPpipig_fits}, with resulting couplings that come out closer to 
    the prediction via $a_2 \to \pi \gamma$ than $a_2\to 3\pi$. 
    In case of the $\eta$, however, the spectrum has only a limited phase-space range that does not allow for a direct extraction in a sufficiently reliable way. Accordingly, we determine $c_{\eta 4 \pi}$ from $U(3)$ symmetry, allowing for a generous variation to account for the associated uncertainties. Table~\ref{Tab:etaPpipig_fits} also displays the branching fractions for $\eta'\to2(\pi^+\pi^-)$ that correspond to the different fit variants. In all cases, the result lies below the BESIII measurement, $\Br[\eta'\to2(\pi^+\pi^-)]=8.56(34)\times 10^{-5}$~\cite{BESIII:2023ceu}, but such a moderate mismatch is to be expected, since the underlying unitarity relation linking $\eta'\to2(\pi^+\pi^-)$ and $\eta'\to\pi^+\pi^-\gamma$, assuming $\rho$ dominance, does not take into account the effect of overlapping $\rho$ bands, as would be required for a precision calculation of this decay channel. For that reason, agreement at the present level serves as another plausibility check for our $\eta'\to 4\pi$ amplitude.

\begin{figure}[t]
        \centering
        \includegraphics[width=0.8\linewidth]{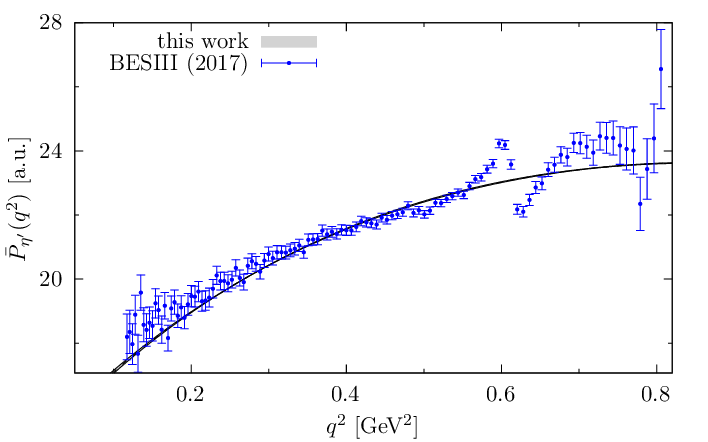}
        \caption{Fit to the $\eta'\to\pi^+\pi^-\gamma$ spectrum from BESIII~\cite{BESIII:2017kyd} with different variants, as detailed in Table~\ref{Tab:etaPpipig_fits}, from which the error band is derived (the central value is defined as the average of all fit variants, the band by the maximal difference). $\bar P_{\eta'}(q^2)$ is defined in Eq.~\eqref{Eq:pipig_Pdef}, removing the effects of the $\rho(770)$ resonance and phase-space functions from the spectrum.}
        \label{Fig:etaP_P}
    \end{figure}

    Given the scalar amplitude in Eq.~\eqref{Eq:etapipig_scfn} with subtracted partial wave of Eq.~\eqref{Eq:f1subtracted} as input, the differential decay width into the $\pi^+\pi^-\gamma$ final state, with $\pi^+\pi^-$ invariant mass $t$, can be expressed as
    \begin{equation}
		\frac{\diff \Gamma^{\etapp}}{\diff t}(t) = \Gamma_0^{\etapp}(t) |\mathcal{F}_{\etapp\pi\pi\gamma}(t,0)|^2,\qquad \Gamma_0^{\etapp}(t) = \frac{\alpha}{\pi^2} \frac{t \sigma_\pi^3(t) (M_{\etapp}^2 - t)^3}{1536 M_{\etapp}}.
	\end{equation}
    Since the experimental spectra of KLOE for $\eta\to\pi^+\pi^-\gamma$~\cite{KLOE:2012rfx} and BESIII for $\eta' \to \pi^+ \pi^- \gamma$~\cite{BESIII:2017kyd} are not normalized to physical units, we fit an overall normalization factor in addition to the left-hand-cut coupling strength $c_{\eta' 4 \pi}$ to the data, while the subtraction constant $A$ is constrained from the RPP values for $\Gamma[\etapp \to \pi^+ \pi^- \gamma]$~\cite{ParticleDataGroup:2024cfk}. Accordingly, we first decompose
    \begin{equation}
    \label{Eq:etapipig_fitrep}
		\mathcal{F}_{\etapp\pi\pi\gamma}(t,0) = A I_A^{\etapp}(t) + I_0^{\etapp}(t),
	\end{equation}
	with
	\begin{align}
		I_A^{\etapp}(t) &=  \frac{1}{96 \pi^2} \int_{4 M_\pi^2}^{\Lambda^2} \diff x\ \sigma_\pi^3(x) \big[F_\pi^V(x)\big]^*   (x+t)  \Omega(t) \Omega(x),\notag\\
		I_0^{\etapp}(t) &=  \frac{1}{96 \pi^2} \int_{4 M_\pi^2}^{\Lambda^2} \diff x\ \sigma_\pi^3(x) \big[F_\pi^V(x)\big]^* \notag \\ 
        &\qquad \times\bigg[\left(\frac{t^2}{\pi}D^{\etapp}(t,x) \Omega(t) + \hat{G}^{\etapp}(t,x) \right) \Omega(x) \notag\\
        &\qquad\qquad + \left(\frac{x^2}{\pi}D^{\etapp}(x,t) \Omega(x) + \hat{G}^{\etapp}(x,t) \right) \Omega(t)\bigg].
	\end{align}
    The integral of the differential decay width within the phase-space region gives the corresponding partial decay width. This provides a condition to determine the subtraction constant $A$ in each step of the fit iteration,  by demanding
    \begin{equation}
    \label{Eq:Asolution}
		\Gamma[\etapp \to \pi^+ \pi^- \gamma] \stackrel{!}{=} A^2 I^{(2)}_{\etapp} + 2 A I^{(1)}_{\etapp} + I^{(0)}_{\etapp},
	\end{equation}
	where
	\begin{align}
		I^{(2)}_{\etapp} &= \int_{4M_\pi^2}^{M_{\etapp}^2} \diff t\ \Gamma_0^{\etapp}(t) \big| I_A^{\etapp}(t) \big|^2,\qquad  I^{(0)}_{\etapp} = \int_{4M_\pi^2}^{M_{\etapp}^2} \diff t\ \Gamma_0^{\etapp}(t) \big| I_0^{\etapp}(t) \big|^2,\notag\\
		I^{(1)}_{\etapp} &= \int_{4M_\pi^2}^{M_{\etapp}^2} \diff t\ \Gamma_0^{\etapp}(t)\, \Re\big[I_A^{\etapp}(t) \big(I_0^{\etapp}\big)^*(t) \big].
	\end{align}
    Note that Eq.~\eqref{Eq:Asolution} gives two solutions for the subtraction constant $A$, we always find a positive and a negative value. For either sign choice, the coupling $c_{\etapp 4 \pi}$ obtains its corresponding sign in the fit to the $\eta' \to \pi^+ \pi^- \gamma$ data. A related sign ambiguity arises in the derivation of the $\eta' \to \pi^+ \pi^- \gamma$ discontinuity of App.~\ref{app:deriv_disc_etapipig}, the choices given here, however, ensure a consistent scheme.
    The spectrum of $\eta' \to \pi^+ \pi^- \gamma$ spectrum~\cite{BESIII:2017kyd} features a prominent isospin-breaking signal due to $\rho$--$\omega$ mixing~\cite{Hanhart:2016pcd}. While being relevant for precision analyses of $\eta' \to \ell^+ \ell^- \gamma$~\cite{Holz:2022smu}, the impact of these isospin-breaking corrections in the space-like region of the TFF is negligible. We, therefore, exclude data of Ref.~\cite{BESIII:2017kyd} in the region $M_\omega \pm 3 \Gamma_{\omega}$ from fits of the representation in Eq.~\eqref{Eq:etapipig_fitrep}, with the $\omega$ mass and width parameters fixed to the values taken from the RPP~\cite{ParticleDataGroup:2024cfk}.

    \begin{table}[t]
	\renewcommand{\arraystretch}{1.3}
	    \centering
        	\begin{tabular}{l l r r r}
            \toprule
            & $(s_c, \Lambda^2)\ [\text{GeV}^2]$ & $A \ [\text{GeV}^{-7}]$ & $\chi^2/\text{dof}$ & $c_{\eta 4\pi}\ [\text{GeV}^{-5}]$\\ \midrule
            \multirow{4}{*}{$c_{\eta'4\pi}-15\,\%$} & $(1,\,2.25)$ & 2330(56) & 1.70 & 354  \\
            & $(1.5,\,2.25)$ & 2326(53) & 1.44 & 345\\
            & $(1,\,6.25)$ & 2381(56) & 1.40 & 354 \\
            & $(1.5,\,6.25)$ & 2326(53) & 1.43 & 345\\ \midrule
            \multirow{4}{*}{$c_{\eta'4\pi}-30\,\%$} & $(1,\,2.25)$ & 2278(56) & 1.39 & 291 \\
            & $(1.5,\,2.25)$ & 2326(53) & 1.44 & 284 \\
            & $(1,\,6.25)$ & 2381(56) & 1.40 & 292 \\
            & $(1.5,\,6.25)$ & 2326(53) & 1.43 & 284\\ \midrule
            \multirow{4}{*}{$c_{\eta'4\pi}-45\,\%$} & $(1,\,2.25)$ & 2432(56) & 1.17  & 229 \\
            & $(1.5,\,2.25)$ & 2359(53) & 1.24 &  223 \\
            & $(1,\,6.25)$ & 2432(56) & 1.17 & 229\\
            & $(1.5,\,6.25)$ & 2359(53) & 1.24 & 223\\ \bottomrule
            \end{tabular}
        \caption{Results of the one-parameter fits to the $\eta\to\pi^+\pi^-\gamma$ data of KLOE~\cite{KLOE:2012rfx}, varying the input coupling from the $\eta'\to\pi^+\pi^-\gamma$ fits in Table~\ref{Tab:etaPpipig_fits}, cut parameter $s_c$ in the underlying $\eta \to 2(\pi^+ \pi^-)$ amplitudes, and cutoff $\Lambda^2$ of the dispersive integrals.}
        \label{Tab:etapipig_fits}
    \renewcommand{\arraystretch}{1.0}
    \end{table}

    The cut parameter $s_c$ in the underlying $\eta' \to 2(\pi^+ \pi^-)$ amplitudes is varied from $1$ to $1.5\GeV^2$. Additionally, the dispersive integrals of the underlying $\eta' \to 2(\pi^+ \pi^-)$ representation in Eq.~\eqref{Eq:etapipirho_inhom} as well as the one connecting to the $\pi^+ \pi^- \gamma$ final state extend up to an integral cutoff $\Lambda^2$ ranging from $(1.5)^2$ to $(2.5)^2\GeV^2$. As input for the pion vector form factor we use $F_\pi^V(s) = P_\pi^V(s) \Omega(s)$, with the Omn\`es function constructed from the $\pi\pi$ $P$-wave phase shift of the (modified) inverse amplitude method as detailed in App.~\ref{app:phaseshift}, and the (linear) polynomial fit to the $\tau^- \to \pi^- \pi^0 \nu_\tau$ data of Ref.~\cite{Belle:2008xpe}. Here, the polynomial $P_\pi^V(s)$ is continued to a constant $P_\pi^V(s_c)$ for $s>s_c$ in the same way as the $\etapp \to 2(\pi^+ \pi^-)$ amplitudes.
    The outcomes of the two-parameter fits for the dispersive variants with different values for the cut parameter $s_c$ and integral cutoff $\Lambda^2$ are listed in Table~\ref{Tab:etaPpipig_fits}. In Fig.~\ref{Fig:etaP_P}, we show the observable
    \begin{equation}
    \label{Eq:pipig_Pdef}
        \bar{P}_{\etapp}(t) = \Bigg(\frac{1}{\Gamma_0^{\etapp}(t)|\Omega(t)|^2} \frac{\diff \Gamma^{\etapp}}{\diff t}\Bigg)^{1/2},
    \end{equation}
    defined in such a way to remove the effects of $\rho$ peak and phase-space factors from the decay spectrum.

    \begin{figure}[t]
        \centering
        \includegraphics[width=0.8\linewidth]{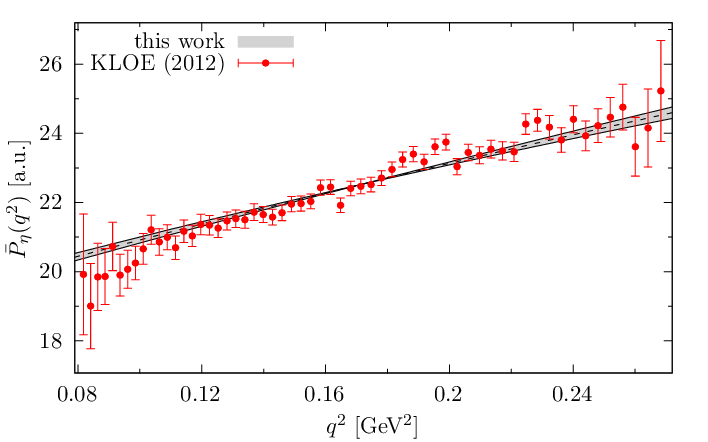}
        \caption{Fit to the $\eta\to\pi^+\pi^-\gamma$ spectrum from KLOE~\cite{KLOE:2012rfx} for the different variants detailed in Table~\ref{Tab:etapipig_fits}, from which the error band is derived (the central value is defined as the average of all fit variants, the band by the maximal difference). $\bar P_{\eta}(q^2)$ is defined in analogy to Eq.~\eqref{Eq:pipig_Pdef}, removing the effects of the $\rho(770)$ resonance and phase-space functions from the spectrum.}
        \label{Fig:eta_P}
    \end{figure}
    
It may be worth commenting on the spectrum in Fig.~\ref{Fig:etaP_P} (as well as the corresponding one for the $\eta$ in Fig.~\ref{Fig:eta_P}) in comparison to prior analyses in the literature.  Originally~\cite{Stollenwerk:2011zz}, the $\pi\pi$ spectra in both radiative decays were described by $P$-wave Omn\`es functions, multiplied by linear polynomials with two free parameters (normalization and slope); additional left-hand cuts due to $a_2$ exchange were shown to induce curvature in $\bar{P}_{\etapp}(t)$~\cite{Kubis:2015sga}, but ultimately failed to describe the experimental data~\cite{BESIII:2017kyd} with sufficient precision, such that a quadratic polynomial with three fit parameters was employed~\cite{Holz:2022hwz}.  Our approach here is different and has \textit{fewer} degrees of freedom: the $\etapp\to\pi^+\pi^-\gamma$ decays are reconstructed via a dispersion relation, based on underlying $\etapp\to2(\pi^+\pi^-)$ amplitudes that come with only one subtraction constant $A$ and one effective coupling $c_{\etapp4\pi}$ for the $a_2$-exchange contribution.  We consider the fact that this approach reproduces the $\pi\pi$ spectra in the radiative $\etapp$ decays to very high accuracy, although maybe not as perfectly as a free three-parameter fit, a highly nontrivial and very convincing validation of our construction.

\begin{figure}[t]
        \centering
        \includegraphics[width=0.49\linewidth]{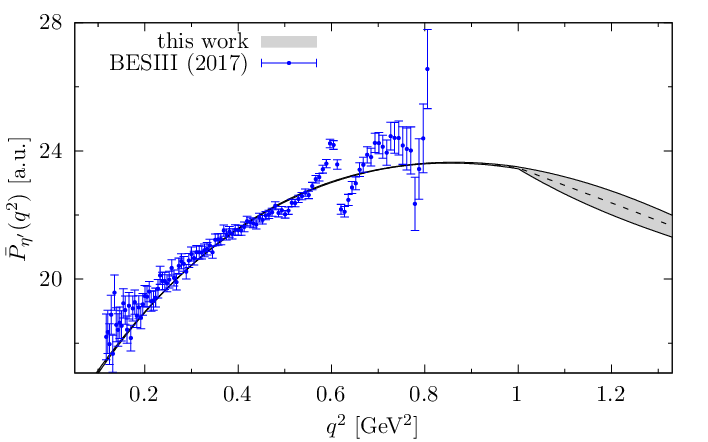}
        \includegraphics[width=0.49\linewidth]{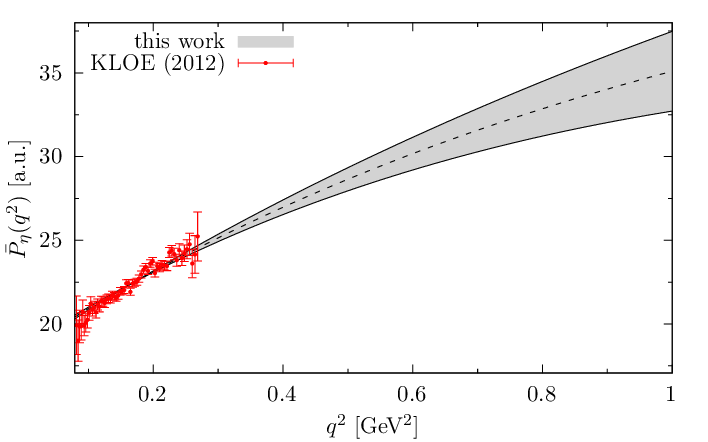}
        \caption{Extrapolation of the fits to the $\etapp\to\pi^+\pi^-\gamma$ spectra beyond the respective phase space. The parts of the spectra in which data are available coincide with Fig.~\ref{Fig:etaP_P} ($\eta'$) and Fig.~\ref{Fig:eta_P} ($\eta$). The bands are derived as in Figs.~\ref{Fig:etaP_P} and \ref{Fig:eta_P}.}
        \label{Fig:eta_P_large}
    \end{figure}

  The same analysis for the $\eta$, see Table~\ref{Tab:etapipig_fits} and Fig.~\ref{Fig:eta_P}, is complicated by the limited phase space. In general, we observe that the fit prefers a smaller effective left-hand-cut coupling than for the $\eta'$, but the sensitivity to the implied curvature is limited. Accordingly, we fix the central value of $c_{\eta 4\pi}$ to $c_{\eta'4\pi}$ reduced by $30\%$, in line with a typical violation of $U(3)$ symmetry, but consider variations by $\pm 15\%$ to account for the associated uncertainties. It is then instructive to also consider the extrapolation of the resulting fit function beyond the respective phase-space boundaries, see Fig.~\ref{Fig:eta_P_large}. For the $\eta'$, one clearly sees the curvature
  in the spectrum, whose high-energy growth is cut at $s_c$. For the $\eta$, the figure illustrates how the available phase space only provides limited sensitivity to the curvature, motivating the additional constraint that arises when restricting the tolerated level of the violation of $U(3)$ symmetry.

\subsection{Analytic continuation to space-like region}
\label{sec:space_like}

Applying the unitarity condition, it is possible to write down a dispersion relation in order to relate the scalar $\etapp \to \pi^+ \pi^- \gamma$ amplitudes of Eq.~\eqref{Eq:etapipig_scfn} with the TFFs through a $\pi^+\pi^-$ intermediate state
\begin{equation}
    \tilde{F}_{\etapp}^{(I=1)}(q_1^2,k^2) = \frac{1}{96\pi^2} \int_{4M_\pi^2}^{\Lambda^2} \diff x\, \frac{x \sigma_\pi^3(x) [F_{\pi}^{V}(x)]^* \mathcal{F}_{\etapp \pi\pi\gamma}(x,k^2)}{x-q_1^2}.
\end{equation}
The dispersion relation above is intentionally kept unsubtracted to ensure the correct asymptotic behavior. A potential violation of the sum rule for the normalization $F_{\etapp \gamma \gamma}$ is to be restored by the addition of the effective-pole pieces $F^{\text{eff}}_{\etapp}$, see Sec.~\ref{sec:isoscalar}. Furthermore, the dispersion integrals are carried out up to a cutoff $\Lambda^2$. In order to facilitate the evaluation in the full space-like $Q_1^2$--$Q_2^2$-plane, it is useful to apply another dispersion relation in the second variable. After symmetrization in both arguments, this enables us to write down the TFFs in a double-spectral representation
 \begin{align}
     F_{\etapp}^{(I=1)}(-Q_1^2,-Q_2^2) &= \frac{1}{\pi^2} \int_{4M_\pi^2}^{\Lambda^2}\diff x\, \diff y\,  \frac{ \rho_{\etapp}(x,y)}{(x+Q_1^2)(y+Q_2^2)} + (Q_1 \leftrightarrow Q_2), \label{Eq:TFF_I1}
\end{align}
     with double-spectral density
\begin{equation} 
    \label{Eq:doubspecdens}
     \rho_{\etapp}(x,y) = \frac{x \sigma_\pi^3(x)}{192 \pi} \Im \Big\{\big[F_\pi^V(x)\big]^*  \mathcal{F}_{\etapp \pi\pi\gamma}(x,y)\Big\}. 
 \end{equation}
 It is this final dispersive representation that we use to obtain our main results for the isovector TFFs in the space-like region. 

\section{Isoscalar and effective-pole contributions}
\label{sec:isoscalar}

In contrast to the elaborate calculation necessary for the isovector contribution described in Sec.~\ref{sec:DR}, the isoscalar TFFs are sufficiently well described by the narrow $\omega$ and $\phi$ resonances 
 \beq
   \label{Eq:TFF_I0}
     F_{\etapp}^{(I=0)}(-Q_1^2,-Q_2^2) = \sum\limits_{V\in \lbrace \omega,\phi\rbrace} \frac{w_{\etapp V \gamma} F_{\etapp \gamma \gamma}  M_V^4}{(M_V^2+Q_1^2)(M_V^2+Q_2^2)}, 
 \eeq
 given that the overall contribution is much smaller and concentrated around the very narrow resonance peaks. The weight factors $w_{\etapp V\gamma}$ are determined phenomenologically via the corresponding decays of $V$ and $\eta'$~\cite{Gan:2020aco}
 \beq
w_{PV\gamma}^2=
 \begin{cases}
 \frac{9M_V^2M_P^3\Gamma[V\to e^+e^-]\Gamma[V\to P\gamma]}{2\alpha(M_V^2-M_P^2)^3\Gamma[P\to\gamma\gamma]} &\text{if}\quad M_V>M_P,\\
 \frac{3M_P^6\Gamma[V\to e^+e^-]\Gamma[P\to V\gamma]}{2\alpha M_V(M_P^2-M_V^2)^3\Gamma[P\to\gamma\gamma]} &\text{if}\quad M_P>M_V,
 \end{cases}
 \eeq
and their signs by comparison to the vector-meson-dominance (VMD) expressions. This strategy has the advantage of automatically accounting for symmetry-breaking terms in an effective Lagrangian approach. Using Eq.~\eqref{F_etagg_etapgg} for $\Gamma[P\to\gamma\gamma]$ and the RPP values for the other branching fractions and decay widths, one obtains 
\begin{align}
w_{\eta\omega\gamma}&=0.099(7), & w_{\eta\phi\gamma}&=-0.188(5),\notag\\
w_{\eta'\omega\gamma}&=0.071(2),& w_{\eta'\phi\gamma}&=0.155(4).
\end{align}
These numbers can vary slightly depending on the treatment of experimental input quantities and vacuum-polarization corrections, e.g., Ref.~\cite{Holz:2022hwz} finds $w_{\eta'\omega\gamma}=0.072(2)$, $w_{\eta'\phi\gamma}=0.158(6)$, but the uncertainties are sufficiently small that they can be neglected in the error propagation, see Ref.~\cite{Holz:2022hwz} for an explicit breakdown in the case of the slope parameter $b_{\eta'}$. 

In general, the sum of isovector and isoscalar contributions constructed so far does not suffice to saturate the TFF normalizations exactly nor to describe the TFFs at virtualities $\gtrsim 1\GeV^2$, both due to the impact of hadronic intermediate states not explicitly included in the representation. To rectify this omission, we add an effective-pole term
 \beq
 \label{eff_pole_A}
     F_{\etapp}^{\text{eff}\,(A)}(-Q_1^2,-Q_2^2) = \frac{g_\text{eff} F_{\etapp \gamma \gamma}   M_{\text{eff}}^4}{(M_{\text{eff}}^2+Q_1^2)(M_{\text{eff}}^2+Q_2^2)},
 \eeq
with coupling constrained to fulfill the normalization exactly and mass parameter fit to singly-virtual space-like TFF data from $e^+e^-\to e^+e^-\etapp$ for $Q^2\geq 5 \GeV^2$. Accordingly, the low-energy TFF remains a prediction, just the transition to the asymptotic region is determined by further data input.  Phenomenologically, the picture that emerges is as follows: for the $\eta$, the sum of the low-energy contributions actually overfulfills the sum rule for the normalization, in such a way that $g_\text{eff}$ becomes negative, in the range $-2\,\%$ to $-13\,\%$, with a mass parameter $M_\text{eff}$ around $(1.3\text{--}2.2)\GeV$. For the $\eta'$, $g_\text{eff}$ is positive, around $5\,\%$, while the effective mass comes out around $M_\text{eff}=1.4\GeV$. In general, the effective-pole contributions thus remain reasonably small, and the mass scales are compatible with the expected contributions of higher intermediate states. 

However, in comparison to the $\pi^0$ case~\cite{Hoferichter:2018dmo,Hoferichter:2018kwz}, we observe that $M_{\text{eff}}$ tends to come out smaller, and the separation of low-energy degrees of freedom and asymptotic constraints is less robust, necessitating a more thorough uncertainty analysis. To this end, we consider a second effective-pole variant
 \beq
 \label{eff_pole_B}
     F_{\etapp}^{\text{eff}\,(B)}(-Q_1^2,-Q_2^2) = \sum\limits_{V\in \lbrace \rho',\rho''\rbrace} \frac{g_V F_{\etapp \gamma \gamma} M_V^4}{(M_V^2+Q_1^2)(M_V^2+Q_2^2)},
\eeq
in which the mass parameters are fixed at the $\rho'$, $\rho''$ masses and the two couplings fit to normalization and singly-virtual data above $5\GeV^2$.  These resonances are expected to subsume the dominant effects not explicitly included in the dispersive representation (together with excited isoscalar resonances in the same mass region), so that this approach should give a perspective on the effective-pole uncertainties complementary to Eq.~\eqref{eff_pole_A}. In the fit to singly-virtual TFF data of one of the effective couplings, the normalization sum rule is being kept fulfilled in every step of the fit iteration
\begin{equation}
\label{Eq:norm_cond}
    F_{\etapp \gamma \gamma} = F_{\etapp}^{(I=1)}(0,0) + F_{\etapp\gamma\gamma} \bigg(w_{\etapp \omega \gamma}+w_{\etapp \phi \gamma}+\sum\limits_{V}g_V\bigg),
\end{equation}
by adjusting the other one accordingly. In case of the $\eta$ TFF, $g_{\rho'}$ is found to in the range $-21\,\%$ to $4\,\%$ and $g_{\rho''}$ in the range $-6\,\%$ to $10\,\%$. For the $\eta'$ TFF, $g_{\rho'}$ is determined to be around $22\,\%$, while $g_{\rho''}$ comes out around $-16\,\%$.
The spread between the two effective-pole variants $(A)$ and $(B)$ will be included in the final uncertainty estimate, see Sec.~\ref{sec:num}.

\section{Matching to short-distance constraints}
\label{sec:SDC}

The leading short-distance constraints are obtained by expanding Eq.~\eqref{TFF_def} around the light cone $x^2=0$. In this way, one obtains the relation~\cite{Lepage:1979zb,Lepage:1980fj,Brodsky:1981rp}
\beq
\label{F_asym_BL}
F_{P\gamma^*\gamma^*}(q_1^2,q_2^2)=-\frac{\bar F^P_\text{asym}}{3}\int_0^1\text{d}u\frac{\phi_P(u)}{u q_1^2+(1-u)q_2^2},
\eeq
where the wave function $\phi_P(u)$ can be expanded in Gegenbauer polynomials, and the leading term in the conformal limit~\cite{Braun:2003rp} becomes $\phi_P(u)=6u(1-u)$. For the pion, the coefficient, $\bar F^\pi_\text{asym}=2F_\pi$, is predicted in terms of the pion decay constant, while for $\etapp$ its value depends on the mixing parameters, see Sec.~\ref{sec:mixing}. In the symmetric asymptotic limit, Eq.~\eqref{F_asym_BL} predicts~\cite{Nesterenko:1982dn,Novikov:1983jt}   
\beq
     \lim\limits_{Q^2\to\infty} Q^2 F_{P \gamma^* \gamma^*}(-Q^2,-Q^2) = \frac{1}{3}\bar{F}^P_{\text{asym}},
 \eeq
a factor three less than in the singly-virtual direction 
 \beq
\lim\limits_{Q^2\to\infty} Q^2 F_{P \gamma^* \gamma^*}(-Q^2,0) = \bar{F}^{P}_{\text{asym}}.
\eeq
The second limit goes beyond a strict operator product expansion~\cite{Gorsky:1987idk,Manohar:1990hu}, resumming higher-order terms into the wave function, which can thus be interpreted as the nonperturbative matrix element in a factorization approach~\cite{Bauer:2002nz,Rothstein:2003wh,Grossman:2015cak}. In addition to the leading result~\eqref{F_asym_BL}, 
$\alpha_s$ corrections~\cite{delAguila:1981nk,Braaten:1982yp} and higher-order terms in the context of QCD sum rules~\cite{Chernyak:1981zz,Chernyak:1983ej,Radyushkin:1996tb,Khodjamirian:1997tk,Agaev:2010aq,Stefanis:2012yw,Agaev:2014wna,Mikhailov:2016klg} were studied in the literature, see Ref.~\cite{Hoferichter:2018kwz} for an estimate of the impact of the $\alpha_s$ corrections on the asymptotic matching for the $\pi^0$ TFF. However, to extend the matching to lower virtualities, arguably, corrections from the finite pseudoscalar mass are likely to generate the most important effect, which naturally changes Eq.~\eqref{F_asym_BL} to~\cite{Hoferichter:2020lap}
\beq
\label{F_asym_BL_mass}
F_{P\gamma^*\gamma^*}(q_1^2,q_2^2)=-\frac{\bar F^P_\text{asym}}{3}\int_0^1\text{d}u\frac{\phi_P(u)}{u q_1^2+(1-u)q_2^2-u(1-u)M_P^2}.
\eeq
These corrections should be retained when reformulating Eq.~\eqref{F_asym_BL} as a dispersion relation~\cite{Khodjamirian:1997tk}.

In general, we follow the approach from Refs.~\cite{Hoferichter:2018dmo,Hoferichter:2018kwz} to rewrite Eq.~\eqref{F_asym_BL} as a double dispersion relation, imposing a lower matching scale $\sm$. In particular, we choose boundary terms in evaluating the double-spectral density 
\beq
\label{double_spectral_density}
\rho^\text{asym}(q_1^2,q_2^2)=-\pi^2 \bar F^P_\text{asym}q_1^2q_2^2\delta''(q_1^2-q_2^2)
\eeq
in such a way that the result vanishes in the singly-virtual limit
\beq
\label{Fasym_massless}
F^\text{asym}_{P\gamma^*\gamma^*}(q_1^2,q_2^2)=\bar F_\text{asym}^P\int_{\sm}^\infty \text{d}x\frac{q_1^2q_2^2}{(x-q_1^2)^2(x-q_2^2)^2}.
\eeq
The motivation for this procedure is that in the singly-virtual limit the dispersive representation has the same asymptotic behavior as Eq.~\eqref{F_asym_BL}, with a coefficient that can be determined by a fit to space-like TFF data measured in $e^+e^-\to e^+e^- P$. With $\bar F_\text{asym}^P$ thus inferred from the data, via a superconvergence relation, also the doubly-virtual contribution is predicted.  

The generalization of the double-spectral density~\eqref{double_spectral_density} to finite pseudoscalar mass was derived in Ref.~\cite{Zanke:2021wiq}. To preserve the behavior of Eq.~\eqref{Fasym_massless} for small virtualities, appropriate subtractions need to be introduced~\cite{Hoferichter:2024bae}, leading to the form
 \begin{align}
 \label{Eq:TFF_asym}
     F_{P}^{\text{asym}}(q_1^2,q_2^2) &= \frac{- \bar{F}_{\text{asym}}^{P}}{M_{P}^4} \int_{2 \sm}^{\infty}\text{d}v \bigg[ \frac{q_2^2}{v-q_1^2} \bigg[\frac{1}{v-q_1^2-q_2^2} - \frac{1}{q_1^2-q_2^2} \bigg] f^{\text{asym}}_{P}(v,q_1^2) + (q_1^2 \leftrightarrow q_2^2) \bigg], \notag \\
     f^{\text{asym}}_{P}(v,q^2) &= \frac{(v-2q^2)^2 - M_{P}^2 v}{\sqrt{(v-2q^2)^2 - 2 M_{P}^2 v +M_{P}^4}} +2 q^2 - v.
 \end{align}
 In this form, $F_{P}^{\text{asym}}(q_1^2,q_2^2)$ reduces to Eq.~\eqref{Fasym_massless} in the limit $M_P\to 0$, and the behavior $F_{P}^{\text{asym}}(q_1^2,q_2^2)=\Order(q_1^2q_2^2)$ for small virtualities is maintained. For the pion, these pseudoscalar mass effects are negligible, but for $P=\etapp$ we observe that keeping the corresponding mass corrections indeed improves the matching to short-distance constraints. 

 For the numerical analysis, we therefore employ an asymptotic contribution in the form~\eqref{Eq:TFF_asym}. Motivated by light-cone sum rules~\cite{Khodjamirian:1997tk,Agaev:2014wna} (see also App.~E of Ref.~\cite{Hoferichter:2018kwz}), we set $\sm=1.5(3)\GeV^2$ for the $\eta'$, while for the $\eta$ we allow for a larger range, $\sm=1.4(4)\GeV^2$, to include TFF variations that display a slightly smoother transition in the doubly-virtual direction. We also investigated alternative formulations in which the asymptotic contribution does not vanish in the singly-virtual direction, similarly to the strategy for the short-distance matching of axial-vector TFFs in Ref.~\cite{Hoferichter:2024bae}, but found no further improvement compared to Eq.~\eqref{Eq:TFF_asym}. 

The asymptotic coefficients $\bar{F}_{\text{asym}}^{P}$ follow from a superconvergence relation
\begin{equation}
    \bar{F}_{\text{asym}}^{P} = F_{P \gamma \gamma}\sum\limits_{V} g_V M_V^2 +\frac{1}{\pi^2} \int_{4M_\pi^2}^{\Lambda^2} \diff x\, \diff y \bigg[\frac{\rho_{P}(x,y)}{x} + \frac{\rho_{P}(x,y)}{y}\bigg],
\end{equation}
written in terms of the double-spectral densities defined in Eq.~\eqref{Eq:doubspecdens}. The sum extends over $V\in \{\omega,\, \phi,\, \text{eff}\}$ in case of effective-pole variant $(A)$, where $g_{\omega/\phi} \equiv w_{P \omega/\phi \gamma}$, and $V\in \{\omega,\, \phi, \rho',\, \rho''\}$ for effective-pole variant $(B)$. The numerical results for $\bar{F}_{\text{asym}}^{\etapp}$ are provided in Eq.~\eqref{barF_asym_num}.

 \begin{figure}[t]
     \centering
     \includegraphics[width=0.9\linewidth]{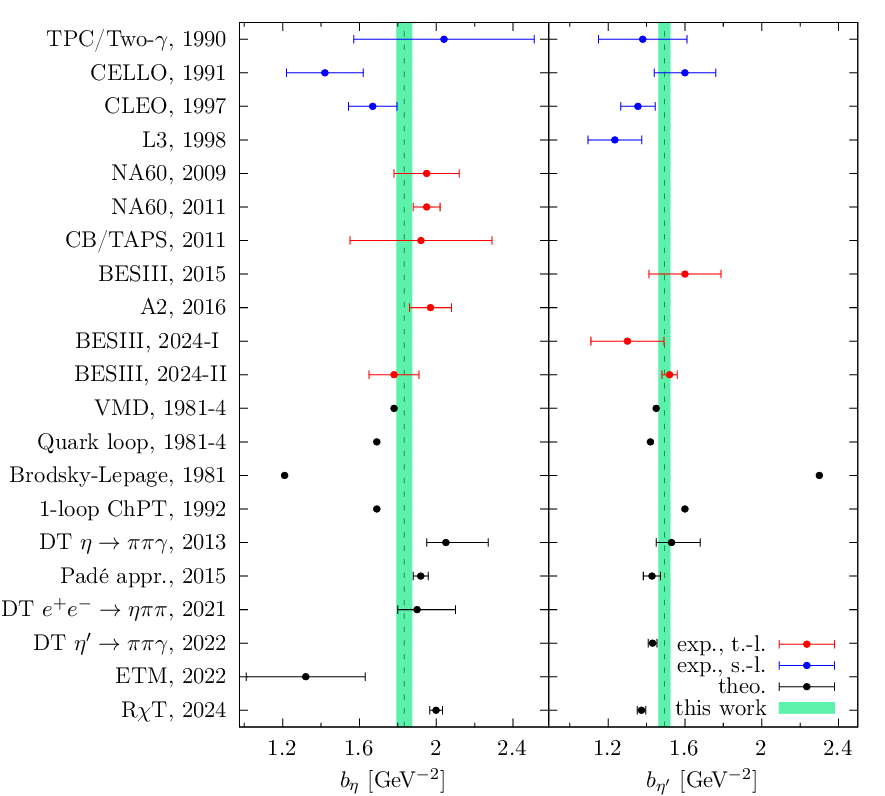}
     \caption{Comparison of the $\eta$ and $\eta'$ slope parameters with previous time-like and space-like experimental measurements as well as theoretical analyses. References are given in the main text.}
     \label{Fig:slopes}
 \end{figure}

\section{Numerical results}
\label{sec:num}

In this section, we discuss our numerical results for the slope parameters $b_{\etapp}$, the $\etapp$ decay constants and mixing angles, the space-like TFFs in singly- and doubly-virtual kinematics, and the $\etapp$-pole contributions to $a_\mu$. In each case, we assess the uncertainties as follows:  the uncertainty in the normalizations of the TFFs is propagated from the RPP values given in Eq.~\eqref{F_etagg_etapgg} (``norm''); for the uncertainty of the dispersive representation (``disp''), we consider different variants for integral cutoffs  $\Lambda$ and cut parameters $s_c$ (and $U(3)$-symmetry violation for the $\eta$), as given in Tables~\ref{Tab:etaPpipig_fits} and~\ref{Tab:etapipig_fits}, assigning the maximal variation as the resulting error; the singly-virtual Brodsky--Lepage (``BL'') limit is described by effective poles, with parameters varied within the fit uncertainties and scanning over the two variants defined in Eqs.~\eqref{eff_pole_A} and~\eqref{eff_pole_B}; for the uncertainty of the asymptotic contribution (``asym'') we vary the threshold parameter $\sm$ as described in Sec.~\ref{sec:SDC}, adding the variation observed when replacing our superconvergence values for $\bar F_\text{asym}^{\etapp}$ by the determination from Ref.~\cite{Bali:2021qem} (see also Ref.~\cite{Ottnad:2017bjt} for an earlier calculation in lattice QCD). All uncertainties are added in quadrature. 

\subsection{Slope parameters}
\label{sec:slope}

The slope parameters of the TFFs are defined as 
\beq
b_{\etapp} \equiv \frac{1}{F_{\etapp\gamma\gamma}} \frac{\partial}{\partial q^2}  F_{\etapp \gamma^* \gamma^*}(q^2,0) \big|_{q^2=0},
\eeq
By construction, the asymptotic part of the TFF representation does not contribute, and we obtain
 \begin{align}
     b_{\eta} &= 1.833\,(16)_\text{norm}\,(36)_\text{disp}\,(9)_\text{BL}\, [41]_\text{tot}\GeV^{-2},\notag\\
     b_{\eta'} &= 1.493\,(10)_\text{norm}\,(30)_\text{disp}\,(6)_\text{BL}\,[32]_\text{tot}\GeV^{-2}.
     \label{Eq:slope}
 \end{align}
 In  both cases, the result is broadly consistent with determinations from space-like experiments~\cite{TPCTwoGamma:1990dho,CELLO:1990klc,CLEO:1997fho,L3:1997ocz}, time-like measurements~\cite{Dzhelyadin:1979za,Dzhelyadin:1980kh,NA60:2009una,Usai:2011zza,Berghauser:2011zz,BESIII:2015zpz,Adlarson:2016hpp,BESIII:2024awu,BESIII:2024pxo}, and earlier theoretical determinations~\cite{Bramon:1981sw,Ametller:1983ec,Pich:1983zk,Brodsky:1981rp,Ametller:1991jv,Escribano:2013kba,Escribano:2015nra,Escribano:2015yup,Hanhart:2013vba,Holz:2022hwz,Estrada:2024cfy}, see Fig.~\ref{Fig:slopes} for an overview. For the $\eta'$, we can compare to a highly optimized dispersive representation for the  singly-virtual TFF, including a fully consistent treatment of isospin breaking due to  
 $\rho$--$\omega$ mixing, which gives $b_{\eta'} = 1.431(23)\GeV^{-2}$~\cite{Holz:2022hwz}, in reasonable agreement with the outcome of the present analysis.

 \subsection[$\etapp$ mixing parameters]{$\boldsymbol{\etapp}$ mixing parameters}
 \label{sec:mixing}

\begin{table}[t]
	\renewcommand{\arraystretch}{1.3}
	    \centering
        \scalebox{0.89}{
	    \begin{tabular}{lrrrrrrrr}
	         \toprule
	         &$\frac{F_8}{F_\pi}$ & $\frac{F_0}{F_\pi}$ & $\theta_8 \ [^\circ]$ & $\theta_0 \ [^\circ]$ & $\Lambda_3$ & $\Lambda_1$ & $K_2\ \big[\text{GeV}^{-2}\big]$ & $\chi^2$\\\midrule
             Ref.~\cite{Bali:2021qem} & $1.25(3)$ & $1.15(3)$ & $-25.8(2.3)$ & $-8.1(1.8)$ &-- & -- & -- & --\\
             Ref.~\cite{Escribano:2015yup} & $1.27(2)$ & $1.14(5)$ & $-21.2(1.9)$ & $-6.9(2.4)$ & $-0.02(7)$ & $0.01(13)$ & $-0.45(57)$ & $2.01$\\
             This work & $1.28(3)$ & $1.19(5)$ & $-22.4(1.0)$ & $-9.0(1.8)$ & $0.01(5)$ & $0.14(12)$ & $0.04(15)$ & $1.79$\\
	         \bottomrule
	    \end{tabular}}
	    \caption{$\etapp$ mixing parameters derived in this work, in comparison to the similar analysis from Ref.~\cite{Escribano:2015yup} and the lattice-QCD calculation from Ref.~\cite{Bali:2021qem}. In all cases, $F_0$ is defined at $\mu_0=1\GeV$.}
	    \label{tab:mixing_parameters}
		\renewcommand{\arraystretch}{1.0}
	\end{table}

 Defining the pseudoscalar decay constants $F_P^a$ by 
 \beq
\Big\langle 0\Big|\bar q\gamma^\mu\gamma_5\frac{\lambda^a}{2}q\Big|P(p)\Big\rangle = i p^\mu F_P^a,
 \eeq
 with Gell-Mann matrices $\lambda^a$ and $\lambda^0=\sqrt{2/3}{\mathds{1}}_3$, we employ the singlet--octet two-angle mixing scheme~\cite{Feldmann:1998vh,Feldmann:1999uf,Escribano:2005qq}
 \beq
\begin{pmatrix}
    F_\eta^8 & F_\eta^0\\
    F_{\eta'}^8 & F_{\eta'}^0
\end{pmatrix}
\equiv\begin{pmatrix}
F_8 \cos \theta_8& -F_0 \sin \theta_0\\
F_8\sin\theta_8 & F_0\cos\theta_0
\end{pmatrix},
 \eeq
which overcomes the limitations of a one-angle scheme at leading order in the chiral and large-$N_c$ expansion. To determine these mixing parameters using as input $F_{\etapp\gamma\gamma}$ from Eq.~\eqref{F_etagg_etapgg} and our superconvergence results for $\bar F_\text{asym}^{\etapp}$, 
\begin{align}
\label{barF_asym_num}
    \bar F_\text{asym}^\eta&=0.186(7)_\text{norm}(7)_\text{disp}(9)_\text{BL} [13]_\text{tot}\GeV,\notag \\
    \bar F_\text{asym}^{\eta'}&=0.264(5)_\text{norm}(5)_\text{disp}(11)_\text{BL} [13]_\text{tot}\GeV,
\end{align}
we follow the strategy put forward in Ref.~\cite{Escribano:2015yup}. First, one has at next-to-leading order in large-$N_c$ ChPT~\cite{Feldmann:1998vh,Feldmann:1999uf,Escribano:2005qq}
\beq
\label{F08_ChPT}
F_8^2=\frac{4F_K^2-F_\pi^2}{3},\qquad F_0^2=\frac{2F_K^2+F_\pi^2}{3}+F_\pi^2\Lambda_1,\qquad 
F_8 F_0\sin(\theta_8-\theta_0)=-\frac{2\sqrt{2}}{3}\big(F_K^2-F_\pi^2\big).
\eeq

Defining the scale-dependent singlet decay constant as $F_0\equiv F_0(\mu_0)$, $\mu_0=1\GeV$, as appropriate for the decomposition of the two-photon decay widths, one then needs to introduce renormalization-group (RG) corrections for the asymptotic coefficients~\cite{Leutwyler:1997yr,Kaiser:2000gs,Agaev:2014wna}, which can be subsumed into 
\beq
F_0(\mu)=F_0(\mu_0)\bigg[1+\frac{2N_f}{\pi \beta_0}\Big(\alpha_s(\mu)-\alpha_s(\mu_0)\Big)\bigg]\equiv F_0(\mu_0)\big[1+\delta(\mu,\mu_0)\big],
\eeq
where $\beta_0=11-2N_f/3$, $N_f$ the number of active quark flavors, and $\delta_\infty\equiv\delta(\infty,\mu_0)=-0.10$~\cite{Escribano:2015nra,Escribano:2015yup}. Introducing weights as 
\beq
C_a=\frac{1}{2}\text{Tr}\big({\mathcal Q}^2\lambda_a\big),\qquad C_8=\frac{1}{6\sqrt{3}},\qquad C_0=\frac{2}{3\sqrt{6}},
\eeq
as well as versions including higher-order contributions (both chiral and large-$N_c$-suppressed corrections)~\cite{Leutwyler:1997yr,Kaiser:2000gs,Escribano:2015yup}
\beq
\bar C_8=C_8\Big(1+\frac{K_2}{3}\big(7\mpi^2-4M_K^2\big)\Big),\qquad 
\bar C_0=C_0\Big(1+\Lambda_3+\frac{K_2}{3}\big(2\mpi^2+M_K^2\big)\Big),
\eeq
one has~\cite{Escribano:2015yup}
\begin{align}
\label{FPgg_FPasym}
\bar F_\text{asym}^P&=12\big(C_8F_P^8+C_0(1+\delta_\infty) F_P^0\big),\notag\\
F_{\eta\gamma\gamma}&=\frac{3}{2\pi^2}\frac{\bar C_8 F^0_{\eta'}-\bar C_0 F_{\eta'}^8}{F_{\eta'}^0F_\eta^8-F_{\eta'}^8F_\eta^0},\qquad 
F_{\eta'\gamma\gamma}=\frac{3}{2\pi^2}\frac{\bar C_8 F^0_{\eta}-\bar C_0 F_{\eta}^8}{F_{\eta}^0F_{\eta'}^8-F_{\eta}^8F_{\eta'}^0}.
\end{align}
\begin{table}[t!]
	\renewcommand{\arraystretch}{1.3}
	    \centering
	    \begin{tabular}{lrrrrrrr}
	         \toprule
	         &$\frac{F_8}{F_\pi}$ & $\frac{F_0}{F_\pi}$ & $\theta_8$ & $\theta_0$ & $\Lambda_3$ & $\Lambda_1$ & $K_2$\\\midrule
             $\frac{F_8}{F_\pi}$ & $1.00$ & $0.07$ & $-0.33$ & $0.01$ & $0.09$ & $0.07$ & $-0.08$\\
             $\frac{F_0}{F_\pi}$ & & $1.00$ & $-0.16$ & $-0.18$ & $0.86$ & $0.92$ & $0.09$\\
             $\theta_8$ & & & $1.00$ & $0.27$ & $0.01$ & $-0.15$ & $-0.65$\\
             $\theta_0$ & & & & $1.00$ & $-0.45$ & $-0.16$ & $-0.20$\\
             $\Lambda_3$ & & & & & $1.00$ & $0.80$ & $-0.13$\\
             $\Lambda_1$ & & & & & & $1.00$ & $0.09$\\
             $K_2$ & & & & & & & $1.00$\\
	         \bottomrule
	    \end{tabular}
	    \caption{Correlation coefficients among the different quantities quoted in Table~\ref{tab:mixing_parameters}.}
	    \label{tab:correlations_mixing}
		\renewcommand{\arraystretch}{1.0}
	\end{table}
  %
%
\begin{figure}[t]
    \centering
    \includegraphics[width=0.9\linewidth]{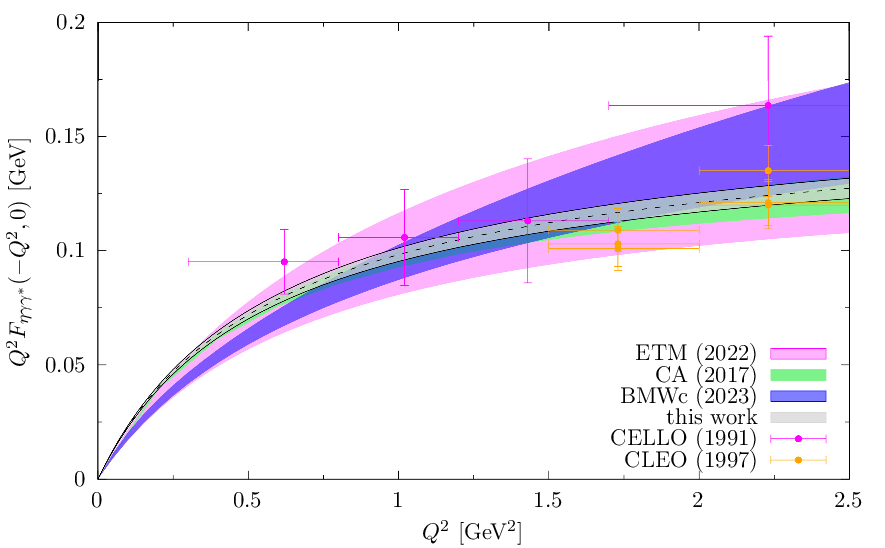}
    \includegraphics[width=0.9\linewidth]{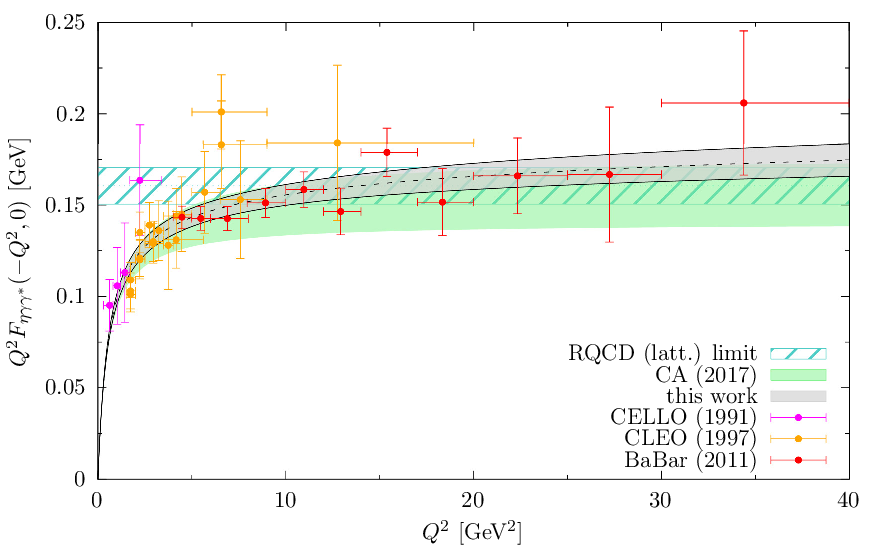}
    \caption{Singly-virtual $\eta$ TFF, in comparison to CA~\cite{Masjuan:2017tvw}, ETM~\cite{ExtendedTwistedMass:2022ofm}, BMWc~\cite{Gerardin:2023naa}, and the data from CELLO~\cite{CELLO:1990klc} and CLEO~\cite{CLEO:1997fho}. The lower figure, displaying a broader range in $Q^2$, also shows the data from BaBar~\cite{BaBar:2011nrp} and the asymptotic value implied by  RQCD~\cite{Bali:2021qem}. Only data with $Q^2\geq 5\GeV^2$ are included in our fit.}
    \label{Fig:eta_singly}
\end{figure}
\begin{figure}[t]
    \centering
    \includegraphics[width=0.9\linewidth]{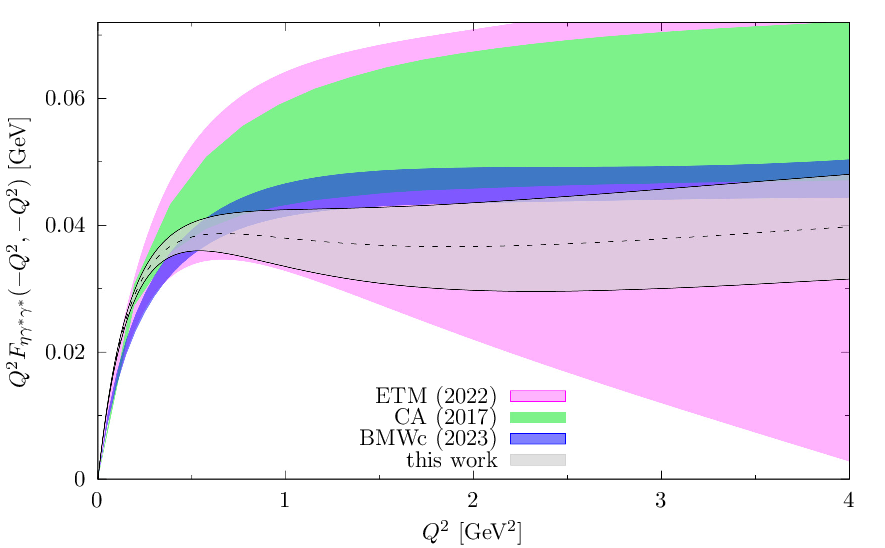}
    \includegraphics[width=0.9\linewidth]{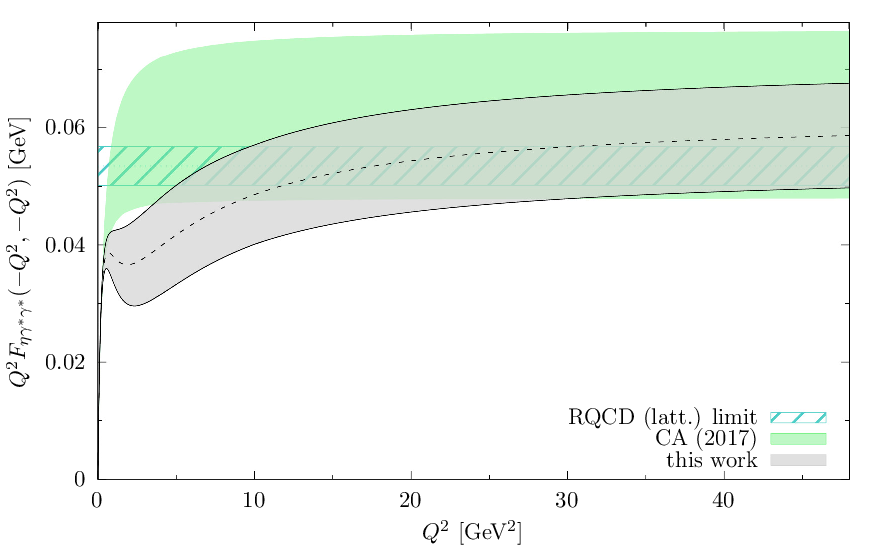}
    \caption{Doubly-virtual $\eta$ TFF, with legends as in Fig.~\ref{Fig:eta_singly}.}
    \label{Fig:eta_doubly}
\end{figure}
%
\begin{figure}[t]
    \centering
    \includegraphics[width=0.9\linewidth]{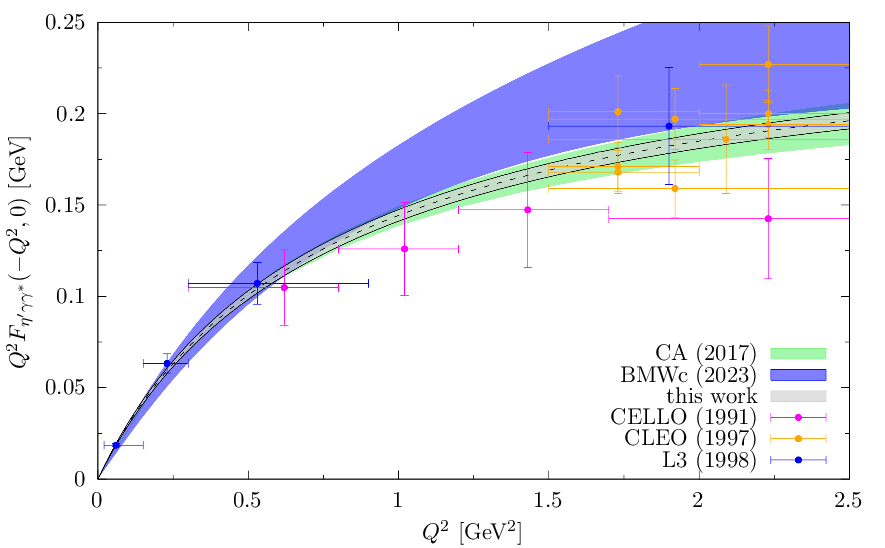}
    \includegraphics[width=0.9\linewidth]{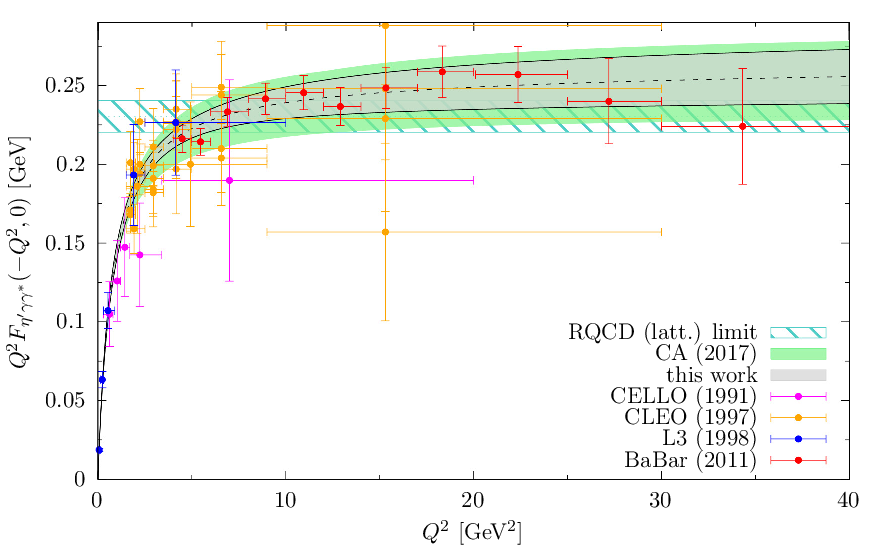}
    \caption{Singly-virtual $\eta'$ TFF, with legends as in Fig.~\ref{Fig:eta_singly}. The L3 data are from Ref.~\cite{L3:1997ocz}.}
    \label{Fig:etap_singly}
\end{figure}
The mixing angles drop out in the combination
\begin{align}
\label{Fgg_Fasym_comb}
\bar F_\text{asym}^\eta F_{\eta\gamma\gamma}+\bar F_\text{asym}^{\eta'} F_{\eta'\gamma\gamma}&=\frac{18}{\pi^2}\Big(C_8\bar C_8+C_0\bar C_0(1+\delta_\infty)\Big)\notag\\
&=\frac{3}{2\pi^2}\bigg[1+\frac{8}{9}\Big((1+\delta_\infty)(1+\Lambda_3)-1\Big)\notag\\
&\qquad+\frac{K_2}{27}\Big(\mpi^2(23+\delta_\infty)+4M_K^2(1+2\delta_\infty)\Big)\bigg],
\end{align}
 which is therefore predicted by the anomaly apart from RG, singlet, and quark-mass corrections, parameterized by $\delta_\infty$, $\Lambda_3$, and $K_2$, respectively. The scale dependence inherent in $\delta_\infty$ and $\Lambda_3$ drops out up to higher orders in the expansion. $K_2$ also describes the iso-symmetric quark-mass dependence of $\pi^0\to\gamma\gamma$, for which constraints from lattice QCD are available, $K_2=-0.13(15)\GeV^{-2}$~\cite{Gerardin:2019vio}. Motivated by this range and the even smaller estimate from Ref.~\cite{Kampf:2009tk}, we set $K_2=0$ with an error $\Delta K_2=0.15\GeV^{-2}$. This constraint, the four conditions in Eq.~\eqref{FPgg_FPasym}, and the three ChPT relations~\eqref{F08_ChPT} then amount to eight equations for the seven unknowns $F_8$, $F_0$, $\theta_8$, $\theta_0$, $\Lambda_3$, $\Lambda_1$, and $K_2$. A $\chi^2$ minimization yields the results collected in Table~\ref{tab:mixing_parameters}, where we followed Ref.~\cite{Escribano:2015yup} and accounted for the uncertainty due to higher chiral orders in Eq.~\eqref{F08_ChPT} by assigning an additional $2.4\%$ uncertainty to $F_K/F_\pi=1.1978(22)$~\cite{Dowdall:2013rya,Bazavov:2017lyh,Miller:2020xhy,ExtendedTwistedMass:2021qui,Cirigliano:2022yyo}. Our results are consistent with Ref.~\cite{Escribano:2015yup}, albeit indicating a slightly larger value for $F_0$ and $\theta_0$ (the former being compensated by a corresponding change in $\Lambda_1$). In the comparison to the lattice-QCD calculation of Ref.~\cite{Bali:2021qem}, the biggest difference occurs in $\theta_8$, but even here the results are compatible, especially, if one adds a scale factor to account for the $\chi^2>1$.\footnote{We disagree with Ref.~\cite{Escribano:2015yup} regarding the number of degrees of freedom, because Eq.~\eqref{Fgg_Fasym_comb} is not an independent constraint. Even for $\text{dof}=1$, however, the resulting $p$-value is still $18\%$.} Due to the various constraints, the uncertainties quoted in Table~\ref{tab:mixing_parameters} are not independent, with the correlations given in Table~\ref{tab:correlations_mixing}. Most correlations are reasonably small, apart from the expected strong correlation among $F_0$ and the singlet corrections $\Lambda_3$, $\Lambda_1$. In addition, $\theta_8$ displays a strong correlation with $K_2$, which drives the change in $\theta_8$ compared to Ref.~\cite{Escribano:2015yup}, while the changes in $F_0$, $\theta_0$ largely derive from the higher values of $\bar F^{\etapp}_\text{asym}$.

\begin{figure}[t]
    \centering
    \includegraphics[width=0.9\linewidth]{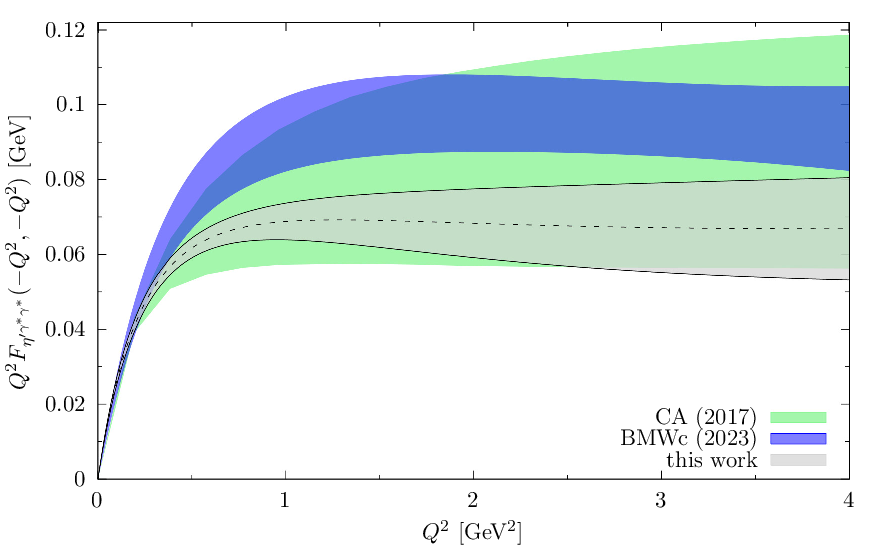}
    \includegraphics[width=0.9\linewidth]{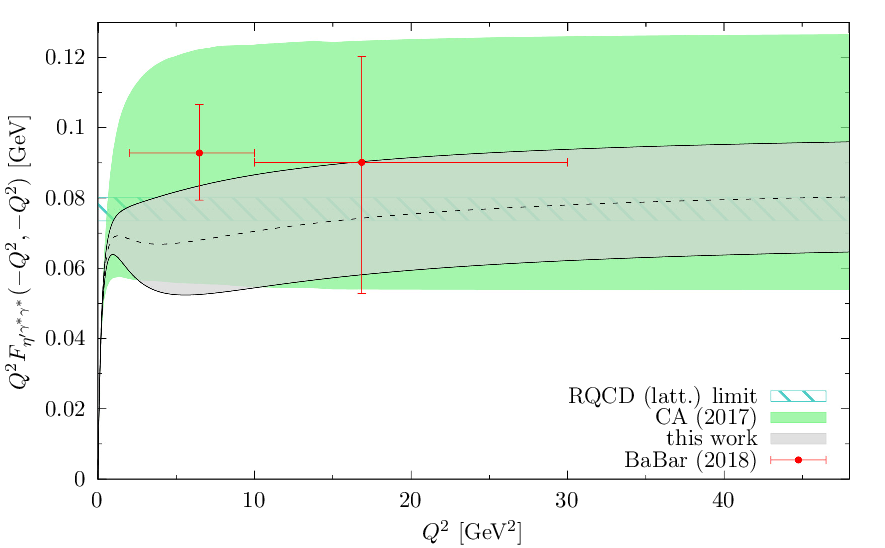}
    \caption{Doubly-virtual $\eta'$ TFF, with legends as in Fig.~\ref{Fig:eta_singly}. The two data points from BaBar~\cite{BaBar:2018zpn} are not included in the fit.}
    \label{Fig:etap_doubly}
\end{figure}
\begin{figure}
    \centering
    \includegraphics[width=0.49\linewidth]{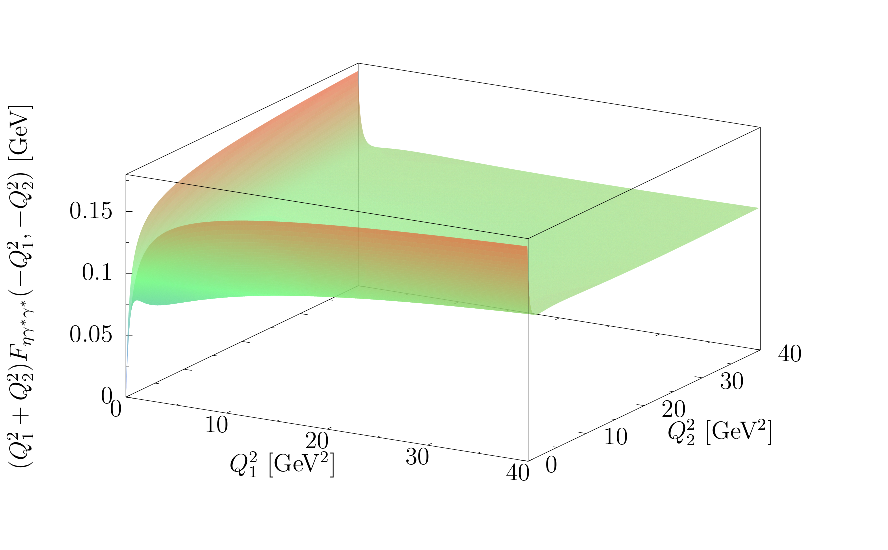}
    \includegraphics[width=0.49\linewidth]{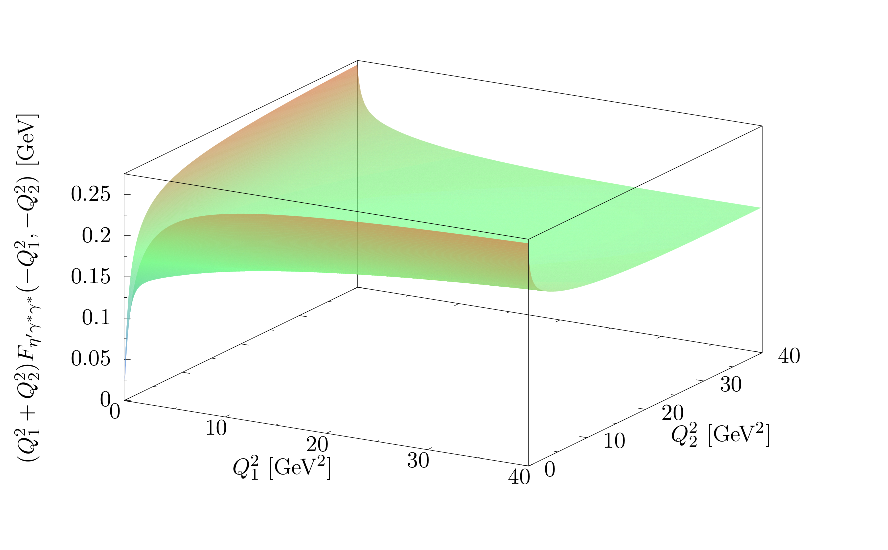}
    \caption{Three-dimensional representation of $(Q_1^2+Q_2^2)F_{\etapp\gamma^*\gamma^*}(-Q_1^2,-Q_2^2)$ for the $\eta$ (left) and $\eta'$ (right) TFFs.}
    \label{Fig:3dplots}
\end{figure}

\subsection{Space-like transition form factors}
\label{sec:TFF_space_like}

\begin{figure}
    \centering
    \includegraphics[width=0.85\linewidth]{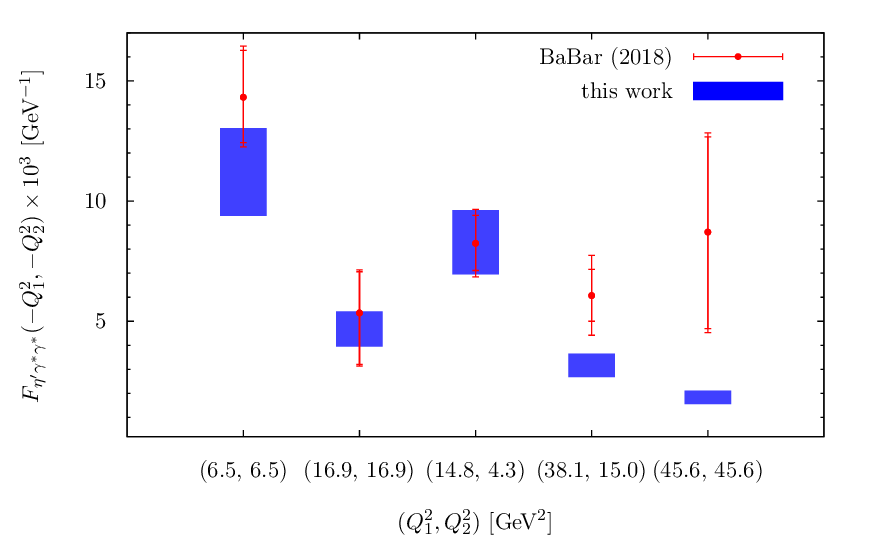}
    \caption{Doubly-virtual $\eta'$ TFF in comparison to the data points from BaBar~\cite{BaBar:2018zpn} with both statistical and total errors.}
    \label{Fig:etap_babar}
\end{figure}

Our results for the space-like TFFs are illustrated in Figs.~\ref{Fig:eta_singly}--\ref{Fig:etap_babar}. First, Fig.~\ref{Fig:eta_singly} shows the comparison for the singly-virtual $\eta$ TFF, in comparison to the available data and selected previous calculations. In general, we observe good agreement, especially for the data with $Q^2\leq 5 \GeV^2$ not included in the fit, while the mismatch to lattice QCD for small virtualities likely reflects the lower value for the normalization $F_{\eta\gamma\gamma}$. In the doubly-virtual direction, see Fig.~\ref{Fig:eta_doubly}, we observe a slower rise of the TFF than in the CA approach, while our asymptotic value even comes out slightly higher. Ultimately, this behavior is driven by the interplay between low-energy dispersive, isoscalar, and the effective-pole contributions, since the negative  effective coupling $g_\text{eff}$ causes the asymptotic value to be saturated more slowly than for an effective pole with opposite sign, and could be scrutinized once additional input from low-energy doubly-virtual data or lattice QCD becomes available. 

The analog plots for the $\eta'$ TFF are shown in Figs.~\ref{Fig:etap_singly} and~\ref{Fig:etap_doubly}, respectively.  In general, we observe again good agreement with previous work as well as the experimental results, although especially in the doubly-virtual direction our curve lies below the one by BMWc. The transition to the asymptotic region indeed proceeds faster than for the $\eta$ TFF, reflecting the fact that $g_\text{eff}$ is positive. Moreover, we found that the mass corrections described in Sec.~\ref{sec:SDC} also tend to lead to a faster increase, affecting the $\eta'$ TFF more strongly than for the $\eta$. We also checked a representation in which part of the singly-virtual TFF is carried by $F^\text{asym}_\eta(q_1^2,q_2^2)$, but the same behavior as for both effective-pole variants remains. Accordingly, the fact that the large size of the combined isovector and isoscalar low-energy contributions to the $\eta$ TFF---and the required compensation by higher intermediate states to reproduce the experimental normalization---enforces a slower transition to the asymptotic form seems to be rather robust among the different interpolations we considered.

Finally, we illustrate the entire $Q_1^2$--$Q_2^2$ range in Fig.~\ref{Fig:3dplots}, again indicating the faster rise to the asymptotic form in the case of the $\eta'$. The numerical results for the $\eta$ and $\eta'$ TFFs in the space-like region corresponding to this figure are provided as ancillary files.
For the $\eta'$, one can also compare to the nondiagonal doubly-virtual data points from BaBar~\cite{BaBar:2018zpn}, see Fig.~\ref{Fig:etap_babar}. Here, some disagreement occurs for the points with the largest values $Q_1^2\simeq 40\GeV^2$, as observed before in the CA approach~\cite{Masjuan:2017tvw}.

\subsection[Pole contributions to $a_\mu$]{Pole contributions to $\boldsymbol{a_\mu}$}
\label{sec:amu}

\begin{figure}[t]
    \centering
    \includegraphics[width=1.\linewidth]{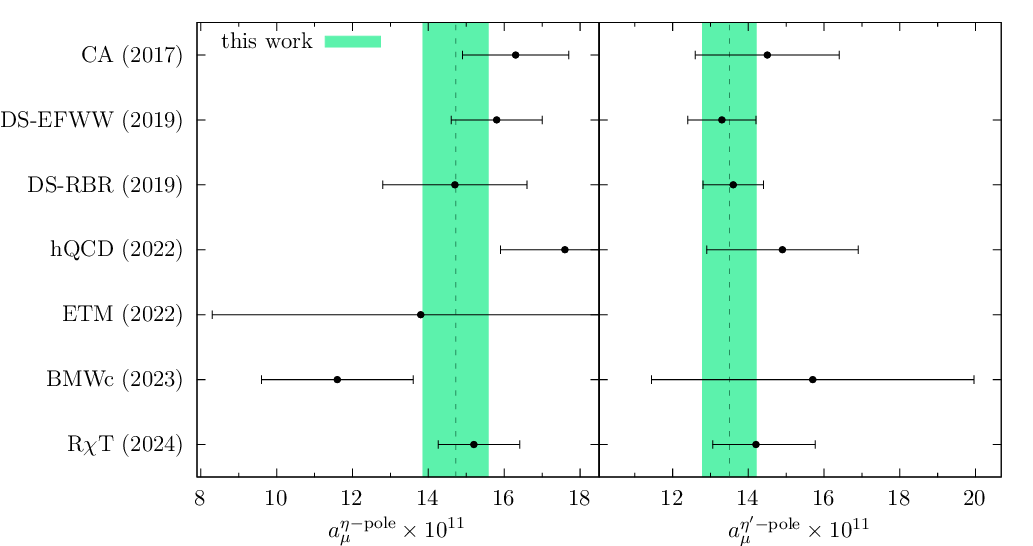}
    \caption{Comparison of the $\eta$ and $\eta'$ pole contributions $a_\mu^{\etapp\text{-pole}}$ between the results of the CA analysis from Ref.~\cite{Masjuan:2017tvw}, two Dyson--Schwinger analyses (DS-EFWW~\cite{Eichmann:2019tjk} and DS-RBR~\cite{Raya:2019dnh}), the holographic-QCD (hQCD) result from Ref.~\cite{Leutgeb:2022lqw}, the lattice-QCD calculations by ETM~\cite{ExtendedTwistedMass:2022ofm} and BMWc~\cite{Gerardin:2023naa}, and the Resonance Chiral Theory (R$\chi$T) analysis  from Ref.~\cite{Estrada:2024cfy} with the results of this work.}
    \label{Fig:pole_conts}
\end{figure}

Using our results for the space-like TFFs as described in Sec.~\ref{sec:TFF_space_like}, the $\etapp$-pole contributions to $a_\mu$ follow from the master formula, Eqs.~\eqref{eq:master_formula} and~\eqref{Eq:Pi_pole}, leading to
 \begin{align}
   a_\mu^{\eta\text{-pole}}&=14.72(56)_\text{norm}(32)_\text{disp}(23)_\text{BL}(54)_\text{asym}[87]_\text{tot}\times 10^{-11},\notag\\
   a_\mu^{\eta'\text{-pole}}&=13.50(48)_\text{norm}(15)_\text{disp}(20)_\text{BL} (48)_\text{asym}[72]_\text{tot}\times 10^{-11},
   \label{Eq:amu}
 \end{align}
where the uncertainties are propagated from the TFFs as before. While our results agree with recent analyses of the pseudoscalar-pole contributions~\cite{Masjuan:2017tvw,ExtendedTwistedMass:2022ofm,Gerardin:2023naa,Czyz:2017veo,Guevara:2018rhj,Estrada:2024cfy,Eichmann:2019tjk,Raya:2019dnh,Hong:2009zw,Leutgeb:2019zpq}, see Fig.~\ref{Fig:pole_conts}, the highly constrained representation for the TFFs translates to reduced uncertainties in $a_\mu^{\etapp\text{-pole}}$. Equation~\eqref{Eq:amu} constitutes the main result of our analysis.

Other definitions of pseudoscalar contributions have been used in the literature, e.g., employing a constant TFF at the singly-virtual vertex~\cite{Melnikov:2003xd}, which apart from meson-mass corrections amounts to a definition in triangle instead of four-point kinematics~\cite{Colangelo:2019uex,Ludtke:2023hvz}. Moreover, definitions of a so-called pseudoscalar-exchange contribution include off-shell-meson effects~\cite{Bartos:2001pg,Jegerlehner:2009ry,Hong:2009zw,Goecke:2010if,Dorokhov:2011zf,Dorokhov:2012qa,Roig:2014uja,Dorokhov:2015psa}, which we do not consider further due to the inherent model dependence. While the central value of the final result~\eqref{Eq:amu} comes out remarkably close to the pioneering calculations $a_\mu^{\eta+\eta'}\simeq 26\times 10^{-11}$ (see Refs.~\cite{Bijnens:2007pz,Jegerlehner:2009ry} for earlier compilations) in the extended Nambu--Jona-Lasinio model~\cite{Bijnens:1995cc,Bijnens:1995xf} and VMD/HLS models~\cite{Hayakawa:1995ps,Hayakawa:1996ki,Hayakawa:1997rq}, the main progress over the last years concerns the precision with which the pseudoscalar-pole contributions can now be evaluated.

\section{Conclusions}
\label{sec:summary}

We presented a comprehensive study of the $\etapp$ TFFs using a dispersive approach, including a number of inputs from both experiment and theory to constrain various properties of the TFFs. The normalizations are determined from
$\etapp\to\gamma\gamma$, the momentum dependence of the isoscalar TFFs via vector-meson couplings that follow from measured branching fractions. The most detailed analysis was performed for the dominant isovector TFFs, for which the unitarity relation was solved including the left-hand-cut singularity due to the $a_2$ resonance. In particular, we detailed how to construct the underlying $\etapp\to2(\pi^+\pi^-)$ amplitude in a way consistent with chiral symmetry, how to numerically solve the required inhomogeneous Muskhelishvili--Omn\`es problem in a stable manner via a carefully chosen path deformation, and how to determine the free parameters from a fit to the $\etapp\to\pi^+\pi^-\gamma$ spectra. The asymptotic behavior of the TFFs was incorporated by matching to the leading result from the light-cone expansion, augmented by the dominant corrections due to $\etapp$ mass effects. Finally, the transition between the low-energy dispersive representations and the short-distance constraints was described by effective poles, with parameters determined by imposing the exact normalization of the resulting representation and by fitting singly-virtual, space-like data measured in $e^+e^-\to e^+e^-\etapp$ for virtualities $Q^2\geq 5 \GeV^2$. For all contributions we performed a comprehensive error analysis, propagating uncertainies from the experimental input quantities as well as theoretical uncertainties from the cutoff parameters in the dispersive representation, the parameterization of the effective poles, and the transition to the asymptotic region.

Our main application concerns the evaluation of the $\etapp$-pole contributions to HLbL scattering, see Eq.~\eqref{Eq:amu} for the main result. In addition, we calculated the slope parameters~\eqref{Eq:slope} and provided the $\eta$--$\eta'$ decay constants and mixing angles that follow from the TFF normalizations together with the asymptotic coefficients determined via a superconvergence relation. Overall, we observed good agreement with previous results, while the highly constrained nature of our representations allows for a reduction in the final  uncertainty. In particular, our calculation, for the first time, quantifies the impact of factorization-breaking contributions generated by the leading $a_2$ left-hand cut, whose implementation we validated by studying the appropriate narrow-width limits. In combination with our previous work for the $\pi^0$, the final result
\beq
\label{amu_PS}
a_\mu^{\text{PS-poles}}=91.2^{+2.9}_{-2.4}\times 10^{-11}
\eeq
concludes a dedicated effort to determine the pseudoscalar-pole contributions to HLbL scattering from a data-driven, dispersive approach. Future applications concern improved Standard-Model predictions for leptonic decays of $\etapp$~\cite{Masjuan:2015cjl,Escribano:2015vjz,Kampf:2018wau,Messerli:2025tbd}, e.g., $\eta\to\mu^+\mu^-$~\cite{Dzhelyadin:1980kj,Abegg:1994wx} and $\eta\to2(\mu^+\mu^-)$, as recently observed for the first time by CMS~\cite{CMS:2023thf}.

While the final uncertainty in Eq.~\eqref{amu_PS} is actually dominated
by the tension between the Belle~\cite{Belle:2012wwz} and BaBar~\cite{BaBar:2009rrj} measurements of the singly-virtual $\pi^0$ TFF at large virtualities, to be clarified by future measurements at  Belle II~\cite{Belle-II:2018jsg}, also several aspects of the $\etapp$ calculation could be improved in future work. This includes additional data input, e.g., for $\eta\to\gamma\gamma$ to be measured in the JLab Primakoff program~\cite{Gan:2014pna} (addressing the inconclusive situation regarding a previous Primakoff measurement~\cite{Browman:1974sj,Rodrigues:2008zza}), the decays $\etapp\to\pi^+\pi^-\gamma$, double-differential data for $e^+e^-\to\etapp\pi^+\pi^-$~\cite{BaBar:2007qju,BaBar:2018erh}, and low-energy, singly-virtual TFF measurements~\cite{BESIII:2020nme}. Moreover, the TFFs in the high-energy, doubly-virtual direction would profit from more precise data~\cite{BaBar:2018zpn}, and, in general, the comparison to lattice-QCD calculations could help corroborate or improve the uncertainties especially for doubly-virtual kinematics. Already the current result~\eqref{amu_PS}, however, meets the precision requirements set by the final result of the Fermilab experiment, and serves as crucial input for a complete dispersive analysis of the HLbL contribution to the anomalous magnetic moment of the muon~\cite{Hoferichter:2024vbu,Hoferichter:2024bae}.


\acknowledgments
We are thankful to Judith Plenter for collaboration at early stages of this project and Pablo S\'anchez-Puertas for helpful discussions regarding Refs.~\cite{Escribano:2015yup,Masjuan:2017tvw}. Further thanks extend to Gurtej Kanwar and Urs Wenger for providing the data of Ref.~\cite{ExtendedTwistedMass:2022ofm}. Financial support by the SNSF (Project Nos.\ 200020\_200553, PCEFP2\_181117, and TMCG-2\_213690) and the DFG through the funds provided to the Sino--German Collaborative
Research Center TRR110 ``Symmetries and the Emergence of Structure in QCD''
(DFG Project-ID 196253076 -- TRR 110), as well as to the Research Unit  ``Photon--photon interactions in the Standard Model and beyond'' (Projektnummer 458854507 -- FOR 5327) is gratefully acknowledged.


\appendix

\section{Left-hand-cut contribution: Feynman rules and couplings}
\label{app:Feynman}

For both the $\eta$ and $\eta'$ case, a contribution to the phenomenological estimation of the curvature term stems from the decays of the $a_2(1320)$. Resonance Lagrangians are employed in order to extract the magnitude of these couplings. In this approach, the tensor meson fields are described by symmetric Hermitian rank-$2$ tensors and arranged in
	\begin{equation}
		\mathcal{T} =
		\begin{pmatrix}
			\frac{1}{\sqrt{6}}f_2^8+ \frac{1}{\sqrt{2}} a_2^0 & a_2^+ & K_2^{*+} \\
			a_2^- & \frac{1}{\sqrt{6}} f_2^8 - \frac{1}{\sqrt{2}} a_2^0 & K_2^{*0} \\
			K_2^{*-} & \bar{K}_2^{*0} & -\frac{2}{\sqrt{6}} f_2^8 
		\end{pmatrix}
		+ \frac{1}{\sqrt{3}} f_2^0\, \mathds{1}_3.
	\end{equation}
	Furthermore, vector mesons are introduced by
	\begin{equation}
		\mathcal{V} =
		\begin{pmatrix}
			\frac{1}{\sqrt{2}} \omega + \frac{1}{\sqrt{2}} \rho^0 & \rho^+ & K^{*+} \\
			\rho^- & \frac{1}{\sqrt{2}} \omega - \frac{1}{\sqrt{2}} \rho^0 & K^{*0} \\
			K^{*-} & \bar{K}^{*0} & \phi
		\end{pmatrix}.
	\end{equation}
	The coupling of tensor, vector, and pseudoscalar mesons is then modeled by the interaction~\cite{Giacosa:2005bw}
	\begin{equation}
    \label{Eq:Lag_TPV}
		\mathcal{L}_\text{TPV} = i c_\text{TPV} \left\langle \mathcal{T}^{ [\mu \nu] \alpha} \left[\Tilde{\mathcal{V}}_{\mu\nu},\, \partial_{\alpha} \Phi \right]\right\rangle,
	\end{equation}
	where
	\begin{equation}
		\mathcal{T}^{ [\mu \nu] \alpha} = \partial^\mu \mathcal{T}^{\nu \alpha} - \partial^{\nu} \mathcal{T}^{\mu \alpha}, \qquad \Tilde{\mathcal{V}}_{\mu\nu} = \frac{1}{2} \epsilon_{\mu\nu\alpha\beta} \left(\partial^\alpha \mathcal{V}^\beta - \partial^\beta \mathcal{V}^\alpha\right),
	\end{equation}
	and the pseudoscalar meson fields are arranged in
	\begin{equation}
		\Phi =
		\begin{pmatrix}
			\frac{1}{\sqrt{2}} \pi^0 + \frac{1}{\sqrt{6}} \eta_8 & \pi^+ & K^+ \\
			\pi^- & -\frac{1}{\sqrt{2}}\pi^0 + \frac{1}{\sqrt{6}} \eta_8 & K^0 \\
			K^- & \bar{K}^0 & - \sqrt{\frac{2}{3}}\, \eta_8
		\end{pmatrix} + \frac{1}{\sqrt{3}} \eta_0\, \mathds{1}_3.
	\end{equation}
	The relevant terms for $a_2 \to 3\pi$ read
	\begin{align}
		\mathcal{L}_\text{TPV} \supset 2\sqrt{2}\, i c_\text{TPV} \epsilon^{\mu \nu \alpha \beta} \bigg\lbrace &\partial_\mu (a_2^+)_{\nu \delta} \big[ \partial_\alpha \rho^-_{\beta}\, \partial^\delta \pi^0 - \partial_\alpha \rho^0_\beta\, \partial^\delta \pi^- \big] \notag\\
		+&\partial_\mu (a_2^0)_{\nu \delta} \big[ \partial_\alpha \rho^+_{\beta}\, \partial^\delta \pi^- - \partial_\alpha \rho^-_\beta\, \partial^\delta \pi^+ \big] \nonumber \\
		-&\partial_\mu (a_2^-)_{\nu \delta} \big[ \partial_\alpha \rho^+_{\beta}\, \partial^\delta \pi^0 - \partial_\alpha \rho^0_\beta\, \partial^\delta \pi^+ \big] \bigg\rbrace.
	\end{align}
	The neutral $\rho$ meson then couples to $a_2^\pm \pi^\mp$ via the Feynman rule
	\begin{equation}
		\label{Eq:vertTPV}
        \raisebox{-1cm}{\includegraphics[scale=1.1]{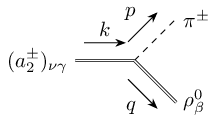}}
		= \pm 2\sqrt{2}\, i c_\text{TPV} \epsilon_{\mu \nu \alpha \beta} k^\mu q^\alpha p_\gamma.
	\end{equation}
	As a spin-$2$ particle, the polarization tensor $\epsilon_{\mu\nu}(k,\lambda)$ is associated to the $a_2(1320)$ with momentum $k$ and polarization $\lambda$. The polarization sum is given by~\cite{Ecker:2007us}
	\begin{equation}
		\sum\limits_{\lambda} \epsilon_{\mu \nu}^*(k,\lambda) \epsilon_{\alpha \beta}(k,\lambda) = P_{\mu\nu,\alpha\beta}(k),
	\end{equation}
	where
	\begin{align}
		P_{\mu\nu,\alpha\beta} &= \frac{1}{2} \left( P_{\mu \alpha} P_{\nu \beta} + P_{\mu \beta} P_{\nu \alpha}\right) - \frac{1}{3} P_{\mu \nu} P_{\alpha \beta}, \notag\\
		P_{\mu \nu}(k) &= g_{\mu \nu} - \frac{k_\mu k_\nu}{M_{a_2}^2}.
	\end{align}
	On the other hand, the coupling of the $\rho$ meson to a pion pair can be expressed as~\cite{Bando:1987br}
	\begin{equation}
		\label{Eq:LVPP}
		\mathcal{L}_\text{VPP} = \frac{i}{\sqrt{2}} c_\text{VPP} \left \langle \mathcal{V}_\mu \left[\Phi,\, \partial^\mu \Phi \right] \right\rangle.
	\end{equation}
	The interaction term for the coupling of a tensor meson to two pseudoscalars is given by
	\begin{equation}
		\label{Eq:ctppint}
		\mathcal{L}_\text{TPP} = \frac{F_\pi^2}{4} c_\text{TPP} \left\langle \mathcal{T}_{\mu\nu} \left[ \lbrace u^\mu,\, u^\nu \rbrace - 2 g^{\mu\nu} \left(u^\rho u_\rho + \chi_+\right)\right] \right\rangle,
	\end{equation}
	where the chiral field in the absence of external sources reduces to $u_\mu = i(u^\dagger \partial_\mu u - u \partial_\mu u^\dagger)$, with $u = \exp\big[i \Phi/(\sqrt{2} F_\pi)\big]$, and $\chi_\pm = u^\dagger \chi u^\dagger \pm u \chi^\dagger u$, $\chi=2 B\, \text{diag}(m_u,m_d,m_s)$. Note that for the tensor field in position space $T_{\mu \nu}(x)$, the matrix element
	\begin{equation}
		\braket{0 | g^{\mu \nu} T_{\mu \nu}(0) | T(k,\lambda)} = \epsilon_{\ \mu}^{\mu} (k,\lambda)=0
	\end{equation}
	vanishes~\cite{Ecker:2007us}. Furthermore, for diagrams with intermediate tensor mesons, the interaction terms proportional to the metric tensor in Eq.~\eqref{Eq:ctppint} would generate terms that do not propagate as a spin-$2$ field and therefore are neglected in the following. The remaining interaction term for the coupling of a tensor meson to two pseudoscalars is given by~\cite{Giacosa:2005bw,Ecker:2007us}
	\begin{equation}
    \label{Eq:Lag_TPP}
		\mathcal{L}_\text{TPP} =  c_\text{TPP} \left\langle \mathcal{T}_{\mu\nu} \partial^\mu \Phi \partial^\nu \Phi \right \rangle.
	\end{equation}
	One finds the following relation of the interaction terms between the octet and singlet pseudoscalars
	\begin{equation}
		\mathcal{L}_{a_2 \eta_0 \pi} = \sqrt{2} \mathcal{L}_{a_2 \eta_8 \pi}.
	\end{equation}
	Therefore, in a single-angle mixing scheme for the $\eta$ and $\eta'$ mesons
	\begin{equation}
		\begin{pmatrix}
			\ket{\eta}\\
			\ket{\eta'}
		\end{pmatrix}
		=
		\begin{pmatrix}
			\cos \theta & - \sin \theta \\
			\sin \theta & \cos \theta
		\end{pmatrix}
		\begin{pmatrix}
			\ket{\eta_8}\\
			\ket{\eta_0}
		\end{pmatrix}
	\end{equation}
	and assuming the mixing angle $\theta = \arcsin\left(- \frac{1}{3} \right)$, one can show that matrix elements involving $\eta$ or $\eta'$ in the asymptotic states can be reduced to
	\begin{align}
		&\bra{a_2 \pi} \left( \mathcal{L}_{a_2 \eta_8 \pi} + \mathcal{L}_{a_2 \eta_0 \pi} \right) \ket{\eta} = \sqrt{2} \braket{a_2 \pi|\mathcal{L}_{a_2 \eta_8 \pi}|\eta_8},\notag\\
		&\bra{a_2 \pi} \left( \mathcal{L}_{a_2 \eta_8 \pi} + \mathcal{L}_{a_2 \eta_0 \pi} \right) \ket{\eta'} = \braket{a_2 \pi|\mathcal{L}_{a_2 \eta_8 \pi}|\eta_8}.
	\end{align}
	The relevant Feynman rule is then given by
	\begin{equation}
    \label{Eq:Feyn_TPP}
        \raisebox{-1cm}{\includegraphics[scale=1.1]{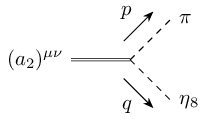}}
		= -i \sqrt{\frac{2}{3}} c_\text{TPP}\, p^\mu q^\nu.
	\end{equation}
	The decay rate $a_2 \to \etapp \pi$ follows according to
	\begin{equation}
		\Gamma[a_2 \to \etapp \pi] = \frac{g_{\etapp}|c_\text{TPP}|^2}{2880 \pi M_{a_2}^7} \lambda^{5/2}\big(M_{a_2}^2,M_{\etapp}^2,M_\pi^2\big), \qquad \text{with}\
		g_{\etapp} =
		\begin{cases}
			&2,\quad \eta \\
			&1,\quad \eta'
		\end{cases},
	\end{equation}
	where the factor $g_{\etapp}$ arises due to the ideal mixing scenario of $\eta$ and $\eta'$ that is considered here. Comparing with the experimental averages of $\Br[a_2 \to \eta \pi] = 14.5(1.2)\, \%$ and $\Br[a_2 \to \eta' \pi] = 5.5(9)\times 10^{-3}$~\cite{ParticleDataGroup:2024cfk}, gives the couplings
	$|c_{a_2 \eta \pi}| = 9.3(4)\GeV^{-1}$ and $|c_{a_2 \eta' \pi}| = 12(1)\GeV^{-1}$.  
    Note that our normalization is such that both couplings would coincide in the limit of a perfectly $U(3)$-symmetric interaction; the symmetry breaking observed here hence supports limiting the difference between $c_{\eta' 4\pi}$ and $c_{\eta 4\pi}$ (which the $\eta$ fit within the limited phase space indicates) to $30\%$ for the central results in Table~\ref{Tab:etapipig_fits}, varied between $15\%$ and $45\%$ to reflect the associated uncertainties.
    
	Furthermore, the width of the decay $\rho \to \pi \pi$ from the interaction in Eq.~\eqref{Eq:LVPP} is given by
	\begin{equation}
    \label{Eq:Gamma_rhoppipi}
		\Gamma[\rho \to \pi \pi] = \frac{|c_\text{VPP}|^2}{48 \pi} M_\rho \left(1 - \frac{4 M_\pi^2}{M_\rho^2} \right)^{3/2} .
	\end{equation}
	Together with the BW parameters $M_\rho = 775.26(23)\MeV$ and $\Gamma_\rho = 149.1(8)\MeV$ from the RPP~\cite{ParticleDataGroup:2024cfk}, the coupling strength can be estimated to be $|c_\text{VPP}|=5.98(2)$. In this case, the coupling actually comes close to the result in a definition in terms of the residue at the pole~\cite{Garcia-Martin:2011nna,Hoferichter:2017ftn,Hoferichter:2023mgy}.

    \section{Left-hand-cut contribution: cross checks of couplings}
    \label{app:cross_check}

    \subsection[$a_2 \to 3\pi$]{ $\boldsymbol{a_2 \to 3\pi}$ }

    On the level of these phenomenological Lagrangians, the matrix element for the decay $a_2^- \to \pi^- \pi^+ \pi^-$ can be written as the sum of two tree-level diagrams
	\begin{equation}
		i \mathcal{M}_{a_2} = \raisebox{-1.cm}{\includegraphics[scale=1.1]{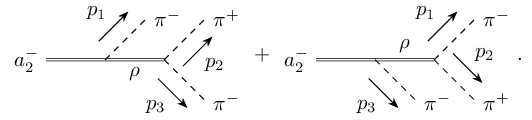}}
	\end{equation}
	Upon replacing the $\rho$ propagators by $\pi\pi$ $P$-wave Omn\`es functions,
	\begin{equation}
		\frac{M_\rho^2}{M_\rho^2 - p^2 - i M_\rho \Gamma_\rho(p^2)} \to \Omega(p^2),
	\end{equation}
	the unpolarized squared matrix element appears as 
    \begin{align}
        \label{Eq:a23pisq}
		\frac{1}{5}&\sum\limits_{\lambda} |\mathcal{M}_{a_2}|^2\\
        &=\frac{|c_\Omega|^2}{5 M_{a_2}^2} \Big[ M_{a_2}^2 (2 M_{\pi}^4+s_{12} s_{23}) - M_{a_2}^4 M_{\pi}^2 - M_{\pi}^6+3 M_{\pi}^2 s_{12} s_{23} - s_{12} s_{23} (s_{12}+s_{23}) \Big] \nonumber \\ 
        &\times \bigg\{\Omega^*(s_{12}) \bigg[M_{a_2}^4 (\Omega(s_{12})+\Omega(s_{23}))-M_{a_2}^2 \big(2 (M_{\pi}^2+s_{12}) \Omega(s_{12})+\Omega(s_{23}) (s_{12}+s_{23})\big)\nonumber\\
        &\qquad+(M_{\pi}^2-s_{12}) \big(M_{\pi}^2 (\Omega(s_{12})-\Omega(s_{23}))-\Omega(s_{12}) s_{12}+\Omega(s_{23}) s_{23}\big)\bigg]\nonumber\\
        &+\Omega^*(s_{23}) \bigg[ M_{a_2}^4 (\Omega(s_{12})+\Omega(s_{23}))-M_{a_2}^2 \big(2  (M_{\pi}^2+s_{23})\Omega(s_{23}) +\Omega(s_{12}) (s_{12}+s_{23})\big)\nonumber\\
        &\qquad-(M_{\pi}^2-s_{23}) \big(M_{\pi}^2 (\Omega(s_{12})-\Omega(s_{23}))-\Omega(s_{12}) s_{12}+\Omega(s_{23}) s_{23}\big)\bigg]\bigg\},\nonumber
    \end{align}
	where the couplings are collected in $c_\Omega = c_\text{TPV} c_\text{VPP}/M_\rho^2$ and the Mandelstam variables are defined as
	\begin{equation}
		s_{12} = (p_1 + p_2)^2,\qquad s_{13} = (p_1 + p_3)^2,\qquad
		s_{23} = (p_2 + p_3)^2.
	\end{equation}
	In order to obtain an estimate of the collective coupling $c_\Omega$, the decay width can be compared to the experimental total BW width $\Gamma_{a_2}=107(5)\MeV$ combined with the experimental branching fraction average $\Br[a_2 \to 3\pi] = 70.1(2.7)\, \%$~\cite{ParticleDataGroup:2024cfk}. For this comparison,
	an average over the initial isospin states and a sum over the final pion state configurations needs to be taken. Starting from the phenomenological Lagrangians, it can be worked out that the decay $a_2^- \to \pi^- 2 \pi^0$ is to be described by the same squared matrix element as the one in Eq.~\eqref{Eq:a23pisq}, as are the decays $a_2^+ \to \pi^+ \pi^- \pi^+ / \pi^+ 2 \pi^0$. Furthermore, a symmetry factor of $1/2$ due to two identical particles being present in the final state needs to be multiplied to the representation of the decay width $\text{d}\Gamma$. Conversely, the decay width for the decay $a_2^0 \to \pi^0 \pi^+ \pi^-$ does not obtain any symmetry factor. Note that the decay via the $\rho$ resonance to three neutral pions is forbidden by $C$-parity. Therefore, the partial decay width for $a_2\to3\pi$ can just be expressed by
	\begin{align}
		\Gamma[a_2 \to 3\pi] &= \Gamma[a_2^- \to \pi^- \pi^+ \pi^-] + \Gamma[a_2^- \to \pi^- 2 \pi^0]= \Gamma[a_2^0 \to \pi^0 \pi^+ \pi^-] \nonumber\\
		&= \frac{1}{(2\pi)^3 32 M_{a_2}^3} \int \text{d}s_{12}\,\text{d}s_{23}\, \left[ \frac{1}{5}\sum\limits_{\lambda} |\mathcal{M}_{a_2}|^2 \right],
	\end{align}
	resulting in $|c_\Omega| = 61(2)\GeV^{-4}$ when compared to the experimental value (employing the Omn\`es function generated from the IAM phase shift), which by means of $c_{\etapp 4\pi} = 2 c_{a_2 \etapp \pi} c_{\Omega}/\sqrt{3}$ would suggest $c_{\eta4\pi}^{(a_2\to 3\pi)} = 660(70)\GeV^{-5}$ and $c_{\eta'4\pi}^{(a_2\to 3\pi)} = 850(70)\GeV^{-5}$.

    \subsection[$a_2\to\pi\gamma$]{$\boldsymbol{a_2\to\pi\gamma}$}

    In a VMD approach the coupling of a photon and vector mesons arises from~\cite{Klingl:1996by}
	\begin{equation}
		\mathcal{L}_{\rho \gamma} = - \frac{\sqrt{2} e}{g_{\rho \gamma}} F^{\mu \nu} \langle {\mathcal Q} \mathcal{V}_{\mu \nu} \rangle \supset \frac{e}{g_{\rho \gamma}} \partial_\mu A_\nu \left(\partial^\mu \rho^\nu - \partial^\nu \rho^\mu \right),
	\end{equation}
	expressed in a manifestly gauge-invariant way. In combination with the interaction in Eq.~\eqref{Eq:vertTPV} the decay $a_2\to\pi\gamma$ is thereby induced,
	\begin{equation}
		i \mathcal{M}_{a_2\to\pi\gamma} = \raisebox{-0.8cm}{\includegraphics[scale=1.1]{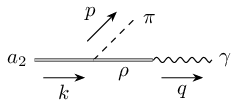}}.
	\end{equation}
	In the configuration $a_2^- \to \pi^- \gamma$, the amplitude reads
	\begin{align}
		i \mathcal{M}_{a_2\to\pi\gamma} = \epsilon^{\nu\gamma}(k)\left(-i \sqrt{8} c_\text{TPV} \epsilon_{\mu\nu\alpha\beta} k^\mu q^\alpha p_\gamma \right) \frac{i \left(g^{\beta\delta} - \frac{q^\beta q^\delta}{M_\rho^2} \right)}{q^2-M_\rho^2} i \frac{e}{g_{\rho \gamma}} \left(q^2 g_{\delta \epsilon} - q_\delta q_\epsilon \right) \epsilon^{\epsilon *}(q).
	\end{align}
	Replacing the $\rho$-propagator with the Omn\`es function, the partial decay width reads
	\begin{equation}
		\Gamma[a_2 \to \pi \gamma] = \frac{e^2 c_\text{TPV}^2}{160 \pi g_{\rho\gamma}^2} |\Omega(q^2)|^2 \frac{\lambda^{5/2}(M_{a_2}^2,M_\pi^2,q^2)}{M_{a_2}^5} \left(\frac{q^2}{M_\rho^2}\right)^2,
	\end{equation}
	which for $q^2=M_\rho^2$, $M_\rho^2 \to 0$ (on-shell limit of external photon), and $g_{\rho \gamma} = c_\text{VPP}$ reduces to
	\begin{equation}
		\Gamma[a_2 \to \pi \gamma] = \frac{e^2 c_\text{TPV}^2}{160 \pi c_\text{VPP}^2} \frac{(M_{a_2}^2-M_\pi^2)^5}{M_{a_2}^5}.
	\end{equation}
    With the experimentally determined partial decay rate $\Gamma[a_2 \to \pi \gamma]=311(25)\keV$~\cite{ParticleDataGroup:2024cfk}, this would imply $c_\text{TPV}=4.0(2)\GeV^{-2}$ or, by means of $c_{\etapp 4\pi} = 2 c_\text{TPV} c_{a_2 \etapp \pi} c_\text{VPP}/(\sqrt{3} M_\rho^2)$, the couplings $c_{\eta4\pi}^{(a_2\to \pi \gamma)} = 430(30)\GeV^{-5}$ and $c_{\eta' 4\pi}^{(a_2\to \pi \gamma)} = 550(20)\GeV^{-5}$. Accordingly, we see that the determination via $a_2 \to \pi \gamma$ tends to be better in line with the fits to the $\etapp\to\pi^+\pi^-\gamma$ spectra discussed in Sec.~\ref{sec:fits} than the one via $a_2\to 3\pi$.
    
	Moreover, comparison to Eq.~(20) of Ref.~\cite{Kubis:2015sga} suggests the matching equation
	\begin{equation}
		\frac{c_T}{F_\pi} \sim \frac{c_\text{TPV}}{c_\text{VPP}} = M_\rho^2 \frac{c_\Omega}{c_\text{VPP}^2}.
		\label{Eq:matchct}
	\end{equation}
	Numerically, this matching condition is not fulfilled very well, only up to a relative factor $1.6$ in $0.65(4)\GeV^{-2}$ vs.\ $1.03(3)\GeV^{-2}$. This mismatch likely reflects the limitations of $\rho$-dominance in $a_2\to3\pi$ due to overlapping $\rho$ bands in the Dalitz plot (cf.\ also Ref.~\cite{Stamen:2022eda}), and indeed the phenomenological determination discussed in Sec.~\ref{sec:fits} comes out closer to 
    the prediction via $a_2\to \pi\gamma$.  
    In contrast, we observe that the matching condition for the tensor-to-two-pseudoscalar-meson coupling in Ref.~\cite{Kubis:2015sga}
	\begin{equation}
		g_T' \sim \frac{F_\pi^2}{4} c_{a_2 \eta' \pi},
		\label{Eq:matchgt}
	\end{equation}
	is fulfilled much better, $25.5(2.3)\MeV$ vs.\ $25.4(2.1)\MeV$. 

    \subsection[Narrow-width approximation for $\etapp \to \pi^+ \pi^- \gamma$]{Narrow-width approximation for $\boldsymbol{\etapp \to \pi^+ \pi^- \gamma}$}

    \begin{table}[tb]
        \centering
        \renewcommand{\arraystretch}{1.3}
        \begin{tabular}{l l}
            \toprule
            Coupling & Definition \\ \midrule
             $c_\text{TPV}$ & Eq.~\eqref{Eq:Lag_TPV}, Eq.~\eqref{Eq:vertTPV}\\
             $c_\text{TPP} \equiv c_{a_2 \etapp \pi}$ & Eq.~\eqref{Eq:Lag_TPP}, Eq.~\eqref{Eq:Feyn_TPP}\\
             $c_\text{VPP}$ & Eq.~\eqref{Eq:LVPP}, Eq.~\eqref{Eq:Gamma_rhoppipi}\\
             $c_{a_2}$ & below Eq.~\eqref{Eq:Fa2_Drho}\\
             $c_{\Omega}$ & below Eq.~\eqref{Eq:a23pisq} \\
             $c_{\etapp \pi \pi \rho}^{a_2}$ & below Eq.~\eqref{Eq:G_not_int}\\
             $c_{\etapp 4 \pi}$ & below Eq.~\eqref{Eq:eta4pi_rhodecay}, Eq.~\eqref{Eq:ceta4pi}\\
             \bottomrule
        \end{tabular}
        \renewcommand{\arraystretch}{1.0}
        \caption{Overview of the various phenomenological couplings, together with their defining equation(s).}
        \label{Tab:coup_overview}
    \end{table}

    In the narrow-resonance approximation of the scalar function of the $\etapp \to \pi^+ \pi^- \gamma$ matrix element in Eq.~\eqref{Eq:etapipig_scfn}, one replaces
	\begin{equation}
		|\Omega(x)|^2 \to \pi \frac{M_\rho^3}{\Gamma_\rho} \delta(x-M_\rho^2),
	\end{equation}
	in the integrand with integration variable $x$. Using the representation $F_\pi^V(s)=(1+\alpha_\pi s)\Omega(s)$ for the pion vector form factor, the scalar function then reduces to
	\begin{align}
	\label{Fetapipirho_narrow_width}
		\mathcal{F}_{\etapp \pi\pi\gamma} (t,0) &= \frac{1}{96\pi} \Bigg\lbrace \Omega(t) \Bigg[ \frac{M_\rho^3}{\Gamma_\rho} \sigma_\pi^3(M_\rho^2) (1+\alpha_\pi M_\rho^2) \Big[ A(2+\alpha M_\rho^2) + A \alpha t  \notag\\
		&\qquad\qquad\qquad+ \frac{c_{\etapp 4 \pi}}{\pi} \Big( t^2  D^{\etapp}(t,M_\rho^2) + M_\rho^2 D^{\etapp}(M_\rho^2,t) \Big) \Big] \notag\\
		&\qquad+ \frac{c_{\etapp 4 \pi}}{\pi} \int_{4M_\pi^2}^{\infty} \diff x\, \sigma_\pi^3(x) (F_\pi^V(x))^* \hat{G}^{\etapp}(x,t) \Bigg] \nonumber \\
		&+ c_{\etapp 4 \pi}\frac{M_\rho^3}{\Gamma_\rho} \sigma_\pi^3(M_\rho^2) (1+\alpha_\pi M_\rho^2) \hat{G}^{\etapp}(t,M_\rho^2) \Bigg\rbrace,
	\end{align}
	where
	\begin{equation}
		\label{Eq:ceta4pi}
		c_{\etapp 4 \pi} = \frac{2}{\sqrt{3}M_\rho^2} c_\text{TPV} c_{a_{2}\etapp\pi} c_\text{VPP} =  \frac{2}{\sqrt{3}} c_{a_{2}\etapp\pi} c_\Omega.
	\end{equation}
	Moreover, when switching off the $a_2$ contribution, the representation~\eqref{Fetapipirho_narrow_width} becomes
	\begin{equation}
		\label{Eq:Fredred}
		\mathcal{F}_{\etapp \pi\pi\gamma} (t,0) =  \frac{1}{96\pi}  \Omega(t) \Bigg\lbrace \frac{M_\rho^3}{\Gamma_\rho} \sigma_\pi^3(M_\rho^2) (1+\alpha_\pi M_\rho^2)\, A \Big[ (2+\alpha M_\rho^2) + \alpha t \Big] \Bigg\rbrace.
	\end{equation}
	For comparison, the representation in Ref.~\cite{Kubis:2015sga} is given by
	\begin{equation}
		\label{Eq:FKP}
		\mathcal{F}_{\etapp\pi\pi\gamma}^\text{\cite{Kubis:2015sga}} (t) = \Omega(t) \left\lbrace A_\Omega (1+\alpha_\Omega t) + g_\Omega \frac{t^2}{\pi} D(t,0) \right\rbrace + g_\Omega \hat{G}(t,0).
	\end{equation}
	Comparing Eqs.~\eqref{Eq:Fredred}~and~\eqref{Eq:FKP} allows for the identification
	\begin{equation}
		\alpha_\Omega \sim \frac{\alpha}{2 + \alpha M_\rho^2} \quad \Leftrightarrow\quad \alpha \sim \frac{2\alpha_\Omega}{1-M_\rho^2 \alpha_\Omega}.
	\end{equation}
	Since $\alpha_\pi M_\rho^2 \sim 0.08\ll1$, we can neglect this correction, obtaining
	\begin{align}
		\label{Eq:matchAg}
		A_\Omega &\sim \frac{1}{96\pi} \frac{M_\rho^3}{\Gamma_\rho} \sigma_\pi^3(M_\rho^2) (1+\alpha_\pi M_\rho^2) (2+\alpha M_\rho^2) A \approx \frac{M_\rho^2}{2c_\text{VPP}^2} (2+\alpha M_\rho^2) A, \notag\\
		g_\Omega &\sim \frac{1}{96\pi} \frac{M_\rho^3}{\Gamma_\rho} \sigma_\pi^3(M_\rho^2) (1+\alpha_\pi M_\rho^2) c_{\etapp 4 \pi} \approx \frac{M_\rho^2}{2c_\text{VPP}^2} c_{\etapp 4\pi}.
	\end{align}
	Employing the matching conditions to Ref.~\cite{Kubis:2015sga}, Eqs.~\eqref{Eq:matchct}~and~\eqref{Eq:matchgt}, the combined coupling that multiplies hat function and left-hand-cut dispersive integral, see, e.g., Eq.~(32) of Ref.~\cite{Kubis:2015sga}, appears as
	\begin{equation}
		g_\Omega=\frac{4 c_T g_T}{\sqrt{3} F_\pi^3} \sim \frac{c_{a_2 \eta' \pi} c_\text{TPV}}{\sqrt{3}c_\text{VPP}} = \frac{M_\rho^2}{\sqrt{3}} \frac{c_{a_2 \eta' \pi} c_\Omega}{c_\text{VPP}^2},
	\end{equation}
	in agreement to the value of $g_\Omega$ extracted via the right-hand side of Eq.~\eqref{Eq:matchAg},
	\begin{equation}
		g_\Omega \sim \frac{M_\rho^2}{2c_\text{VPP}^2} c_{\etapp 4\pi} = \frac{ M_\rho^2}{\sqrt{3}} \frac{c_{a_2 \eta' \pi} c_\Omega}{c_\text{VPP}^2},
	\end{equation}
	via $c_{\etapp 4\pi}$ defined in Eq.~\eqref{Eq:ceta4pi}, and thereby serving as a strong consistency check on our calculation. An overview of the various couplings is given in Table~\ref{Tab:coup_overview}.
    
\begin{table}[t]
	\renewcommand{\arraystretch}{1.3}
	    \centering
	    \begin{tabular}{c c c c c}
	         \toprule
	         $s_\text{cut}\ [\GeV^2]$ & $\bar{l}_2-\bar{l}_1$ & $\hat{l}_s \ [\GeV^{-2}]$ & $\hat{l}_\pi  \ [\GeV^{-2}]$ &$\chi^2 \ / \ \text{dof}$ \\
	         \midrule
	         1 & $4.73(3)$ & $1.45(3)$ & $420(20)$ & $282/472 \approx 0.6$ \\
	         1.69 & $4.47(3)$ & $1.74(3)$ & $560(20)$ & $874/827 \approx 1.1$ \\
	         \bottomrule
	    \end{tabular}
	    \caption{Fits of the modified IAM phase to the Bern phase~\cite{Caprini:2011ky} up to two different cutoff values $s_\text{cut}$.}
	    \label{Tab:IAM_fits}
		\renewcommand{\arraystretch}{1.0}
	\end{table}    

    \section{$\boldsymbol{\pi\pi}$ $\boldsymbol{P}$-wave phase shift}\label{app:phaseshift}

   The following is presented as an addition to the phase-shift construction found in Ref.~\cite{Holz:2015tcg}.
    In $SU(2)$ ChPT, the $\Order(p^2)$ and $\Order(p^4)$ $\pi\pi \to \pi\pi$ scattering amplitudes projected onto the $P$ partial wave appear as~\cite{Gasser:1983yg,Dax:2018rvs}
    \begin{align}
    \label{Eq:appChPTamplitudes}
        t_2(s)&=\frac{s\sigma^2}{96\pi F^2},\qquad \sigma \equiv \sigma_\pi(s)=\sqrt{1-\frac{4M_\pi^2}{s}}, \notag\\
        t_4(s)&=\frac{t_{2}(s)}{48\pi^2F^2}\bigg\lbrace s \left(\bar{l}_2 - \bar{l}_1+\frac{1}{3}\right)-\frac{15}{2}M_\pi^2
         -\frac{M_\pi^4}{2s}\Big[41-2L_\sigma\big(73-25\sigma^2\big) \notag\\
                 &\qquad+3L_\sigma^2\big(5-32\sigma^2+3\sigma^4\big) \Big]\bigg\rbrace+i\sigma\,\big[t_2(s)\big]^2,
\end{align}
respectively, where 
\begin{equation}
    L_\sigma=\frac{1}{\sigma^2}\left(\frac{1}{2\sigma}\log\frac{1+\sigma}{1-\sigma}-1\right),
\end{equation}
$F$ is the pion decay constant in the chiral limit, $\bar{l}_i$ are low-energy constants (LECs), and $s$ the $\pi\pi$ invariant mass squared. The unitarized scattering amplitude  can then be written as~\cite{Dobado:1989qm,Truong:1991gv,Dobado:1992ha}
\begin{equation}
\label{Eq:appIAMamplitude}
    t_\text{IAM}(s) = \frac{\big[t_2(s)\big]^2}{t_2(s) - t_4(s)}.
\end{equation}
This form, in principle, allows for an extraction of the $\pi\pi$ $P$-wave phase shift once values for the LECs are inserted. In order to enforce the desired convergence of the phase shift to $\pi$ for $s\to \infty$, however, we work with an approximation of the two-loop amplitude~\cite{Bijnens:1997vq,Niehus:2020gmf}, and add $\Order(p^6)$ inspired terms by hand,
\begin{equation}
\label{Eq:appO6inspired}
    t_4(s) \mapsto t_4(s) + \frac{t_{2}(s)}{48\pi^2F^2}\big(\hat{l}_s s^2 + \hat{l}_\pi M_\pi^4\big),
\end{equation}
introducing two additional free parameters $\hat{l}_s$ and $\hat{l}_\pi$. Asymptotically, the corresponding phase shift behaves as
	\begin{equation}
	    \delta_1^1(s) = \pi - \frac{2}{\sqrt{1 + 4 \pi^2 \hat{l}_s^2 s^2} + 2 \pi \hat{l}_s s} + \mathcal{O}(s^{-3}).
	\end{equation}
Hence, the modified IAM phase converges with $1/s$ to $\pi$. We treat the combination of LECs $\bar{l}_2-\bar{l}_1$ as well as $\hat{l}_s$ and $\hat{l}_\pi$ as free parameters. These are then fit to the solution of the Roy equations of $\pi \pi$ scattering (``Bern phase'')~\cite{Caprini:2011ky}, while taking the value of the pion decay constant in the chiral limit from the ratio $F_\pi/F = 1.062 (7)$~\cite{FlavourLatticeAveragingGroupFLAG:2021npn,MILC:2010hzw,Borsanyi:2012zv,BMW:2013fzj,Boyle:2015exm,Beane:2011zm}. The results of the fits up to two different cutoff values are given in Table~\ref{Tab:IAM_fits}. As a consistency check, one can also consider the $\rho$ pole parameters via analytic continuation to the second Riemann sheet, see Table~\ref{Tab:rhopole_comparison}, which shows reasonable agreement with previous analyses.

\begin{table}[t]
		\renewcommand{\arraystretch}{1.3}
	    \centering
	    \begin{tabular}{l c c}
	        \toprule
	         \ & $M_\rho \ [\MeV]$  & $\Gamma_\rho \ [\MeV]$  \\ \midrule 
	         $s_\text{cut} = 1 \GeV^2$ & 758.8 & 140.0 \\
	         $s_\text{cut} = 1.69 \GeV^2$ & 759.3 & 138.7 \\ \midrule
	         Madrid (GKPY)~\cite{Garcia-Martin:2011iqs} & $763.7\, (1.6)$ & $146.4 \, (2.2)$ \\
	         Bern (Roy)~\cite{Caprini:2011ky} & $762.4 \, (1.8)$ & $145.2 \,(2.8)$ \\ \bottomrule
	    \end{tabular}
	    \caption{Comparison of the $\rho$ pole positions extracted from the modified IAM phase fits up to different cutoff values to more sophisticated analyses.}
	    \label{Tab:rhopole_comparison}
	    \renewcommand{\arraystretch}{1.0}
	\end{table}

\section{Derivation of the  $\boldsymbol{\etapp \to \pi^+ \pi^- \gamma^*}$ discontinuity}\label{app:deriv_disc_etapipig}

	By means of the unitarity condition, in our approximation, the discontinuity of the $\etapp \to \pi^+ \pi^- \gamma^*$ amplitude in the photon virtuality appears as
	\begin{align}
		&\operatorname{disc}_{k^2} \mathcal{M}\big[\etapp(q)\to\pi^+(p_1)\pi^-(p_2)\gamma^*(k)\big]=
		i (2\pi)^4 \int \text{d}\Phi_2(k;l_1,l_2)\notag\\ &\qquad\times\Big(\mathcal{M}\big[\gamma^*(k) \to \pi^+(l_1) \pi^-(l_2)\big]\Big)^*\mathcal{M}\big[\etapp \to \pi^+(p_1) \pi^-(p_2) \pi^+(l_1) \pi^-(l_2)\big],
	\end{align}
	in terms of the matrix elements
	\begin{align}
		\mathcal{M}^*\big[\gamma^*(k) \to \pi^+(l_1) \pi^-(l_2)\big] &= e \epsilon_\mu^*(k) (l_1 - l_2)^\mu  \big[F_\pi^V(k^2)\big]^*,\notag\\
		 \mathcal{M}\big[\etapp \to \pi^+(p_1) \pi^-(p_2) \pi^+(l_1) \pi^-(l_2)\big] &= \epsilon_{\nu \rho \sigma \alpha} p_1^\nu p_2^\rho l_1^\sigma l_2^\alpha f_\text{aux}(t,k^2),
	\end{align}
	with the Lorentz invariants $t=(p_1+p_2)^2$, $k^2=(l_1+l_2)^2$, and the auxiliary function
	\begin{equation}
		f_\text{aux}(t,k^2) = f_1^{\etapp}(t,k^2) \Omega(k^2) + f_1^{\etapp}(k^2,t) \Omega(t).
	\end{equation}
	The unitarity condition then appears as
	\begin{equation}
		\operatorname{disc}_{k^2} \mathcal{M}\big[\etapp(q)\to\pi^+(p_1)\pi^-(p_2)\gamma^*(k)\big] = i (2\pi)^4 e\, \big[F_\pi^{V}(k^2)\big]^* f_\text{aux}(t,k^2)\, \epsilon_\mu^*(k) p_1^\nu p_2^\rho\, P^\mu_{\ \nu \rho},
	\end{equation}
	where the phase space integral
	\begin{align}
		P^\mu_{\ \nu \rho} &= \epsilon_{\nu \rho \sigma \alpha} \int \diff\Phi_2(k;l_1,l_2)\, (l_1 - l_2)^\mu l_1^\sigma l_2^\alpha
		= \frac{1}{2} \epsilon_{\nu \rho \sigma \alpha} k^\alpha\int \diff \Phi_2(k;l_1,l_2)\, (l_1 - l_2)^\mu (l_1 - l_2)^\sigma \nonumber  \\
		&\equiv \frac{1}{2(2\pi)^6} \epsilon_{\nu \rho \sigma \alpha} k^\alpha \tilde{P}^{\mu \sigma},
	\end{align}
	needs to evaluated. The reduced integral $\tilde{P}^{\mu \sigma}$ can be written as
	\begin{align}
		\tilde{P}^{\mu \sigma} &= \int \frac{\text{d}^3l_1 \text{d}^4 l_2}{ 2 l_1^0}\, \delta(l_2^2-M_\pi^2) \theta(l_2^0) \delta^{(4)}(k-l_1-l_2)\, (l_1 - l_2)^\mu (l_1 - l_2)^\sigma \notag\\
		&= \int \frac{\text{d}^3 l}{2 (M_\pi^2 + |{\mathbf l}|^2)^{1/2}} \, \delta((k-l)^2-M_\pi^2) \theta(k^0-l^0)\, (2l-k)^\mu (2l-k)^\sigma,
	\end{align}
	with tensor decomposition
	\begin{equation}
		\tilde{P}^{\mu \sigma} = g^{\mu \sigma} S_g + \frac{k^\mu k^\sigma}{k^2} S_k.
	\end{equation}
	Contracting this equation with $k_\mu k_\sigma$ and $g_{\mu\sigma}$ gives a system of two equations that can be solved for
	\begin{equation}
		S_g = \frac{1}{3} \left(g_{\mu \sigma} - \frac{k_\mu k_\sigma}{k^2} \right) \tilde{P}^{\mu \sigma}, \qquad
		S_k = -\frac{1}{3} \left(g_{\mu \sigma} - 4\frac{k_\mu k_\sigma}{k^2} \right) \tilde{P}^{\mu \sigma}.
	\end{equation}
	In the virtual photon rest frame $\mathbf{k}=\boldsymbol{0}$, the integrands in the two integrals above assume a convenient form, and evaluate to
	\begin{equation}
		S_g = -S_k = \frac{\pi(k^2-4M_\pi^2)^{3/2}}{6\sqrt{k^2}}.
	\end{equation}
	Therefore,
	\begin{equation}
		\tilde{P}^{\mu \sigma} = \frac{\pi}{6} \left(g^{\mu \sigma} - \frac{k_\mu k_\sigma}{k^2} \right) k^2 \sigma_\pi^3(k^2),\qquad
		P_{\mu\nu\rho} = \frac{\pi}{12 (2\pi)^6} \epsilon_{\mu \nu \rho \sigma} k^\sigma.
	\end{equation}
	Expressing the matrix element for $\etapp \to \pi^+ \pi^- \gamma^*$ in terms of a scalar function
	\begin{equation}
		\mathcal{M}\big[\etapp(q)\to\pi^+(p_1)\pi^-(p_2)\gamma^*(k)\big] = e \epsilon_{\mu \nu \rho \sigma} \epsilon^{\mu *}(k) p_1^\nu p_2^\rho k^\sigma \mathcal{F}^{\etapp \pi\pi\gamma}(t,k^2),
	\end{equation}
	the unitarity relation implies
	\begin{equation}
		\operatorname{disc}_{k^2} \mathcal{F}_{\etapp  \pi\pi\gamma}(t,k^2) = \frac{i}{48 \pi} k^2 \sigma_\pi^3(k^2) \big[F_\pi^{V}(k^2)\big]^* f_\text{aux}(t,k^2),
	\end{equation}
	completing the derivation of Eq.~\eqref{etapipig_disc}.


\bibliographystyle{apsrev4-1_mod_2}
\bibliography{amu}

\end{document}